\documentclass{iopjournal}

\usepackage{mathpazo} 
\usepackage{gensymb}
\usepackage{mathptmx}
\usepackage{graphicx}
\usepackage{subfigure}
\usepackage{color}
\usepackage[T1]{fontenc}
\RequirePackage{siunitx}
\RequirePackage{xspace}%
\RequirePackage{upgreek}%
\RequirePackage{amsmath}%
\RequirePackage{amssymb}%
\RequirePackage{wasysym}%
\RequirePackage[mathcal]{euscript}
\RequirePackage{eurosym}
\RequirePackage{textcomp}
\RequirePackage{accents}

\usepackage[
    sorting=none,
    backend=biber,
    style=numeric-comp,
    giveninits=true,
    maxbibnames=10,
    minbibnames=1,
    doi=true,
    url=false,
    isbn=false
]{biblatex}

\DefineBibliographyStrings{english}{
  andothers = {\mkbibemph{et\addabbrvspace al\adddot}},
}
\DeclareNameAlias{default}{family-given}

\DeclareDelimFormat[bib]{multinamedelim}{\addcomma\space}
\DeclareDelimFormat[bib]{finalnamedelim}{\addspace and\space}

\DeclareFieldFormat{doi}{%
  \mkbibparens{doi\addcolon\space
  \href{https://doi.org/#1}{#1}}%
}
\DeclareDelimFormat[bib]{multinamedelim}{\addcomma\space}
\DeclareDelimFormat[bib]{finalnamedelim}{\addspace and\space}

\DeclareFieldFormat[article]{title}{#1}
\DeclareFieldFormat[article]{journaltitle}{\mkbibemph{#1}}
\DeclareFieldFormat[article]{shortjournal}{\mkbibemph{#1}}
\DeclareFieldFormat[article]{volume}{\mkbibbold{#1}}
\DeclareFieldFormat[article]{pages}{#1}
\DeclareFieldFormat{doi}{%
  \mkbibparens{doi\addcolon\space
  \href{https://doi.org/#1}{#1}}%
}

\DeclareFieldFormat[article]{title}{#1}
\DeclareFieldFormat[article]{journaltitle}{\mkbibemph{#1}}
\DeclareFieldFormat[article]{shortjournal}{\mkbibemph{#1}}
\DeclareFieldFormat[article]{volume}{\mkbibbold{#1}}
\DeclareFieldFormat[article]{pages}{#1}

\DeclareBibliographyDriver{article}{%
  \usebibmacro{bibindex}%
  \usebibmacro{begentry}%
  \printnames{author}%
  \setunit{\addspace}%
  \printfield{year}%
  \setunit{\addspace}%
  \printfield{title}%
  \setunit{\addspace}%
  \iffieldundef{shortjournal}
    {\printfield{journaltitle}}
    {\printfield{shortjournal}}%
  \setunit{\addspace}%
  \printfield{volume}%
  \setunit{\addspace}%
  \printfield{pages}%
  \setunit{\addspace}%
  \printfield{doi}%
  \usebibmacro{finentry}%
}

\DeclareFieldFormat[book]{title}{\mkbibemph{#1}}
\DeclareFieldFormat[book]{volume}{vol\addspace #1}

\DeclareBibliographyDriver{book}{%
  \usebibmacro{bibindex}%
  \usebibmacro{begentry}%
  \ifnameundef{author}
    {\printnames{editor}%
     \setunit{\addspace}%
     \printtext{\mkbibparens{ed}}}
    {\printnames{author}}%
  \setunit{\addspace}%
  \printfield{year}%
  \setunit{\addspace}%
  \printfield{title}%
  \setunit{\addspace}%
  \printfield{volume}%
  \setunit{\addspace}%
  \iflistundef{publisher}
    {}
    {\printtext{\mkbibparens{\printlist{publisher}}}}%
  \setunit{\addspace}%
  \printfield{doi}%
  \usebibmacro{finentry}%
}

\DeclareFieldFormat[inproceedings]{title}{#1}
\DeclareFieldFormat[inproceedings]{booktitle}{\mkbibemph{#1}}
\DeclareFieldFormat[inproceedings]{volume}{vol\addspace #1}
\DeclareFieldFormat[inproceedings]{pages}{pp\addspace #1}

\DeclareBibliographyDriver{inproceedings}{%
  \usebibmacro{bibindex}%
  \usebibmacro{begentry}%
  \printnames{author}%
  \setunit{\addspace}%
  \printfield{year}%
  \setunit{\addspace}%
  \printfield{title}%
  \setunit{\addspace}%
  \printfield{booktitle}%
  \setunit{\addspace}%
  \printlist{location}%
  \setunit{\addspace}%
  \printfield{volume}%
  \setunit{\addspace}%
  \ifnameundef{editor}
    {}
    {\printtext{ed\addspace}\printnames{editor}}%
  \setunit{\addspace}%
  \iffieldundef{publisher}
    {}
    {\printtext{\mkbibparens{\printlist{publisher}}}}%
  \setunit{\addspace}%
  \printfield{doi}%
  \usebibmacro{finentry}%
}

\DeclareFieldFormat[incollection]{title}{#1}
\DeclareFieldFormat[incollection]{booktitle}{\mkbibemph{#1}}
\DeclareFieldFormat[article]{volume}{\mkbibbold{#1}}
\DeclareFieldFormat[article]{pages}{#1}

\DeclareBibliographyDriver{incollection}{%
  \usebibmacro{bibindex}%
  \usebibmacro{begentry}%
  \printnames{author}%
  \setunit{\addspace}%
  \printfield{year}%
  \setunit{\addspace}%
  \printfield{title}%
  \setunit{\addspace}%
  \printfield{booktitle}%
  \setunit{\addspace}%
  \printfield{series}%
  \setunit{\addspace}%
  \printfield{volume}%
  \setunit{\addspace}%
  \ifnameundef{editor}
    {}
    {\printtext{ed\addspace}\printnames{editor}}%
  \setunit{\addspace}%
  \iffieldundef{publisher}
    {}
    {\printtext{\mkbibparens{\printlist{publisher}}}}%
  \setunit{\addspace}%
  \printfield{pages}%
  \setunit{\addspace}%
  \printfield{doi}%
  \usebibmacro{finentry}%
}

\usepackage{showexpl} 

\begin{document}


\title{Towards Ultra Scalability of Non-Volatile Magnetic Tunnel Junctions with a 3D Storage Layer}

\author{Nuno Caçoilo$^{1,2*}$\orcid{0000-0001-8847-2034}, Shunsuke Fukami$^{1,3,4,5,6}$\orcid{0000-0001-5750-2990}, Olivier Fruchart$^{2}$\orcid{0000-0001-7717-5229} and Ioan-Lucian Prejbeanu$^{2*}$\orcid{0000-0001-6577-032X}}

\affil{$^1$Laboratory for Nanoelectronics and Spintronics, RIEC, Tohoku University, Sendai 980-8577, Japan}

\affil{$^2$SPINTEC, Université Grenoble Alpes, CNRS, CEA, Grenoble INP, IRIG-SPINTEC, Grenoble, 38000, France}


\affil{$^3$Advanced Institute for Materials Research (WPI-AIMR), Tohoku University, Sendai 980-8577, Japan}

\affil{$^4$Center for Science and Innovation in Spintronics, Tohoku University, Sendai 980-8572, Japan}

\affil{$^5$Center for Innovative Integrated Electronic Systems, Tohoku University, Sendai 980-0845, Japan}

\affil{$^6$Inamori Research Institute for Science, Kyoto 600-8411, Japan}

\affil{$^*$Author to whom any correspondence should be addressed.}

\email{nuno.cacoilo@tohoku.ac.jp and ioan-lucian.prejbeanu@cea.fr}


\begin{abstract}
{The perpendicular Spin Transfer Torque Magnetic Random Access Memory is one of the most promising emerging non-volatile memory technologies, based on ultra-thin magnetic tunnel junctions. However, as these devices are limited by their thermal stability factor at technological nodes smaller than \SI{20}{\nano\meter}, their scalability is compromised. A possible solution to this limitation relies on taking advantage of the shape anisotropy, by increasing substantially the thickness of the storage layer. Thanks to the combination of a vertical aspect-ratio and enhanced volume, high thermal stability can be maintained at sub-\SI{10}{\nano\meter} nodes. Here, we present the technological advancements and understanding of the magnetisation reversal that led to faster switching speeds at reduced switching voltage, providing a viable approach for dense arrays of ultra-small magnetic tunnel junctions.}
\end{abstract}

\tableofcontents

\newpage

\newcommand{\HK}{\mu_0 H^\textrm{eff}_\textrm{K}}
\newcommand{\HKK}{\textbf{H}_\textrm{K}}
\newcommand{\td}{$\tau_\textrm{D}$}
\newcommand{\kbt}{\textit{k}_{\textrm{B}}\textit{T}}
\newcommand{\RA}{R\times A}
\newcommand{\Rp}{R_\textrm{P}}
\newcommand{\Rap}{R_\textrm{AP}}
\newcommand{\kv}{K_\textrm{V}}
\newcommand{\ks}{k_\textrm{s}}
\newcommand{\Ms}{M_\textrm{s}}
\newcommand{\eunix}{\epsilon_\textrm{uniaxial}}
\newcommand{\edemag}{\epsilon_\textrm{demag}}
\newcommand{\einter}{\epsilon_\textrm{interfacial}}
\newcommand{\eex}{\epsilon_\textrm{exchange}}
\newcommand{\esys}{\epsilon_\textrm{system}}
\newcommand{\Heff}{\textbf{H}_\textrm{eff}}
\newcommand{\Hext}{\textbf{H}_\textrm{ext}}
\newcommand{\Hdemag}{\textbf{H}_\textrm{demag}}
\renewcommand{\hbar}{\mathchar'26\mkern-9mu h}
\newcommand{\eb}{E_\textrm{B}}
\newcommand{\Hstray}{\textbf{H}_\textrm{stray}}

\section{Introduction: The need for non-volatility for computing}
\label{section:introduction}
\begin{figure}[h!]
\centering
\includegraphics{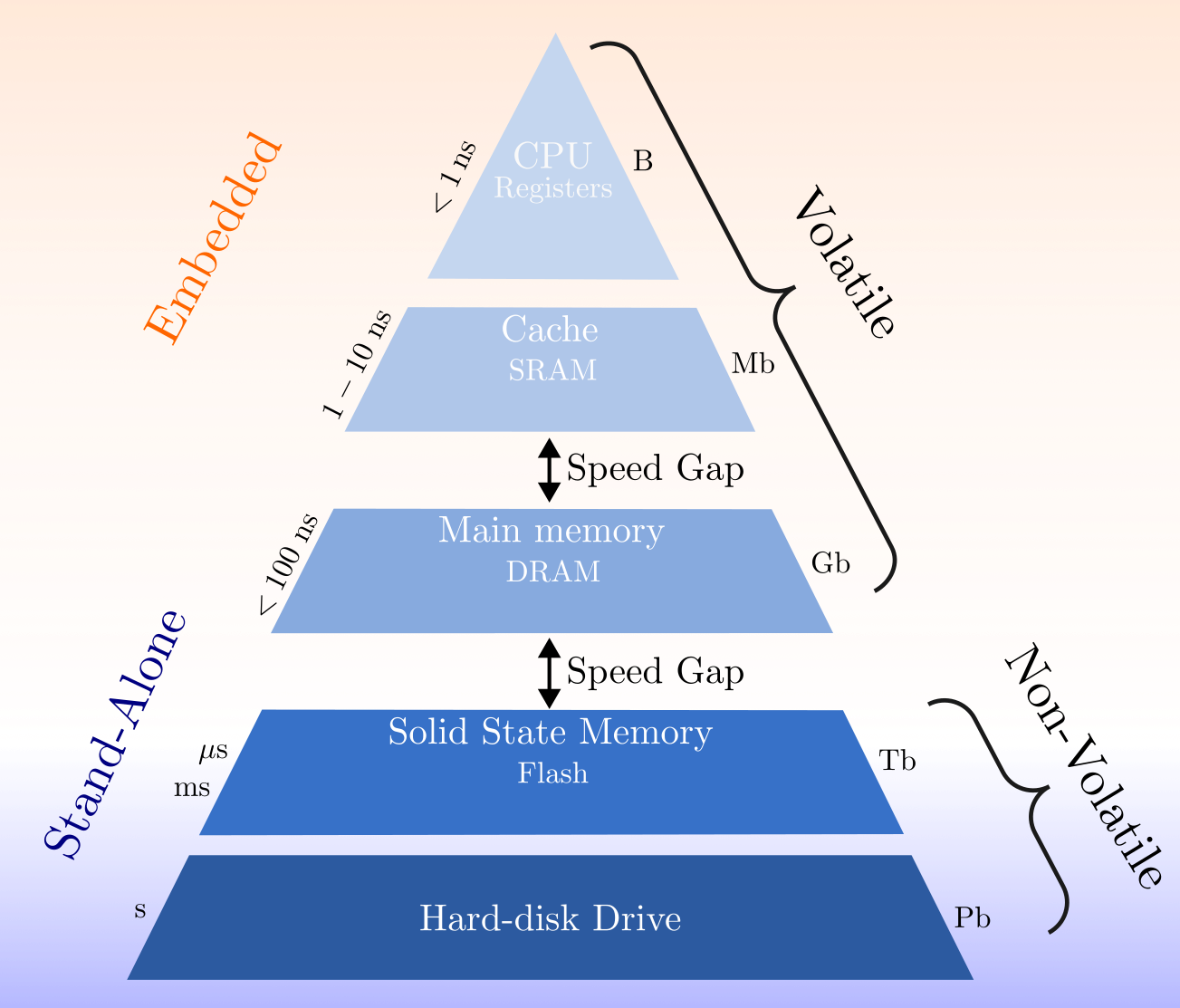}
\caption{Common memory hierarchy in which the operation speed and density are evaluated.}
\label{fig:ch1-figure 1}
\end{figure}

The need for an improvement in data handling has been growing exponentially during the last decade. Nowadays the memory market has needs for higher storage capacity, data treatment speed and reduced energy consumption. To address these requirements, it is necessary to improve certain portions of the so-called memory hierarchy, depicted in figure \ref{fig:ch1-figure 1}. 

The memory hierarchy is commonly represented in a pyramidal shape, where at the bottom the density is large but the operation speed is relatively slow. As we go up in hierarchy the memory density is reduced, but the operation speed increases. Additionally, each memory type can be categorized as stand-alone (an external component to the main unit) or embedded (part of the main unit). Both categories differ in operation speed and density. The stand-alone category offers high density, at the expense of a lower processing speed. Large parts of FLASH and DRAM (dynamic-RAM) fall in this category. These are different in the sense that the FLASH is \textbf{non-volatile}, as it retains its information while turned off, and the DRAM is \textbf{volatile}, so constant power needs to be supplied. The SRAM (Static RAM) is a volatile memory with faster operation speed than the DRAM, although not as dense, falling in the embedded branch. Additionally, there is a large number of applications that make use of embedded DRAM (e-DRAM) and embedded FLASH (e-FLASH).  Combining writing speed and density, a considerable gap between both FLASH and DRAM, and between the DRAM and SRAM appears. This opens avenues for \textbf{emerging}, non-volatile, memory devices.



Nowadays different emerging non-volatile memories tackle the DRAM and SRAM replacement, with superior density and endurance than FLASH. The promising candidates include phase-change RAM (PCRAM), resistive RAM (ReRAM), ferroelectric RAM (FeRAM) and magnetic RAM (MRAM). Although the physical writing mechanisms in these non-volatile memories are different, they all share a similar aspect: after a certain external input, the device shows a measurable change in resistance, from which one can read a deterministic state 1 or 0. Although none of these non-volatile memories shows a global optimal set of parameters, for certain applications, they can replace specific memory devices. An example is the MRAM technology, with potential for NOR-FLASH, e-FLASH and SRAM replacement in the near future, and possibly DRAM in the far future (due to its industrial maturity), thanks to its high endurance, low writing current/voltage, scalability, and CMOS compatibility. Nowadays with several players in the market, the MRAM draws attention as its properties become more competitive with conventional memory technologies. 


\subsection{The Magnetic Random Access Memory}

The building block of the MRAM is the magnetic tunnel junction (MTJ), one of the most relevant structures in spintronics, whose main characteristic is the interplay of charge and spin of the flowing electrons with the magnetisation of the magnetic layer. This structure is composed, in its simplest implementation, of two ferromagnetic layers spaced by a thin insulating layer. While one of the ferromagnetic layers is designed to remain with a fixed magnetisation (fixed layer or reference layer), the other ferromagnetic layer has a magnetisation that can be switched, hence called free layer (or storage layer). The insulating layer placed between these two ferromagnets is called tunnel barrier since, when thin enough (usually below 2 nm), it allows the tunnelling of electrons \cite{apalkov_magnetoresistive_2016}. The combination of these layers allows to perform the required operations of a non-volatile memory device: readout of the deterministic binary state 1 and 0, storage of the information for a certain amount of time (application dependent) and the writing of the binary state.


\begin{figure}[h!]
\centering
\includegraphics{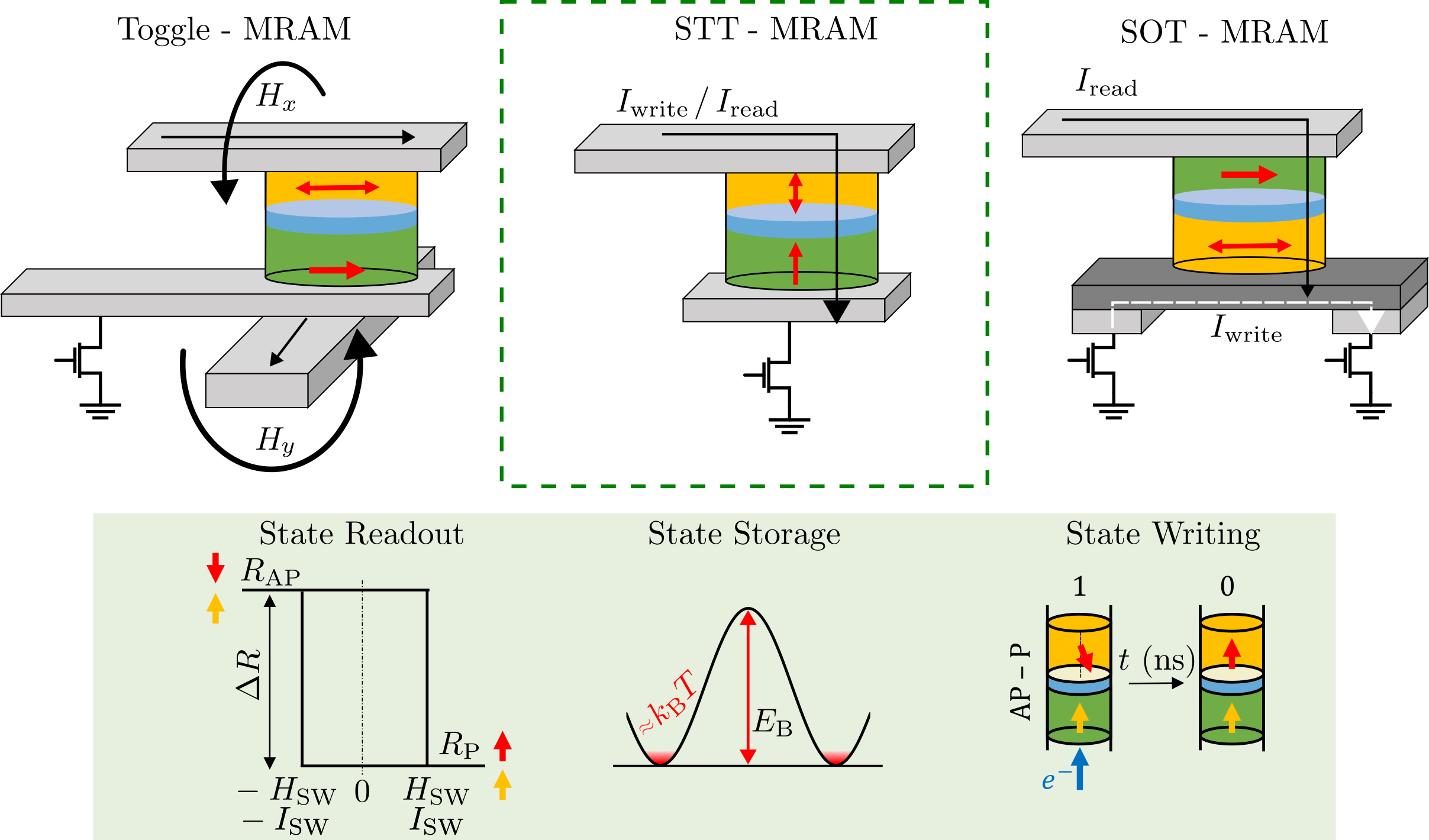}
\caption{Schematics of different families of MRAM designs. In all situations the reference layer is shown with green, while the storage layer is shown with yellow.}
\label{fig:ch1-figure 2}
\end{figure}

During the last years, to meet the increasing demands of the microelectronic industry, the MRAM expanded into different branches (examples are shown in Figure \ref{fig:ch1-figure 2}), targeting specific sectors of the memory hierarchy. These differ in architectures, magnetic properties and writing mechanisms. In a first generation, the writing process used a magnetic field generated by crossed write lines superimposed to the cell to reverse the magnetic state (Toggle-MRAM). However, this has constraints in scalability, as the current density increases to maintain the same magnitude of magnetic field at smaller nodes \cite{apalkov_magnetoresistive_2016}. A second generation makes use of the spin polarised current to switch the magnetisation of the storage layer. This uses the spin transfer torque (STT) effect, bringing scalability and low power consumption at increased bit densities, as the current to write scales down with the device dimension \cite{locatelli_spin-torque_2014}, and so does decrease the requirement of associated transistors. Inside this MRAM branch, there has been a significant improvement during the last years. Firstly, the magnetisation of both storage and reference layers were in the plane of the device. In that case, significant storage capability was obtained using a patterned elliptical shape and thicker storage layer ($\approx$ 1.5 - \SI{2}{nm}). Later, it was shown that the use of a perpendicular orientation of the magnetisation allows for smaller switching current, lower cell variations, alongside larger storage capability at lower technological nodes \cite{apalkov_magnetoresistive_2016}. These counterparts are labelled as in-plane and perpendicular STT-MRAM. The perpendicular STT-MRAM is a somewhat mature technology, with several players on the market \cite{dieny_opportunities_2020, endoh_recent_2020, choe_recent_2023}. Additionally, other memory architectures have been under development in the last years. Emphasis is given to spin orbit torque (SOT) MRAM, which shows promise for SRAM replacement thanks to its high-speed magnetisation reversal and possible unlimited endurance, as the writing and reading paths are separated and thus do not compromise the endurance of the MTJ by breakdown due to the large current required for writing \cite{krizakova_spin-orbit_2022}. Although these devices are inherently different, they all obey the necessity of a state readout, storage and writing. Moving forward, the focus will be given to the STT-MRAM branch, and a first introduction to its characteristics is shown below.

\subsubsection{State readout: Tunnel Magnetoresistance}

A common trait between the different emerging non-volatile memories is that, after a certain external input, there is a measurable change in the resistance of the device. For the case of the magnetic tunnel junction, its resistance is associated with the relative orientation of the two ferromagnetic layers. When the magnetisation of these layers is parallel to each other, the device has a low resistance state ($\Rp$) and, if anti-parallel, the device has a high resistance state ($\Rap$) (Figure \ref{fig:ch1-figure 2}). The difference in resistance state defines the tunnel magnetoresistance (TMR):

\begin{equation}
\textrm{TMR} = \frac{\Rap-\Rp}{\Rp}.
\end{equation}

The readout of the magnetic state is easier for large TMR values. Several improvements in the quality of interfaces and materials have contributed to the increase of TMR through the last years. Namely, the crystallisation of the layers may contribute to the filtering of spin polarised electrons based on their orbital symmetry, increasing the TMR. Theoretical calculations predict TMR ratios has large as 1600\% for ideal MgO-based MTJs \cite{butler_spin-dependent_2001, mathon_theory_2001}, while values exceeding 600\% have been experimentally demonstrated in devices at room temperature \cite{ikeda_tunnel_2008, scheike_631_2023}. In technologically relevant devices, TMR ratios above 200\% can be achieved, which provides a sufficient readout margin.

\subsubsection{Non-volatile storage}

In addition to the necessary high TMR, the memory device needs to retain its information for a certain amount of time. Shorter retention times (from seconds to hours) may be sufficient for SRAM- or DRAM-like applications, whereas several years are necessary for embedded non-volatile memory and FLASH-like storage applications. The data retention in an MRAM cell relies on the magnetic properties of the storage layer, characterised by a thermal stability factor $\Delta$. This parameter is given by the ratio between the energy barrier ($\eb$) it needs to overcome to flip the magnetisation from one state to the other, and the thermal activation energy $\kbt$ (at a working temperature \textit{T}) \cite{brown_thermal_1963} 

\begin{equation}
    \Delta = \frac{\eb}{\kbt}.
    \label{eq:ch1:delta}
\end{equation}

The energy barrier originates from magnetic anisotropy, which defines the preferred orientation of the magnetisation. In the storage layer of an MTJ, the total density of magnetic anisotropy energy contains the contribution of different terms, including the volumic uniaxial anisotropy ($\epsilon_\textrm{uniaxial}$), the interfacial anisotropy ($\epsilon_\textrm{interfacial}$) and the volumic shape anisotropy, which is related to the demagnetising field ($\epsilon_\textrm{demag}$). These three terms are listed sequentially on the right-hand side of the following equation:

\begin{equation}
    \varepsilon_\text{system} = \left[- K_\text{V} -\frac{k_{\text{s}}}{L} + \frac{1}{2}\mu_0M_\text{s}^2\mathcal{N}\right](\hat{n} \cdot \hat{m})^2,
    \label{ch1:eq:e_system}
\end{equation}
with a volume anisotropy constant $\kv$ [J/m$^3$], surface anisotropy constant $k_\mathrm{s}$ [J/m$^2$], demagnetising tensor $\mathcal{N}$, spontaneous magnetisation $M_\mathrm{s}$ and thickness $L$. Here, different energy terms are related with the system easy axis through $(\hat{n}\cdot\hat{m})^2$, with $\hat{n}$ the unitary vector of the easy axis direction and $\hat{m}$ the unitary vector defining the direction of the magnetisation.

The interfacial, or surface anisotropy, is the primary source of perpendicular magnetic anisotropy in the current generation of commercial perpendicular STT-MRAM. It originates at the interface of a non-magnetic material and a ferromagnet, inducing a preferred perpendicular orientation for the magnetisation when the magnetic layer is thin enough. Examples of such are Co/Pt \cite{mangin_current-induced_2006, nistor_ptcooxide_2009, mizunuma_mgo_2009} and FeCo(B)/MgO \cite{ikeda_perpendicular-anisotropy_2010, worledge_spin_2011,yang_first-principles_2011, baumann_origin_2015}. The latter played a central role in the development of perpendicular STT-MRAM, enabling improved scalability and higher thermal stability at reduced switching currents. This anisotropy can be also interpreted as a uniaxial anisotropy source (like $\kv$), as the perpendicular orientation is favoured. Nonetheless, $K_\textrm{V}$ usually addresses bulk-like anisotropies, such as magnetocrystalline, in which the crystal lattice of the material prefers different magnetisation orientations. This is the case, for instance, in FePt and FePd L1$_0$ phase alloys \cite{de_person_magnetic_2007, barabash_first-principles_2009}.

Interfacial and bulk anisotropies tend to align the magnetisation perpendicular to the film plane. These contributions compete with the demagnetising energy, which is related to the geometry of the ferromagnet (described by the unitary demagnetising tensor $\mathcal{N}$) and tends to promote magnetic configurations that reduce resulting stray fields. However, the short-range exchange interaction ($\propto \epsilon_\textrm{exchange}$) penalises spatial variations of the magnetisation by favouring the parallel alignment of neighbouring magnetic moments. The competition between these different energy contributions determines whether the magnetisation remains uniform or forms non-uniform configurations. 

Within the macrospin approximation, the exchange interaction is considered to be strong enough to maintain a spatially uniform magnetisation. The magnetic moments will respond coherently to an external stimulus and behave collectively as a single magnetic moment. Because no spatial variation of the magnetisation is considered, the exchange energy can be omitted from the energy barrier calculation. For a macrospin approximation at a zero applied magnetic field, the energy barrier is then, from Equation \ref{ch1:eq:e_system},

\begin{equation}
    E_\textrm{B} = \delta\varepsilon_\textrm{system} = K_\textrm{eff}V,
\end{equation}
and the thermal stability introduced in Equation \ref{eq:ch1:delta} becomes 

\begin{equation}
    \Delta = \frac{K_\textrm{eff}V}{\kbt}.
    \label{eq:ch1:delta_keff}
\end{equation}

Here, $K_\textrm{eff}$ is an effective anisotropy energy density from the different anisotropy contributions, and \textit{V} the volume of the ferromagnet. This macrospin approximation remains valid only while the formation of a non-uniform magnetic configuration is energetically unfavourable, which is the case below a certain critical diameter $d_\textrm{c}$. This critical diameter can be estimated from the exchange stiffness $A_\textrm{ex}$ [J/m] and the effective anisotropy density energy $K_\textrm{eff}$ [J/m$^3$] of the storage layer, as \cite{chaves-oflynn_thermal_2015, bouquin_size_2018}

\begin{equation}
    d_\textrm{c} = \frac{16}{\pi}\sqrt{\frac{A_\text{ex}}{K_\text{eff}}}.
    \label{eq:critical-diameter}
\end{equation}

Above this diameter, the magnetisation reversal occurs in practice with the creation of a magnetic domain wall, limiting the stability of the device to a fraction of that expected for a macrospin \cite{sato_properties_2014}. 

\subsubsection{State writing: Spin Transfer Torque}
Before presenting the effect of the spin transfer torque in the magnetisation dynamics, it is necessary to understand that, as any other piece of magnetic material, the storage layer feels an internal effective magnetic field $\Heff$. This effective field should not be confused with the effective anisotropy energy $K_\textrm{eff}$ introduced previously. While $K_\textrm{eff}$ characterises the energy barrier at zero applied field resulting from the combined anisotropy and demagnetising contributions, $\Heff$ contains all internal and external magnetic-field contributions governing the magnetisation dynamics. These include externally applied magnetic fields $\Hext$, the intrinsic anisotropy field \textbf{H}$_\textrm{K}$, the self-demagnetising field $\Hdemag$ and stray fields \textbf{H}$_\textrm{stray}$ generated by adjacent uncompensated magnetic layers:

\begin{equation}
    \Heff = \Hext + \HKK + \Hdemag + \Hstray.
    \label{ch1:eq:heff}
\end{equation}

The motion of the storage-layer magnetisation, modelled within the macrospin approximation, can be described using either the Landau--Lifshitz (LL) or the Landau--Lifshitz--Gilbert (LLG) formulation. The two formulations are mathematically equivalent after an appropriate transformation of the gyromagnetic ratio and damping parameter. However, when additional current-induced torques are included, conversion between the two forms also transforms their damping-like and field-like components \cite{thiaville_micromagnetic_2005}. Throughout this review, we adopt the Gilbert convention, for which the dynamics of the normalised storage-layer magnetisation is written as
\begin{equation}
\frac{\mathrm{d}\mathbf{m}_\mathrm{SL}}{\mathrm{d}t}
=
\underbrace{-|\gamma|\mu_0\mathbf{m}_\mathrm{SL}\times\mathbf{H}_\mathrm{eff}}_{\Gamma_\textrm{precession}}
+
\underbrace{\alpha\mathbf{m}_\mathrm{SL}\times\frac{\mathrm{d}\mathbf{m}_\mathrm{SL}}{\mathrm{d}t}}_{\Gamma_\textrm{damping}},
\label{ch1:eq:LLG}
\end{equation}
where $\mathbf{m}_\mathrm{SL}=\mathbf{M}_\mathrm{SL}/M_\mathrm{s}$ is the normalised magnetisation vector of the storage layer, $\alpha$ is the dimensionless Gilbert damping parameter, $|\gamma|$ is the magnitude of the gyromagnetic ratio and $\mu_0$ is the vacuum permeability \cite{apalkov_magnetoresistive_2016}. The first term $\Gamma_\textrm{precession}$ describes the energy-conserving precession of the magnetisation around $\mathbf{H}_\mathrm{eff}$, whereas the Gilbert damping term $\Gamma_\textrm{damping}$ dissipates energy and progressively aligns the magnetisation with $\mathbf{H}_\mathrm{eff}$.

\begin{figure}[h]
\centering
\includegraphics{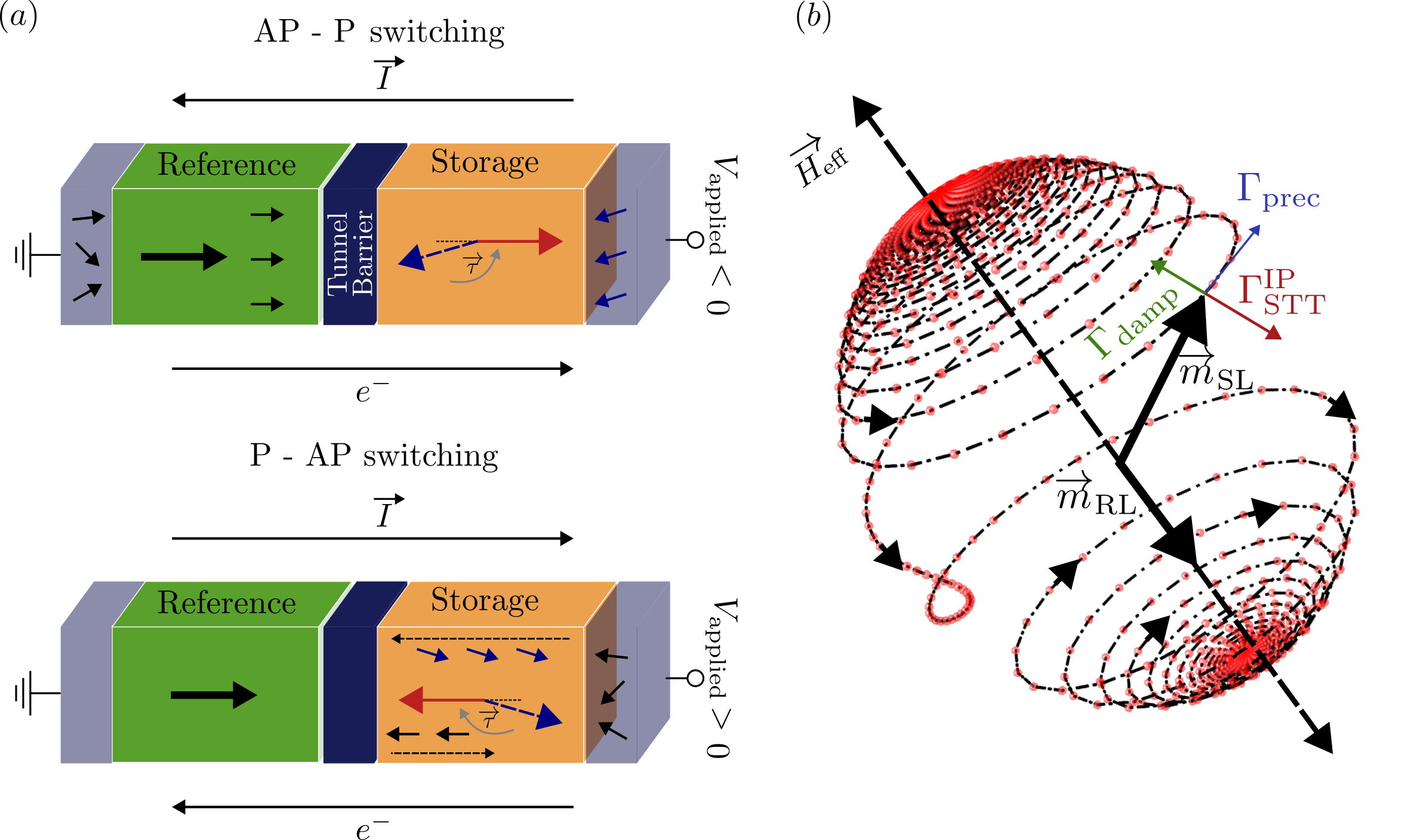}
\caption{(\textit{a}) Schematics of the spin-transfer-torque dynamics in a MTJ for an AP - P and P - AP switching. The initial and final orientations of the magnetisation in the storage layer are respectively blue and red. (\textit{b}) Trajectory of magnetisation switching induced by current described by the LLGS equation.}
\label{fig:ch1-figure 3}
\end{figure}

Current-induced magnetisation switching was first predicted independently by Slonczewski \cite{slonczewski_current-driven_1996} and Berger \cite{berger_emission_1996}. The general mechanism in a perpendicular MTJ is illustrated in Figure \ref{fig:ch1-figure 3} (\textit{a}) (top panel) for antiparallel (AP) to parallel (P) switching and (bottom panel) for P-AP switching. In this conceptual example, the MTJ is simplified to a tunnel barrier sandwiched between the reference and storage layers. In a perfectly parallel or anti-parallel configuration, the two magnetisations are collinear and there is no spin-transfer torque effect. At finite temperature, however, thermal fluctuations continuously misalign the storage and reference layers magnetisations from their equilibrium orientation, creating the small initial misalignment angle represented in figure \ref{fig:ch1-figure 3} ($a$). This initial misalignment provides the transverse component required to initiate the reversal dynamics. For AP-P switching, electrons flow from the reference layer towards the storage layer and become spin-polarised by the reference layer magnetisation. After tunnelling through the tunnel barrier, the component of their spin angular momentum transverse to the storage layer magnetisation is transferred. Conservation of angular momentum gives rise to a spin-transfer torque acting on the storage layer magnetisation. When this torque is large enough to overcome the damping, it drives the magnetisation from the AP to the P state. For the situation of P-AP switching, on the other hand, there is no torque exerted on the storage layer for an initial P configuration and electrons flowing from the reference layer, as both layers have similar magnetisation orientation. Thus, to achieve a P-AP switching, it is necessary to reverse the current/electron flow. At the tunnel barrier, the electrons polarised by the storage layer and with spin opposite to those of the reference layer have a higher probability of being backscattered, thereby accumulating in the storage layer and exerting a torque on its magnetisation, leading to its reversal.

The spin-transfer-torque contributions can be added to Equation~\ref{ch1:eq:LLG}, resulting in the Landau--Lifshitz--Gilbert--Slonczewski (LLGS) equation:
\begin{equation}
\frac{\mathrm{d}\mathbf{m}_\mathrm{SL}}{\mathrm{d}t}
= -|\gamma|\mu_0\mathbf{m}_\mathrm{SL}\times\mathbf{H}_\mathrm{eff}
+
\alpha\mathbf{m}_\mathrm{SL}\times\frac{\mathrm{d}\mathbf{m}_\mathrm{SL}}{\mathrm{d}t}
+
\Gamma_\mathrm{STT}^{\mathrm{IP}}
+
\Gamma_\mathrm{STT}^{\mathrm{OOP}} .
\label{ch1:eq:LLGS_general}
\end{equation}
In the Gilbert convention, the in-plane and out-of-plane Slonczewski torques are written as
\begin{equation}
\begin{aligned}
\Gamma_\mathrm{STT}^{\mathrm{IP}}
&=
-|\gamma|\eta_\mathrm{STT}\frac{\hbar I}{2|e|M_\mathrm{s}V}\mathbf{m}_\mathrm{SL}\times
\left(
\mathbf{m}_\mathrm{SL}
\times
\mathbf{m}_\mathrm{RL}
\right),\\
\Gamma_\mathrm{STT}^{\mathrm{OOP}}
&=
|\gamma|\eta'_\mathrm{STT}\frac{\hbar I}{2|e|M_\mathrm{s}V}\mathbf{m}_\mathrm{SL}\times\mathbf{m}_\mathrm{RL}.
\end{aligned}
\label{ch1:eq:gamma_stt}
\end{equation}
Here, $I$ is the applied current, $V$ is the storage-layer volume, $|e|$ is the magnitude of the electron charge, $\hbar$ is the reduced Planck constant, and $\mathbf{m}_\mathrm{RL}$ is the unit vector defining the reference-layer magnetisation. The coefficients $\eta_\mathrm{STT}$ and $\eta'_\mathrm{STT}$ describe the efficiencies of the in-plane and out-of-plane torque components, respectively. For the in-plane component, its efficiency is related to the TMR ratio through \cite{sun_spin-torque_2013}
\begin{equation}
\eta_\mathrm{STT} = \sqrt{\frac{\mathrm{TMR}\left(\mathrm{TMR}+2\right)}{2\left(\mathrm{TMR}+1\right)}}\;.
\end{equation}

Although the LL and LLG formulations are mathematically equivalent, converting the LLGS equation into an explicit LL form transforms the torque coefficients and mixes the damping-like and field-like components through terms proportional to $\alpha$ \cite{thiaville_micromagnetic_2005}. Maintaining the Gilbert form ensures consistency with the micromagnetic formulation used in \ref{section-micromagnetics}. The in-plane torque $\Gamma_\mathrm{STT}^{\mathrm{IP}}$ is the principal contribution responsible for STT-driven switching. It lies in the plane defined by $\mathbf{m}_\mathrm{SL}$ and $\mathbf{m}_\mathrm{RL}$ and describes the transfer of spin angular momentum to the storage-layer magnetisation. It is commonly referred to as the damping-like torque because it has the same vector symmetry as the Gilbert damping term $\Gamma_\textrm{damping}$. Depending on the current polarity, it drives $\mathbf{m}_\mathrm{SL}$ towards (or away) from the spin-polarisation direction imposed by $\mathbf{m}_\mathrm{RL}$, favouring either the parallel or antiparallel configuration. The out-of-plane torque $\Gamma_\mathrm{STT}^{\mathrm{OOP}}$ is commonly referred to as the field-like torque because it induces precession of $\mathbf{m}_\mathrm{SL}$ around $\mathbf{m}_\mathrm{RL}$. Its magnitude is generally smaller than that of the damping-like torque and, although it can modify the precessional trajectory of the magnetisation, it generally only has a limited influence on the switching time. It is therefore commonly neglected when modelling STT-driven perpendicular MTJs \cite{ralph_spin_2008, oh_bias-voltage_2009, timopheev_respective_2015}. 

Neglecting the field-like contribution, the LLGS equation can be reduced to
\begin{equation}
\begin{aligned}
\frac{\mathrm{d}\mathbf{m}_\mathrm{SL}}{\mathrm{d}t}
={}&
-|\gamma|\mu_0
\mathbf{m}_\mathrm{SL}
\times
\mathbf{H}_\mathrm{eff}
+
\alpha
\mathbf{m}_\mathrm{SL}
\times
\frac{\mathrm{d}\mathbf{m}_\mathrm{SL}}{\mathrm{d}t}\\
&-
|\gamma|\eta_\mathrm{STT}
\frac{\hbar I}{2|e|M_\mathrm{s}V}
\mathbf{m}_\mathrm{SL}
\times
\left(
\mathbf{m}_\mathrm{SL}
\times
\mathbf{m}_\mathrm{RL}
\right).
\end{aligned}
\label{ch1:eq:LLGs}
\end{equation}

Depending on the current direction, the damping-like torque may either reinforce or oppose the intrinsic damping. When it opposes and exceeds the intrinsic damping, the initial deviation of the storage-layer magnetisation is amplified, increasing the precession amplitude until magnetisation switching occurs, as illustrated in Figure~\ref{fig:ch1-figure 3} (\textit{b}). Within the macrospin approximation, switching in a perpendicular MTJ occurs above a critical current $I_\mathrm{c0}$ \cite{apalkov_magnetoresistive_2016}

\begin{equation}
I_{\text{c}0} =  \frac{4|e|}{\hbar}\frac{\alpha \eb}{\eta_\textrm{STT}}\;.
\label{eq:ico_expanded}
\end{equation}


\subsection{Lifting the stability constraints of ultra-small MTJ}

\begin{figure}[h]
    \centering
    \includegraphics{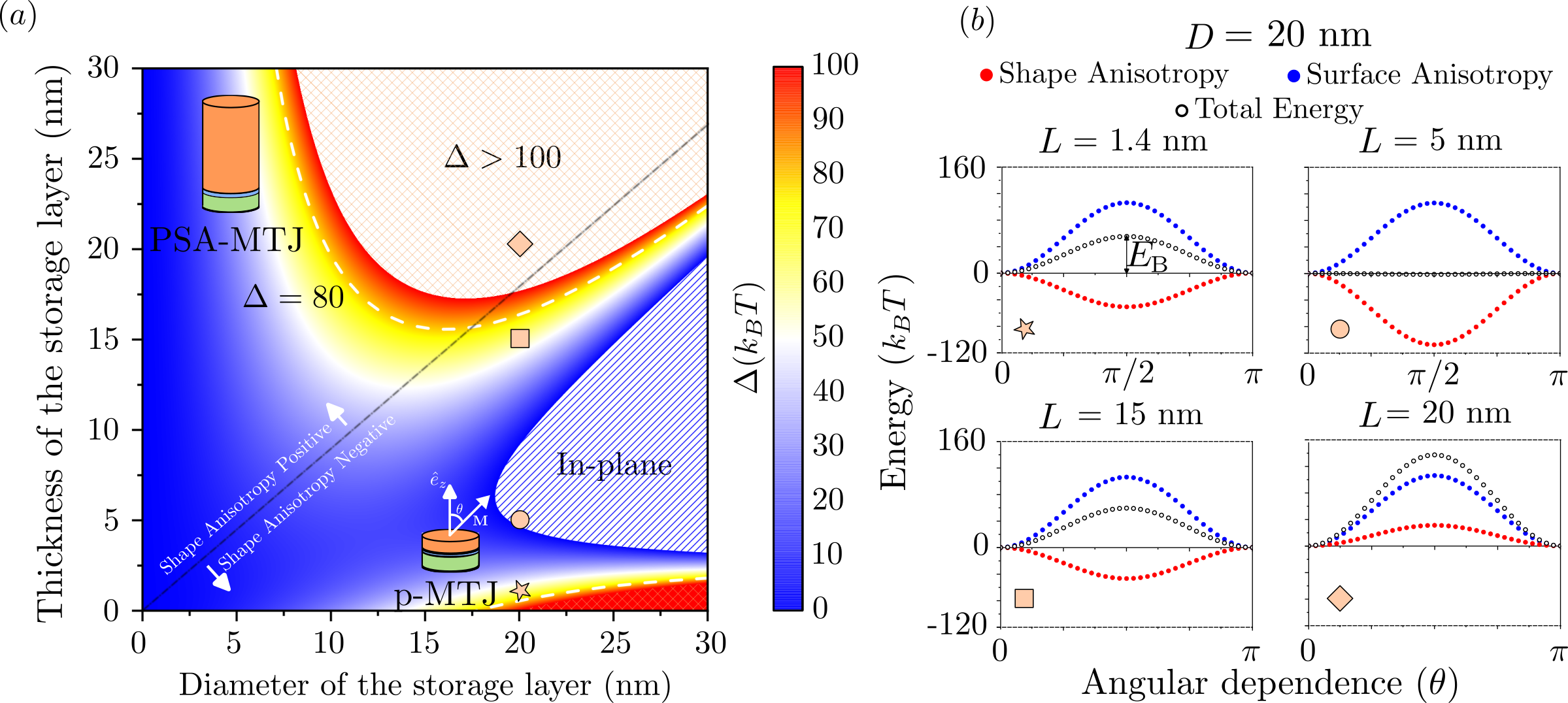}
    \caption{(\textit{a}) Stability diagram for different dimensions of the storage layer for a single FeCo(B)/MgO interface with a surface anisotropy constant of \SI{1.4}{\milli\joule\per\meter\squared} for an  interfacial thickness of \SI{1.4}{\nano\meter} with a spontaneous magnetisation $\Ms$ =  \SI{1}{\mega\ampere\per\meter}. The dashed gray line emphasises the perpendicular and in-plane nature of the shape anisotropy, perpendicular for an aspect ratio above $\tau_\textrm{AR}>0.89$. The orange pattern shows a region with an energy barrier greater than 100 $\kbt$ and the blue pattern shows the region where the magnetisation has a preferential in-plane orientation. Orientation of magnetisation shown for the case of the p-MTJ (perpendicular for $\theta = 0\degree$ and in-plane for $\theta = \pi/2$). Reprinted figure with permission from \cite{cacoilo_dipole-coupled_2024}. Copyright 2024 by the American Physical Society. (\textit{b}) Evolution of the different contributions to the total energy of the system for a \SI{20}{\nano\meter} diameter and increasing thickness as a function of the angle between the magnetisation and the normal to the plane: surface anisotropy (blue circles), shape anisotropy (red circles) and total energy (open black circles) for increasing thickness of \SI{1.4}{\nano\meter}, \SI{5}{\nano\meter}, \SI{15}{\nano\meter} and \SI{20}{\nano\meter}.}
    \label{fig:ch1-figure 4}
\end{figure}

Although the STT-MRAM is currently the leading MRAM-branch for industrial applications, it still suffers from some major challenges, mainly when its MTJ lateral dimension goes to sub-\SI{20}{nm} dimensions. As the device diameter shrinks, its thermal stability $\Delta$ reduces and, consequently, its retention time \cite{thomas_perpendicular_2014, sato_magnetic_2017, yoshida_size_2019}. This reduction can be understood considering the energy of the system present in Equation \ref{ch1:eq:e_system} , where the normal vector states that the energy is minimised for a perpendicular magnetisation [system axis are shown in Figure \ref{fig:ch1-figure 4} (\textit{a})]:

\begin{equation}
    E(\theta) = V \left[-\frac{1}{2}\mu_0M_\textrm{s}^2\left(\frac{3\mathcal{N}_{zz} -1 }{2}\right) + \frac{k_\textrm{s}}{L} \right]\sin^2{\theta}.
    \label{eq:ch2 - EB}
\end{equation}

In a conventional flat MTJ, the storage layer is approximately between 1.1--\SI{1.4}{\nano\meter} thick, with diameters typically ranging from \SI{30}{\nano\meter} to \SI{80}{\nano\meter}. Its small aspect-ratio (AR), $L/D$ $\approx 10^{-2}$, results in $\mathcal{N}_{zz} \approx 1$ so that the geometrical factor associated with the demagnetising energy $-(3\mathcal{N}_{zz} -1)/2$ is approximately $-1$. Thus, the demagnetising energy favours an in-plane orientation of the magnetisation. However, for a sufficiently thin storage layer, the interfacial anisotropy $k_\textrm{s}$ can overcome the demagnetising contribution, resulting in an effective positive anisotropy that stabilises the perpendicular orientation. Neglecting the small volume anisotropy contribution $K_\textrm{V}$, the energy barrier in this limit can be expressed as 

\begin{equation}
    E_\textrm{B} = K_\textrm{eff}V \approx \left( k_\textrm{s} - \frac{1}{2}\mu_0M_\textrm{s}^2L\right)A,
\end{equation}

where $A$ is the cross-sectional area of the storage layer. The resulting energy barrier therefore scales with the junction area and decreases quadratically with its lateral dimension. Adding a second source of interfacial anisotropy, for example through an MgO capping layer  \cite{sato_perpendicular-anisotropy_2012}, increases the effective anisotropy coefficient but does not modify its areal scaling. Therefore, maintaining sufficient thermal stability remains challenging at dimensions below \SI{20}{\nano\meter}. 

A promising answer to this limitation lies in the modulation of the shape anisotropy of the storage layer. In conventional low aspect-ratio storage layers, the shape anisotropy favours an in-plane magnetisation and reduces the energy barrier. Unlike the intrinsic areal scaling of $k_\textrm{s}$, this contribution can be engineered through the geometry of the storage layer itself. Increasing its aspect-ratio reduces the negative contribution of the shape anisotropy and, at sufficiently large aspect-ratios, causes it to favour a perpendicular magnetisation, further increasing the energy barrier. To quantify this geometrical dependence, we focus on the perpendicular component of the demagnetising tensor $\mathcal{N}$$_{zz}$. For a cylindrical storage layer, $\mathcal{N}$$_{zz}$ can be calculated through a Fourier transform formalism \cite{beleggia_computation_2003, tandon_computation_2004, beleggia_general_2005} as

\begin{equation}
    \mathcal{N}_{zz}^{\mathrm{cylinder}} = 1 + \frac{4}{3\pi\tau_\textrm{AR}} - F_{0}\left(-\frac{1}{\tau_\textrm{AR}^2}\right),
\label{ch2:eq:demag}
\end{equation}
where $\tau_\textrm{AR}$  is the aspect-ratio of the cylinder given by $L/D$ and $F_0$ is a function shown for convenience, with $_2F^1$  a hypergeometric function (which can be related to elliptical integrals through \cite{beleggia_demagnetization_2005})

\begin{equation*}
    F_0(x) = {_{2}F_1}\left( -\frac{1}{2}, \frac{1}{2}, 2, x\right) \approx 1-\frac{x}{8}-\frac{x^2}{64}+\mathcal{O}(x^3).
\end{equation*}

The impact of different geometries of the storage layer is shown by the stability diagram of Figure \ref{fig:ch1-figure 4} ($a$). In the low-thickness region, the diagram reproduces the previously discussed scalability limitation of conventional p-MTJs, with the thermal stability strongly decreasing below a lateral dimension of \SI{15}{\nano\meter}. Consistent with the geometrical dependence introduced above, increasing the storage layer thickness modifies $\mathcal{N}$$_{zz}$: it first reduces the negative, in-plane contribution of the shape anisotropy and ultimately causes the shape anisotropy itself to favour a perpendicular orientation of the magnetisation. The dashed line in the stability diagram marks the cross-over between the in-plane and perpendicular shape anisotropy favoured orientation magnetisation. An illustration of the separated contributions of the surface and shape anisotropy are shown in Figure \ref{fig:ch1-figure 4} (\textit{b}) for a pillar diameter of \SI{20}{\nano\meter}.

For a storage layer thickness of \SI{1.4}{\nano\meter}, the total energy is minimised at $\theta = 0$ and $\theta = \pi$, indicating a perpendicular easy axis. Although the shape anisotropy contribution alone favours an in-plane magnetisation, with an energy minimum at $\theta = \pi/2$, the interfacial anisotropy is sufficiently strong at this small thickness to stabilise the perpendicular orientation. As the thickness increases, there is a non-monotonous contribution of shape anisotropy $V$($\mathcal{N}$$_{xx}$ - $\mathcal{N}$$_{zz}$). For instance, at a thickness of \SI{5}{\nano\meter}, the magnitude of the in-plane shape anisotropy energy is sizeable, enough to compete with the interfacial anisotropy, highlighting the cross-over between perpendicular and in-plane easy-axis. If the thickness is further increased: first the magnetisation turns in-plane because of the increased relative contribution of the shape anisotropy promoting the in-plane direction, followed by a recovery of the perpendicular orientation. A quick stability enhancement is observed when the shape anisotropy approaches the promoting of the perpendicular direction, and eventually turned perpendicular at $\tau_\textrm{AR} >$ 0.89. Now turning to lower diameters, which was the motivation of the 3D storage layer concept, it appears that the reduction of stability can be compensated for by suitably increasing the thickness of the storage layer, making use of the positive contribution of the shape anisotropy and the volume involved.

The control of the shape anisotropy through the enhancement of the storage layer aspect ratio is the corner stone in this novel concept, for which it is expected to retain perpendicular orientation at diameters as low as \SI{5}{\nano\meter} \cite{watanabe_shape_2018, perrissin_highly_2018}. Although at first sight this appears as a feasible approach for ultra-small high density MRAM, there are still several open questions that we aim to tackle in this manuscript. In Section \ref{section-micromagnetics}, we will explore the reversal dynamics of such high aspect-ratio elements using micromagnetic simulations. This will provide us with a set of guidelines to obtain a coherent and fast magnetisation reversal. Based on these results, we will present the process flow to obtain ultra-small magnetic tunnel junctions with high aspect-ratio in Section \ref{section-fabrication}, followed by experimental voltage and field measurements. In Section \ref{section-perspectives}, as perspectives, focus is given to the large neighbouring inter-cell crosstalk at large bit densities, where prospects for fast and coherent reversal are discussed.
\renewcommand{\hbar}{\mathchar'26\mkern-9mu h}

\section{Magnetisation reversal of high aspect-ratio storage layer}
\label{section-micromagnetics}

Two physical phenomena are expected to modify the switching behaviour of a high aspect ratio storage layer compared to the conventional (thin) storage layer in p-STT-MRAM. Firstly, the magnetisation reversal may become non-uniform along the thickness of the pillar, deviating from the coherent macrospin described in the previous section. Secondly, the spin-transfer torque is an interfacial effect. In a thick storage layer, the torque initially acts predominantly on the magnetisation near the tunnel-barrier interface, and its effect is expected to propagate through the remainder of the magnetic layer. In this section we investigate both matters with micromagnetic simulations, instead of the simplified macrospin model. 

\subsection{Simulations with single high aspect-ratio layer}

In contrast to the macrospin approach, also called coherent reversal and in which all the atomic magnetic moments are considered as a single magnetic moment, the micromagnetic theory uses a continuum description of the magnetisation of the ferromagnetic material,
\begin{equation}
    \textbf{M}(\textbf{r}, t) = M_\textrm{s} \textbf{m}(\mathbf{r}, t)\;,
\end{equation}
where \textbf{m} is the unitary vector defining the orientation of the magnetisation at every location in space \textbf{r} \cite{brown_criterion_1957}. This brings a different notation to the energy terms shown in the previous section. In this framework, exchange energy is written as:
\begin{equation}
    E_\textrm{exchange} = \int_V A_\textrm{ex} \left\{ [\nabla m_x(\mathbf{r})]^2 + [\nabla m_y(\mathbf{r})]^2 + [\nabla m_z(\mathbf{r})]^2\right\}\textrm{d}\textbf{r}\;,
\end{equation}
with $A_\textrm{ex}$ the exchange stiffness $\left[\textrm{J/m}\right]$. The uniaxial anisotropy energy may contain both volume and interfacial contributions:
\begin{equation}
    E_\textrm{uniaxial} = \int_V K_\textrm{u} \left\{ 1 - [\mathbf{u_k}\cdot \mathbf{m}(\mathbf{r}, t)]^2 \right\} \textrm{d}\textbf{V} + \int_S k_\textrm{s} \left\{ 1 - [\mathbf{u_{k}}\cdot \mathbf{m}(\mathbf{r}, t)]^2 \right\} \textrm{d}\textbf{S}\;,
\end{equation}
where $K_\textrm{u}$ is the volume uniaxial anisotropy constant $\left[\textrm{J/m}^3\right]$ and $k_\textrm{s}$ is the surface anisotropy constant $\left[\textrm{J/m}^2\right]$. The unit vector $u_\textrm{k}$ defines the corresponding easy-axis direction. Note the integration along the volume of the sample, which marks the notation difference between $E_\textrm{uniaxial}$, the integrated energy [J], and $\varepsilon_\textrm{uniaxial}$, a volume density of energy $\left[\textrm{J/m}^3\right]$. The same principle is applied to the shape anisotropy energy
\begin{equation}
    E_\textrm{shape} = -\frac{1}{2}\mu_0 M_\textrm{s}\int_V \mathbf{m}(\mathbf{r}, t) \cdot \textbf{H}_\textrm{demag} (\textbf{r}, t) \textrm{d}\textbf{r}\;,
\end{equation}
where $\textbf{H}_\textrm{demag} (\textbf{r}, t)$ is a time and position dependent demagnetising field and to the Zeeman energy (effect of an applied external field)
\begin{equation}
    E_\textrm{Zeeman} = -\mu_0 M_\textrm{s}\int_V \mathbf{m}(\mathbf{r}, t) \cdot \textbf{H}_\textrm{applied} (\textbf{r}, t) \textrm{d}\textbf{r}\;,
\end{equation}
where $\textbf{H}_\textrm{applied} (\textbf{r}, t)$ is an applied magnetic field that can be space and time dependent. 

Based on these different energy contributions, we can define an effective magnetic field $\textbf{H}_\textrm{eff}$, as the variational derivative of the Gibbs free energy density $\varepsilon_\textrm{total}$
\begin{equation}
    \textbf{H}_\textrm{eff} = -\frac{1}{\mu_0M_\textrm{s}}\frac{\delta\varepsilon_\textrm{total}}{\delta\textbf{m}}
    \label{eq: Gibbs}
\end{equation}
where $\varepsilon_\textrm{total}$ encompasses the energy densities present in the magnetic system
\begin{equation}
    \varepsilon_\textrm{total} = \varepsilon_\textrm{Zeeman} + \varepsilon_\textrm{shape} + \varepsilon_\textrm{uniaxial} + \varepsilon_\textrm{exchange}\;.
\end{equation}
Through the variational principle, we can compute the effective magnetic field as 
\begin{equation}
\mathbf{H}_{\textrm{eff}} = \frac{2A_\textrm{ex}}{\mu_0M_\textrm{s}}\nabla^2\textbf{m} + \frac{2K_\textrm{u}}{\mu_0 M_\textrm{s}}(\mathbf{u_k}\cdot \mathbf{m})\mathbf{u_k} + \mathbf{H}_\textrm{applied} + \mathbf{H}_{\textrm{demag}}\;.  
\label{eq:ch2-Heff}
\end{equation}
Equilibrium states must satisfy the following conditions, which express the absence of torque at the boundary of the system (surface) and in its volume:
\begin{equation}
    \frac{\partial \textbf{m}}{\partial \textbf{n}} = 0
\end{equation}
\begin{equation}
    \textbf{m}(\textbf{r}, t) \times \mathbf{H}_\textrm{eff} (\textbf{r}, t) = 0\;.
\end{equation}
with $\textbf{n}$ the unit vector normal to the surface of the magnetic body.

These conditions are the basis of the Brown formalism and are thus defined as Brown’s equations \cite{brown_thermal_1963, hubert_magnetic_1998}.
To understand the dynamical behaviour of the magnetisation during the reversal process, the LLGS equation is implemented in this micromagnetic framework as
\begin{equation}
    \partial_t \textbf{m}_\textrm{SL} = - \gamma \mu_0 (\textbf{m}_\textrm{SL} \times \textbf{H}_\textrm{eff}) + \alpha(\textbf{m}_\textrm{SL} \times \partial_t \textbf{m}_\textrm{SL}) + \Gamma_\textrm{STT}^\textrm{IP},
    \label{ch2:eq:full-numeric}
\end{equation}
and
\begin{equation}
\label{eq:STT}
\Gamma_\textrm{STT}^\textrm{IP} = - |\gamma| a_\parallel V_\textrm{applied} \mathbf{m}_\textrm{SL}\times(\mathbf{m}_\textrm{SL}\times \mathbf{m}_\textrm{RL})
\end{equation}
with $V_\textrm{applied}$ the applied voltage across the tunnel barrier, $\mathbf{m}_\textrm{SL}$ the normalised magnetisation vector of the storage layer, $\mathbf{m}_\textrm{RL}$ the normalised magnetisation vector of the reference layer and $a_\parallel$ a pre-factor of the damping-like torque that is linked to equation \ref{ch1:eq:gamma_stt} as
\begin{equation}
    a_\parallel = \frac{\hbar}{2|e|}\frac{\eta_\textrm{STT}}{R\times A}\frac{1}{M_\textrm{s}t_\textrm{SL}},
    \label{ch2:eq:apara}
\end{equation}
with $R\times A$ the resistance $\times$ area of the  tunnel barrier [$\SI{}{\ohm\per\micro\meter\squared}$].

In the simulations reported in this section, we restrict ourselves to the effect of the damping-like torque, as the contribution of the field-like torque is usually accounted only as a small contribution to the STT effect in usual flat layers. 


We use a finite difference method to solve numerically equation \ref{ch2:eq:full-numeric} \cite{buda_micromagnetic_2002}, from which the system is meshed as  elementary prismatic units (referred to as unitary cells), with dimensions $l_{x,y,z}$. The size of this elementary cell needs to be smaller than the typical length scale at which the magnetisation can vary, so that the system is faithfully described in the continuous model of micromagnetism. In most cases, this condition is met using a cell size smaller than both the exchange length $\gamma_\textrm{ex}$ (indicative of a Néel domain-wall type) and $\gamma_\textrm{B}$ (indicative of a Bloch wall) \cite{abo_definition_2013, stano_magnetic_2018}. These length scales define the competition between the exchange and, respectively, the dipolar energy and the uniaxial anisotropy.
Besides computing the dynamics of the magnetic system based on the LLGS equation, the so-called string method allows to compute the minimum energy path (MEP) \cite{forster_energy_2003, e_energy_2003, chaves-oflynn_stability_2010, carilli_truncation-based_2015} between the two stable states (in our case parallel and anti-parallel). 

\begin{figure}[h]
    \centering
    \includegraphics[width=1\linewidth]{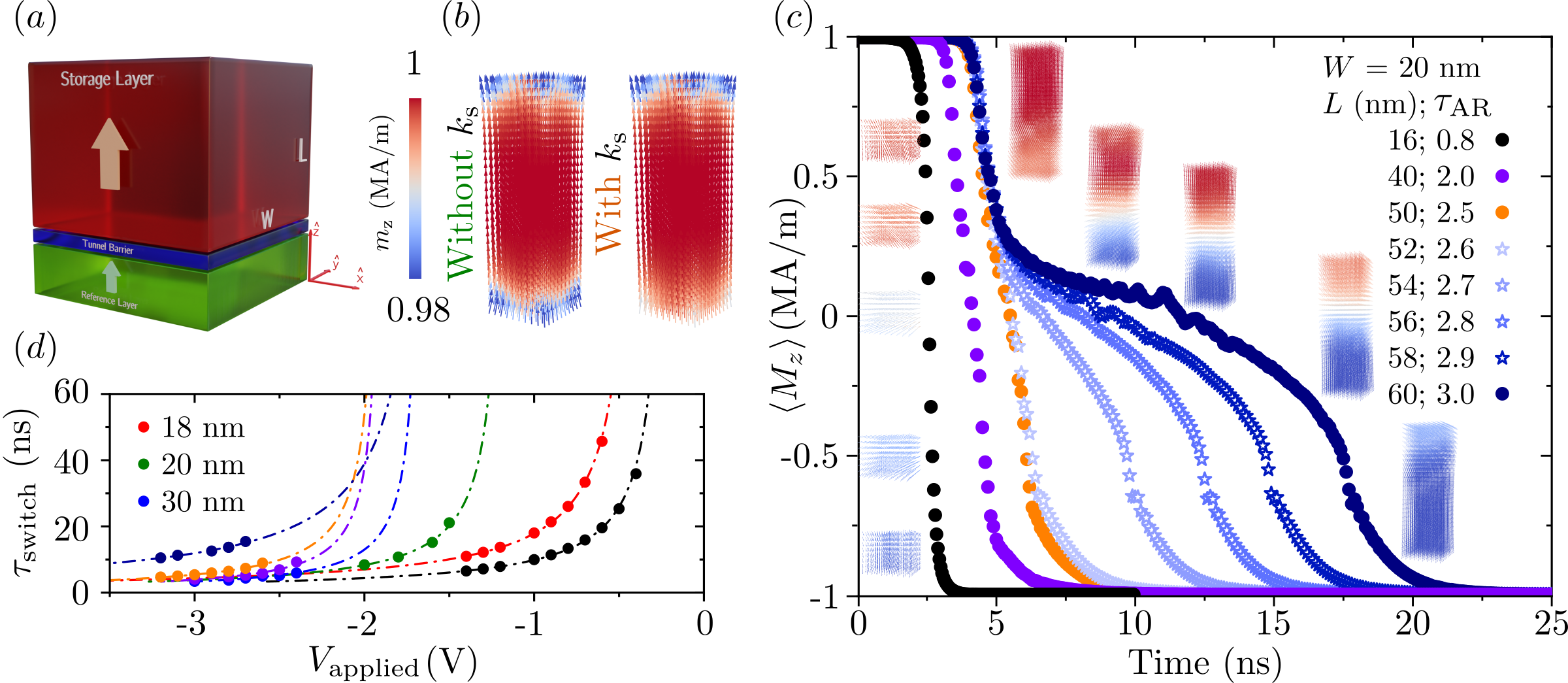}
    \caption{($a$) 3D schematics of the studied prism with thickness \textit{L} and base width \textit{W}. The reference layer is shown with a green colour, the tunnel barrier with a blue colour and the storage layer with a red colour. ($b$) 3D equilibrium relaxed state of a magnetic prism with and without surface anisotropy. The colour indicates the normalised magnitude of the magnetisation of each elementary cell along +$\hat{z}$. ($c$) Time trace of the mean perpendicular magnetisation $\langle M_z\rangle$ for a storage layer with fixed base width \SI{20}{\nano\meter} and various vertical aspect-ratios, for an applied voltage of \SI{-3}{\volt}. Insets of 3D snapshots show the nature of the reversal dynamics, with a colour scale ranging between -- 1 and \SI{1}{\mega\ampere\per\meter}. ($d$) Dependence of the switching time on the applied voltage for different thickness of the storage layer. Reprinted figure with permission from \cite{cacoilo_spin-torque-triggered_2021}. Copyright 2024 by the American Physical Society.}
    \label{fig:ch2-figure1}
\end{figure}

Since the elementary cell is cubic, we start with the simulation of a storage layer with a rectangular base of width \textit{W} = \SI{20}{\nano\meter} and increasing thickness \textit{L}. An example of this geometry is shown in Figure \ref{fig:ch2-figure1} (\textit{a}). In our simulations we consider that the reference layer (green) is strongly pinned along $+\hat{z}$ and its magnetisation fully compensated so that no stray field is present in the modelled storage layer. This allows us to neglect both its own dynamics and  its dipolar influence on the magnetisation of the storage layer. The tunnel barrier, shown in blue for clarity, is not explicitly included in the micromagnetic mesh, which contains only the storage layer. The interfacial anisotropy and spin-transfer torque effect associated with the MgO/CoFeB interface are instead applied to the storage layer's cells adjacent to the tunnel barrier. One of the main differences with thin layers, such as flat perpendicular-MTJ, is that we cannot consider the surface anisotropy and spin-transfer torque to have the same magnitude throughout the entire thickness of the storage layer. The former is represented by an effective uniaxial volume anisotropy $K_\textrm{u}$ distributed over the computational cells closest to the MgO/storage-layer interface. Its magnitude is assumed to decay exponentially through the thickness of the storage layer  
\begin{equation}
K_\textrm{u}(z) = K_\textrm{0}\exp\left({-\frac{z}{l_{\textrm{K}_\textrm{u}}}}\right),
\end{equation}
where $z = 0$ corresponds to the interface and $l_{K_\textrm{u}}$ is the decay length. Since $K_\textrm{u}$ is taken as a volume-energy density in each computational cell, it needs to be calculated at the centre of that cell. For a cell \textit{n}, its centre is calculated at 
\begin{equation}
    z_n = (n+1/2)l_z,
\end{equation}
where $l_\textrm{z}$ is the cell size along the storage layer thickness. The value of the coefficient $K_0$ is adjusted so that the discrete sum of the volume-density-based energy of $K_\textrm{u}$ results in $\ks$
\begin{equation}
\ks = \sum_{n=0}^{\textrm{N}_\textrm{z}-1} K_\textrm{u}[(n+1/2)l_\textrm{z}]l_\textrm{z}
= K_0 \sum_{n=0}^{\textrm{N}_\textrm{z}-1} l_\textrm{z} \exp \left[ - \frac{(n+1/2)l_\text{z}}{l_{\textrm{K}_\textrm{u}}}\right]
\label{eq:ko}
\end{equation}
with $N_z$ the number of computational cells layers along the thickness ($\hat{z}$).

Similar to the implementation of surface anisotropy, the effect of the STT is also introduced considered as decaying along the thickness and away from the interface. Because spin-polarised electrons are injected into the storage layer at the tunnel-barrier interface, their spin polarisation is expected to decay as they interact with the local magnetic moments \cite{grollier_magnetic_2011, chshiev_analytical_2015}. To account for the spatially non-uniform torque, the damping-like STT coefficient is assumed to decay exponentially away from the interface according to
\begin{equation}
a_\parallel = a_0\exp{\left(-\frac{z}{l_\textrm{STT}}\right)},
\end{equation}
where $z = 0$ corresponds to the defined interface, $l_\textrm{STT}$ the length scale of the spin-transfer-torque decay and $a_0$ the torque coefficient at the interface. 

In the macrospin model, for a storage layer thickness $L$, the thickness-averaged $\langle a_\parallel \rangle$ is given by equation \ref{ch2:eq:apara}. The value of the coefficient $a_0$ can be obtained knowing that the averaged sum of each plane of computational cells along the storage layer thickness will result in the macrospin value of $\langle a_\parallel \rangle$:
\begin{equation}
\langle a_\parallel \rangle = \frac{1}{N_z} \sum_{n=0}^{N_z-1} a_\parallel[(n+1/2)l_\textrm{z}]
=\frac{a_0}{N}  \sum_{n=0}^{N_z-1} \exp \left[-\frac{(n+1/2)l_\textrm{z}}{l_\textrm{STT}}\right].
\label{eq:ao}
\end{equation}
This normalisation preserves the total thickness-averaged torque, while allowing its local magnitude to decrease away from the tunnel-barrier interface. 

We now study the nature of the magnetisation reversal under applied voltage for different storage layer thickness, ranging from \SI{16}{\nano\meter} to \SI{60}{\nano\meter}. Here, we use a cell size of $l_\textrm{(x,y,z)}$ = \SI{2}{\nano\meter}. The contribution of the surface anisotropy is taken as $K_0 = \SI{0.61}{\mega\joule\per\cubic\meter}$ at $z = 0$ considering $l_{K_\textrm{u}} = \SI{1}{\nano\meter}$ and $k_\textrm{s} = \SI{1.4}{\milli\joule\per\meter\squared}$. For the transport properties, we assume a TMR of 100$\%$ at a low $R \times A$ product of \SI{1}{\ohm\per\micro\meter\squared} for the calculation of $\langle a_\parallel \rangle$. The decay length $l_\textrm{STT}$ is taken as \SI{1}{\nano\meter}, yielding $a_0 = \SI{60}{\milli\tesla\per\volt}$ at $z = 0$. We start with both initial magnetisations parallel to each other (Figure \ref{fig:ch2-figure1}) and, as the simulations are performed without thermal fluctuations, a small initial misalignment angle of 1$\degree$ with respect to the $\hat{z}$ orientation is introduced in the magnetisation of the storage layer to provide a finite initial spin-transfer torque. As an example, figure  \ref{fig:ch2-figure1} (\textit{c}) shows the time trace $M (z)$ for an applied voltage of \SI{-3}{\volt} for different aspect-ratios. Upon increase of the storage layer thickness, the time to start reversing the magnetisation increases. Moreover, it comes along the emergence of a plateau in the magnetisation evolution, in contrast with the sharp decay shown at a thickness of \SI{16}{\nano\meter}. For this moderate thickness, the preservation of a mostly uniform magnetisation at any time of the reversal is confirmed by the 3D snapshots along the trace of the temporal evolution (see the vertical series on the left-hand side of the graph). In the large-thickness regime, 3D snapshots reveals that magnetisation switching proceeds via the nucleation and propagation of a transverse domain-wall, an expected and well-established reversal mode in elongated magnetic nanostructures \cite{thiaville_micromagnetic_2005, stano_magnetic_2018}. Its formation is associated with a slowing down of the dynamics around $\langle M_\textrm{z}\rangle = 0$. 

\begin{figure} [h]
    \centering
    \includegraphics[width=1\linewidth]{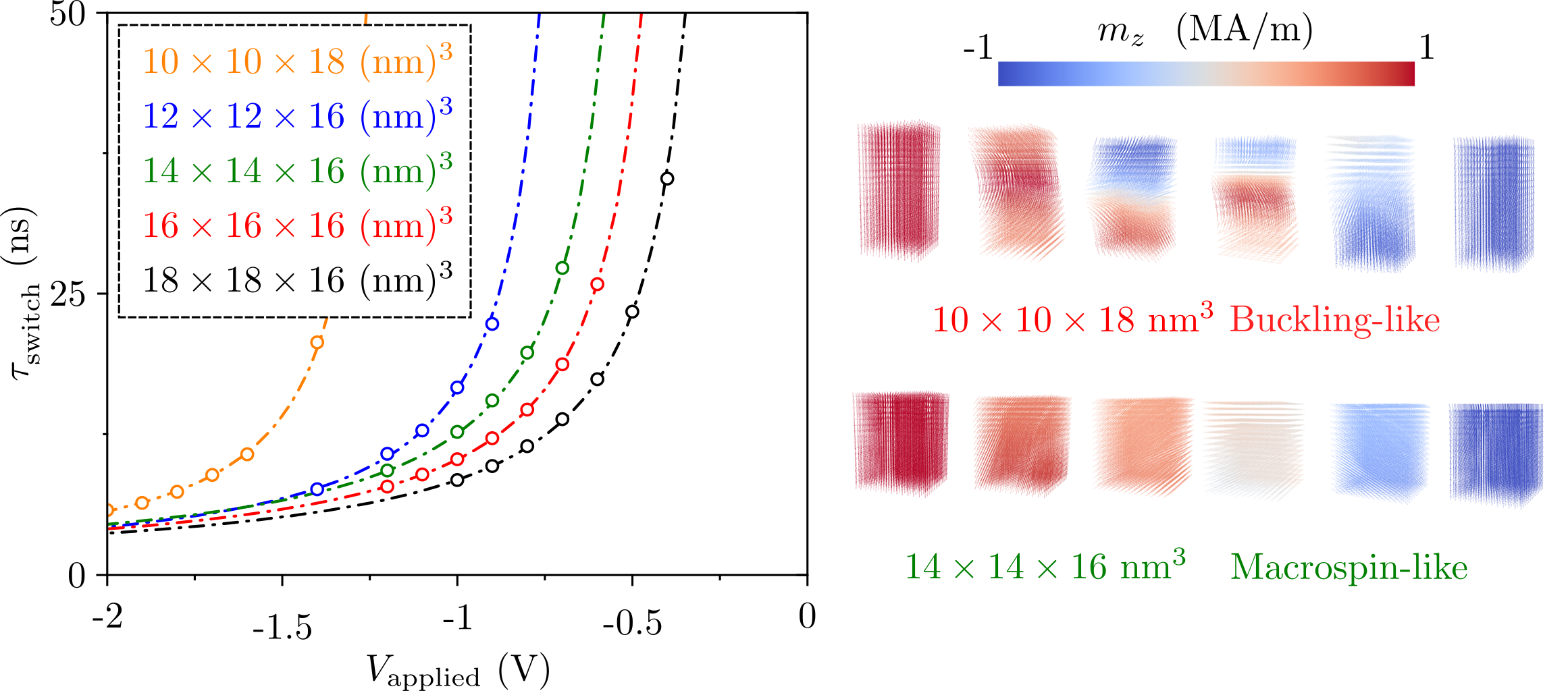}
    \caption{($a$) Switching time dependence on the applied voltage for a different set of storage layer aspect-ratio and ($b$) set of 3D snapshots associated with a buckling-like reversal for the larger aspect-ratio and macrospin-like reversal for a limiting aspect-ratio. Reprinted figure with permission from \cite{cacoilo_spin-torque-triggered_2021}. Copyright 2024 by the American Physical Society.}
    \label{fig:ch2-figure2}
\end{figure}

The voltage-dependent curves of figure Figure \ref{fig:ch2-figure1} (\textit{d}) show that increasing the magnitude of the applied voltage reduces the switching time $\tau_\textrm{switch}$, defined as the time it takes to reverse 90$\%$ of the magnetisation. They also reveal that the voltage necessary to induce magnetisation reversal increases with increasing storage-layer thickness. For the thicker layers considered, this voltage approaches or exceeds the voltage range compatible with reliable operation voltages. 

In the MRAM design, the geometry would be selected to provide a thermal stability and retention time required by the targeted application. It is therefore important to link how switching voltage and switching time depend on the storage layer aspect-ratio for a comparable thermal stability. However, the results presented above do not establish whether the non-coherent reversal originates from the increased aspect-ratio, or simply from the increase in energy barrier. To separate these effects, we consider different storage layers aspect-ratios, but with a comparable stability of 80 $k_\textrm{B}T$. As shown in \cite{cacoilo_spin-torque-triggered_2021}, using a base width ranging from 10 to \SI{18}{nm}, there is a clear modification of the magnetisation reversal, depicted by the increase in switching voltage [Figure \ref{fig:ch2-figure2} (\textit{a})] and the 3D snapshots [Figure \ref{fig:ch2-figure2} (\textit{b})], which show a buckling-like reversal at larger aspect-ratio. Additionally, it is observed that, by decreasing the storage layer width, although its thickness is similar (focus on \SI{16}{\nano\meter} thickness), the required voltage to reverse the magnetic layer is increased. This can be seen as an increase in the effective anisotropy field $\HK$ that plays a role in the critical switching voltage $V_\textrm{c0}$, as will be seen in the following subsection.

These results show that the reversal dynamics are governed not only by the thermal stability, but also by the storage-layer aspect-ratio. Indeed, layers with comparable stability exhibit different reversal mechanisms and switching times when their geometry is changed. These conclusions are taken considering only the effect of a spatially-decaying interfacial damping-like torque but neglects spin-transfer torque associated with magnetisation gradients. This simplified approach might fail to correctly describe a non-homogeneous magnetisation switching, such as a nucleation-propagation domain-wall-mediated reversal, as it neglects bulk spin-transfer torques arising from spatial gradients of the magnetisation. When these are considered, the reversal might become asymmetric with respect to the initial direction of the magnetisation. For instance, for an AP-P reversal, both the electron flow and the bulk STT would drive a domain-wall away from the interface, effectively reversing the magnetic layer. However, for the P-AP case (which is the convention described during this section) the electron flows toward the tunnel barrier in the FL, \textit{i.e}., opposing the propagation of a domain-wall. The latter may slow down or even remain in dynamical equilibrium at a certain distance from the interface, preventing the full reversal of the magnetisation. Although this behaviour could only be grasped considering bulk-STT, the simpler description using the decaying STT is expected to remain qualitatively valid when the reversal is close to coherent. Self-consistent calculations considering the spin accumulation throughout the structure and its coupling to the local magnetisation through \textit{s-d} exchange interaction have additionally shown that the field-like torque can become comparable to the damping-like torque. Together with the non-uniform flower states induced by the pillar geometry, this torque excites edge ferromagnetic-resonance modes that contribute to the magnetisation reversal \cite{Natalia_spin_accumulation}. From a device-perspective, aspect-ratios closer to the transition between in-plane and perpendicular shape anisotropy may therefore be preferable for avoiding non-uniform reversals that bring higher switching voltages, longer switching times and possibly cause write failures depending on the polarity of the current.

\subsection{Increased performance through an increase in $\mu_0H_\textrm{K}^\textrm{eff}$}

In the previous subsection, the lateral dimension and thickness were varied simultaneously to compare different aspect-ratios at approximately similar energy barriers to isolate the influence of the geometry on the reversal dynamics. In a practical MRAM device, however, the lateral dimensions is more or less fixed, constrained by the targeted technological node. Reducing the aspect-ratio must therefore come by decreasing accordingly the thickness (which might come associated with a crossover between perpendicular to in-plane favoured shape anisotropy). The resulting challenge is then to preserve the energy barrier required for thermal stability and data retention in a way different from simply increasing the thickness of the storage layer. A way to address this challenge is by increasing $K_\textrm{eff}$ and thus lowering the required magnetic volume for a given stability. Indeed, for coherent rotation the energy barrier is given by:

\begin{equation}
    \eb = K_\textrm{eff}V = \frac{M_\textrm{s} \mu_0 H^\textrm{eff}_\textrm{K}V}{2}
\end{equation}
where $\HK$ [T] is an effective anisotropy field coming from the combination of the energy terms shown in equation \ref{ch1:eq:e_system}. For the remaining of this document we will use this term rather than its related energy value $K_\textrm{eff}$. It is then obvious that, by increasing $\HK$, we can decrease the layer aspect-ratio without penalty in the stability. This can be done using an additional surface anisotropy ($k_\textrm{s2}$), as commonly used in double-MgO flat perpendicular MTJ \cite{sato_comprehensive_2013}.

\begin{figure}[h!]
    \centering
    \includegraphics[width=1\linewidth]{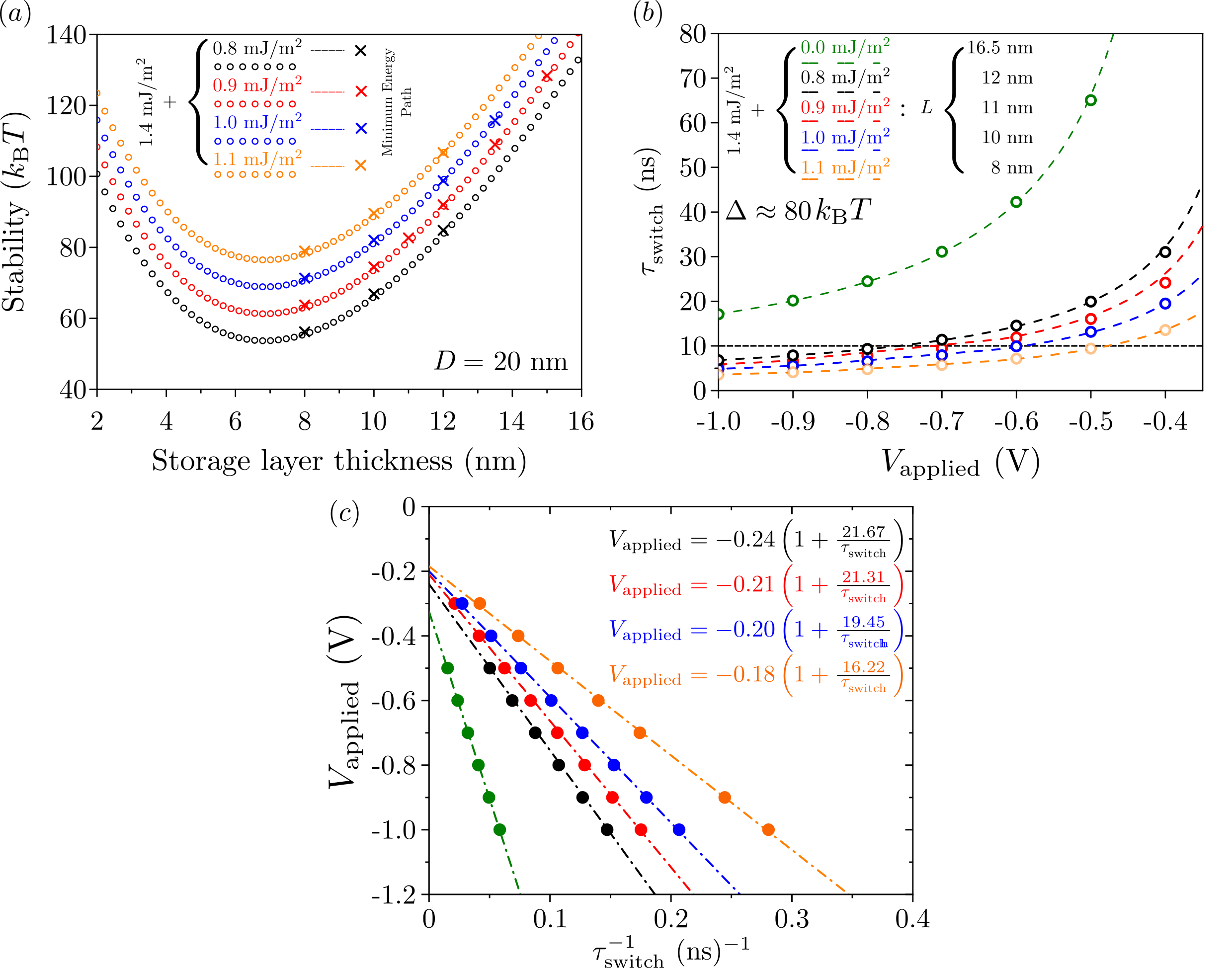}
    \caption{($a$) Stability values obtained from the macrospin model, superimposed with several minimum energy path results for a fixed \SI{20}{nm} diameter and increasing thickness, for different values of $k_\textrm{s2}$. (\textit{b}) Dependence of the switching time on the applied voltage for storage layer with a fixed stability of \SI{80}{\kbt} and different aspect-ratios. (\textit{c}) Linear dependency of the applied voltage on the inverse of the switching time and (\textit{d}) schematics of the reduction of the critical voltage with the reduced aspect-ratio.}
    \label{fig:ch2-figure3}
\end{figure}

We implemented this second interface in the simulations and calculated for which thickness the stability is similar for different values of $k_\textrm{s2}$. Minimum energy path simulations are performed, and the results are superimposed to those of macrospin simulations in Figure \ref{fig:ch2-figure3} (\textit{a}). Both approaches match well, as expected for these nearly coherent-reversal situations. By fixing the diameter of the cylinder to \SI{20}{nm} and increasing the magnitude of $k_{s2}$, a similar stability of \SI{80}{\kbt} can be achieved for thinner layers. Using the interfacial nature of both STT and surface anisotropy (both $k_{s1}$ and $k_{s2}$), the dependence of the switching time on the applied voltage (and $\HK$) is shown in Figure \ref{fig:ch2-figure3} (\textit{b}). It is observed that the switching time decreases sharply for smaller vertical aspect-ratio. Moreover, through the linearity of the inverse of the switching time versus the applied voltage [Figure \ref{fig:ch2-figure3} (\textit{c})], it is possible to extract key features from the STT-driven dynamics. Usually measured as a function of the pulse width $\tau_\textrm{P}$, results in the ballistic regime provide a linear dependency between $\tau_\textrm{P}^{-1}$ and the applied voltage, predicted by the conservation of angular momentum \cite{sun_spin-current_2000, koch_time-resolved_2004, garello_ultrafast_2014, worledge_theory_2017, sun_spin-transfer_2022}
\begin{equation}
    \frac{1}{\tau_\textrm{P}}  = \mathcal{A} (V_\textrm{applied} - V_{c0}),
\end{equation}
where $\mathcal{A}$ is a constant that expresses the angular momentum conservation. 

It is possible to derive a more direct equation to link the material parameters with the device properties through \cite{bedau_spin-transfer_2010, sun_spin-transfer_2022}
\begin{equation}
    V_\textrm{applied} = V_\textrm{c0}\left( 1 + \frac{\tau_\textrm{D}}{\tau_\textrm{P}}\right)
    \label{eq: linear_vc0}
\end{equation}
where $\tau_\textrm{D}$ is a characteristic time for switching. Both $V_\textrm{c0}$ and $\tau_\textrm{D}$ are related to the materials parameters through
\begin{equation}
    \tau_\textrm{D} = \frac{1+\alpha^2}{\alpha\gamma \mu_0 H_\textrm{K}^\textrm{eff}}\ln{\left(\frac{2}{\theta_0}\right)}
\end{equation}
with $\theta_0$ an initial average angle of tilt that can be related to the energy barrier through $\theta_0 \approx 1 / \left( 2\sqrt{\pi \Delta} \right)$ and
\begin{equation}
    V_\textrm{c0} = \frac{\alpha \mu_0 H_\textrm{K}^\textrm{eff}}{a_\parallel},
    \label{ch2:eq:vc0-hkeff}
\end{equation}
which is taken as the voltage required to reverse the magnetic layer under an infinitely long pulse ($\tau_\textrm{P}\rightarrow\infty$). For small voltage pulses, this value will grow larger, depending on the value of $\tau_\textrm{D}$. These equations show that, if both the values of $\HK$ and $a_\parallel$ are kept constant, there should be no dependence on the aspect-ratio of the storage layer. Indeed, from figure \ref{fig:ch2-figure3} (\textit{c}) it is observed that as $\HK$ is increased (aspect-ratio is reduced), the value of $\tau_\textrm{D}$ is reduced. This implies that, at shorter pulse lengths, the increase of $V_\textrm{c0}$ is not excessive. However, at the same time, $V_\textrm{c0}$ seems to be reduced as the aspect-ratio is reduced. We believe that, for thinner layers, although the magnitude of $a_\parallel$ at the interface is the same, as the layer grows thinner, its effect is increased. Moreover, this increase seems to outgrow the increase in $\HK$ for this specific case, leading to a lower $V_\textrm{c0}$ for larger $\HK$. Thus, reducing the thickness of the storage layer leads to an improvement of the switching time and switching voltage. Although this is made possible through the increase in $\HK$, it is not the only parameter involved, as the switching voltage is consequently reduced for thinner layers. Nonetheless, this should no longer be the case for layers thin enough that the STT is acting on the entire volume, which can cause an additional increase in switching voltage.

\subsection{Capping of stability at low nodes}

\begin{figure}[h!]
    \centering
    \includegraphics[width=1\linewidth]{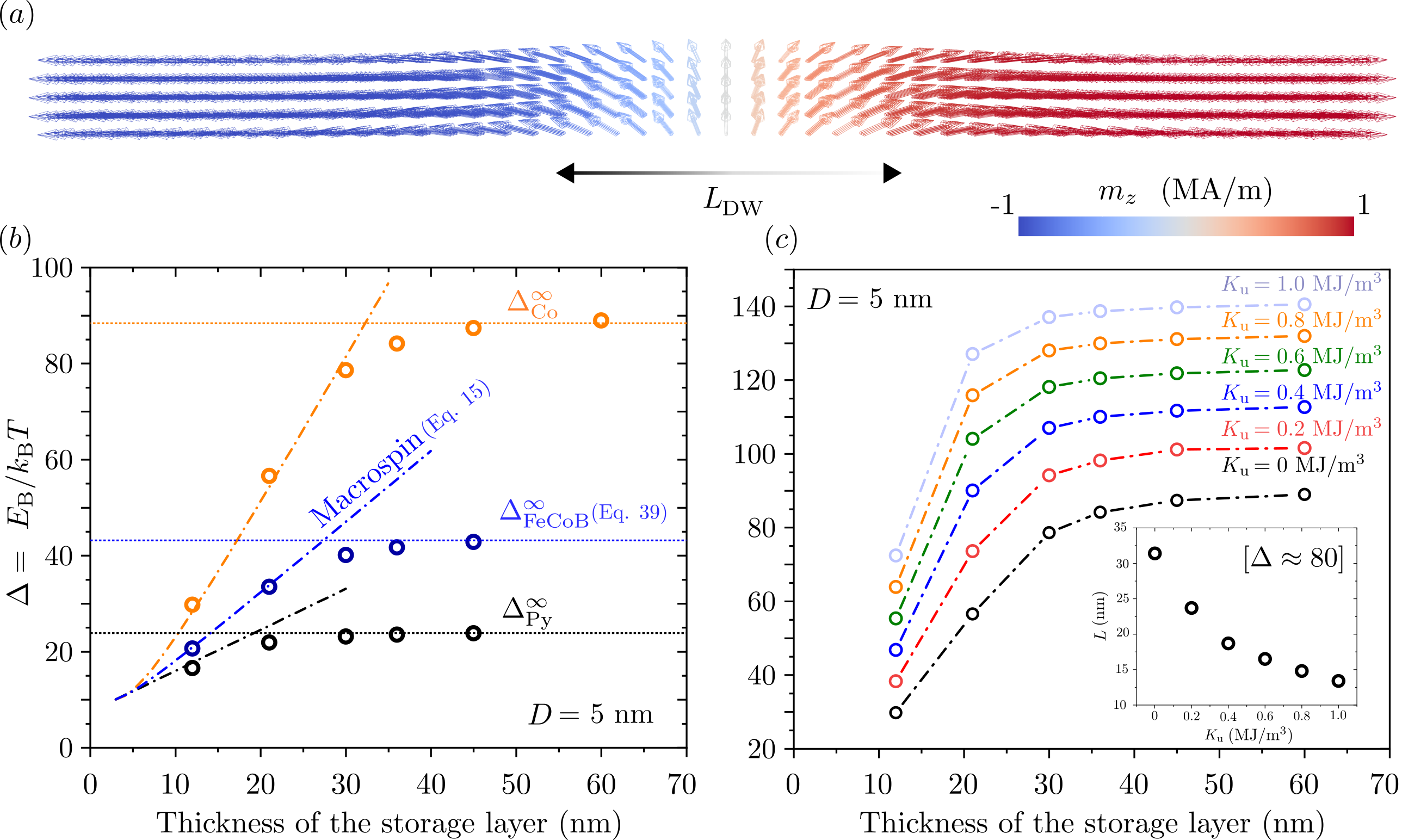}
    \caption{(\textit{a}) Example of a transverse domain-wall switching with domain-wall width $L_\textrm{DW}$. (\textit{b}) Thermal Stability factor of a \SI{5}{\nano\meter} diameter storage layer with different materials, Py (in black), FeCo(B) (in blue) and Co (in orange). The limit $\Delta^\infty$ is shown with a dotted line, with the macrospin approximation shown with a dashed line and micromagnetic MEP results shown with open circles. Reprinted figure with permission from \cite{cacoilo_dipole-coupled_2024}. Copyright 2024 by the American Physical Society. (\textit{c}) Thermal stability factor for a Co pillar with increasing uniaxial anisotropy. The thickness at which a thermal stability of $\SI{80}{\kbt} $ is attained for each value of $K_\textrm{u}$ is shown in the inset of this figure.}
    \label{fig:ch2-figure4}
\end{figure}

At this point, the results indicate that reducing the storage layer aspect-ratio allows faster switching times and lower switching voltages. A second source of perpendicular anisotropy, such as $k_\textrm{s2}$, can compensate for the reduced contribution of the perpendicular shape anisotropy and allow for thinner layers while preserving the thermal stability. Nonetheless, if the aspect-ratio remains close to 1, the downsize capability of the storage layer is ultimately limited. Increasing the aspect-ratio can restore perpendicular shape anisotropy, but it has its limits. At large aspect-ratios, the reversal mechanism becomes non-uniform. The energy barrier is then governed by the energy required to nucleate a domain wall, and no longer scales with volume \cite{perrissin_highly_2018, perrissin_perpendicular_2019}. This upper stability limit is calculated approximately using Taylor expansions in \cite{perrissin_highly_2018}, considering a tail-to-tail domain-wall formed along the thickness of the storage layer [Figure \ref{fig:ch2-figure4} (\textit{a})]
\begin{equation}
    \Delta^{\mathrm{DW}} = \frac{\mu_0M_\textrm{s}^2}{2k_BT}\frac{\pi D^2}{4}\left(\frac{D}{2}+L_{\mathrm{DW}} + \frac{2L_{\mathrm{DW}}^2}{D + 2L_\mathrm{DW}}\right)\;,
    \label{eq:ch2:DW_EB}
\end{equation}
where $L_\mathrm{DW}$ is a measure of the width of the domain wall, with an asymptotic value of $\sqrt{\frac{4A_\textrm{ex}}{\mu_0 M_\textrm{s}^2}}$ for diameters smaller than the exchange length. 

This capping in stability is illustrated in figure \ref{fig:ch2-figure4} ($b$), for the approximate solution of a pillar with \SI{5}{nm} diameter for the case of NiFe (Py) ($M_\textrm{s}$ = \SI{0.789}{MA/m}, $A_\textrm{ex}$ = \SI{8}{pJ/m}), FeCo(B) ($M_\textrm{s}$ = \SI{1}{MA/m}, $A_\textrm{ex}$ = \SI{15}{pJ/m}) and Co ($M_\textrm{s}$ = \SI{1.446}{MA/m}, $A_\textrm{ex} = \SI{30}{pJ/m}$) (dashed line) followed by minimum energy path calculations (open circles). The energy barrier increases with $M_\textrm{s}$, as expected from the asymptotic scaling with $M_\textrm{s}^2$ in equation \ref{eq:ch2:DW_EB}. Thus, at small diameter and for materials with moderate magnetisation, perpendicular shape anisotropy alone is insufficient to reach the  thermal stability required for long-term data storage, even at large aspect-ratios. Even when considering the high saturation magnetisation of Co, achieving this target requires an aspect-ratio larger than 4 and, as discussed in the previous section, such large aspect-ratios pose significant challenges. 

A viable route towards reducing the storage layer thickness is to introduce an additional source of uniaxial anisotropy, such as magnetocrystalline. Figure \ref{fig:ch2-figure4} ($c$) shows that, for a thick layer of Co, we can shift the capping in stability towards larger aspect-ratio. Consequently, a larger stability can be achieved at lower thicknesses, as illustrated in the inset for a fixed stability of 80 $\kbt$. Candidates are alloys such as FePd or MnAl \cite{takeuchi_nanometer-thin_2022}. MnAl is particularly attractive because it has a small $\Ms$, so the negative impact of the shape anisotropy could be heavily reduced. Another approach is to introduce multiple CoFeB/MgO interfaces throughout the storage layer. Such structures have already demonstrated an increase in stability in conventional p-MTJs \cite{nishioka_novel_2019}, and were later extended towards smaller diameters \cite{jinnai_high-performance_2020, nishioka_enhancement_2021}. The reduced aspect-ratio of these structures is expected to favour faster switching speeds, making them more suited towards applications that require short write times, potentially including SRAM-like embedded memories. On the other hand, thicker storage layers that use perpendicular shape anisotropy may exhibit a slower switching time, but their tolerance to elevated operating temperatures remains attractive for applications that don't require such short writing times, such as the ones for automotive applications and potentially DRAM-like storage. These approaches therefore involve different compromises between thermal stability, write speed and fabrication complexity. Magnetocrystalline-anisotropy solutions require precise control of alloy composition, crystalline texture and thermal processing, whereas multilayered approaches require the deposition and control of several high-quality interfaces. Structures with high-aspect-ratio use comparatively conventional magnetic materials. Even if these approaches require different precision on the control of the deposited materials, their patterning towards ultra-scaled dimensions brings high demands on etching, sidewall cleaning and dimensional control. Thus, the viability of these approaches depends, not only on their magnetic and electrical properties, but also on whether the required structure can be fabricated reliably at the targeted dimension, which motivates the fabrication section in the following section. 

\section{Fabrication and characterisation of ultra-small MTJ}
\label{section-fabrication}

Following the guidelines from the last section for fast switching and low switching voltage, we now turn to the fabrication of these ultra-small MTJ and their characterisation.

\subsection{Process flow for achieving ultra-small dimensions}

\begin{figure}[h!]
    \centering
    \includegraphics[width=1\linewidth]{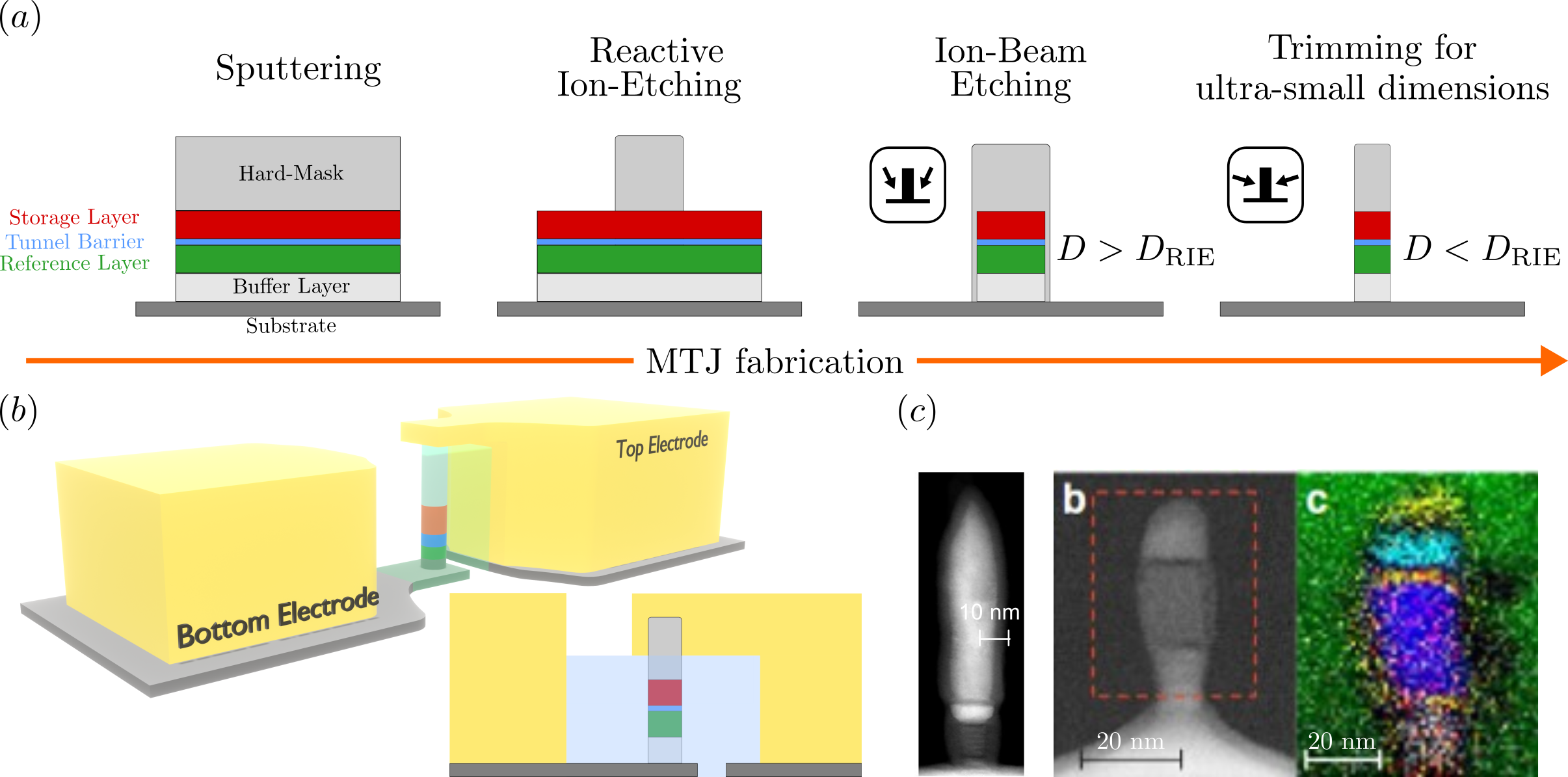}
    \caption{($a$) Schematics of the fabrication of the magnetic tunnel junction, from the sputtering of the layers to the etching of the pillar and (\textit{b}) Schematics of a final encapsulated electrical device after definition of both bottom and top electrodes. (\textit{c}) High-angle angular dark-field scanning electron microscopy imaging of a process pillar (left side) at the facilities of SPINTEC, France and (right side)  at the facilities of Tohoku University, Japan, with associated element mapping using electron energy-loss spectroscopy, with emphasis on the thick storage layer (dark blue) \cite{watanabe_shape_2018}, reproduced with permission from Springer Nature.}
    \label{fig:ch3-figure 1}
\end{figure}

Conventional MTJ process begins with the deposition of a conductive hard-mask layer, typically by magnetron sputtering, on top of the magnetic stack [Figure \ref{fig:ch3-figure 1} (\textit{a})]. In research institutions, the initial pillar diameter is commonly defined by electron-beam lithography (for sub-micron devices), an electron sensitive photoresist is exposed and developed. Afterwards, the pattern is transferred to the hard-mask through a chemically selective etch step, reactive ion-etching (RIE) [Figure \ref{fig:ch3-figure 1} (\textit{b})]. During this step, the reaction products are volatile and bond to the reactive species in the chamber, being effectively pumped out of the chamber and limiting material redeposition. The patterned hard-mask must preserve its lateral dimension (a good vertical shape) and withstand the following physical etching used to transfer its shape to the remaining material stack. These steps are then crucial for the correct fabrication and scaling of the device, as it will limit the final junction diameter and device-device variability. After the hard-mask definition, the pattern is transferred to the MTJ using ion-beam etching (IBE) [Figure \ref{fig:ch3-figure 1} (\textit{c})]. During this step, etched material can redeposit around the pillar sidewalls. This redeposition is particularly detrimental, as it can shunt the device when redeposited around the tunnel barrier. The critical selection of the etching angle may allow for etching without excessive redeposition of conductive elements. For the convention adopted here, with 0$\degree$ corresponding to an etching normal to the film plane, 30-40 $\degree$ etching angle is an optimal choice. Reproducibility can be improved through an additional and final trimming step at a grazing angle (almost horizontal etching, preventing significant vertical etching of the pillar), removing remaining conductive redeposits from the sidewalls. Grazing incidence can also be used to purposely reduce the lateral dimension of the pillar, down to \SI{10}{\nano\meter} [Figure \ref{fig:ch3-figure 1} (\textit{c})]. This approach can be somewhat effective for isolated pillars, but it cannot be used for dense arrays, due to the shadow effect from neighbouring devices \cite{kim_integration_2011}. Although different etch-less approaches have been proposed to circumvent this issue, for instance the use of RIE for the different layers of the stack \cite{sato_1t-1mtj_2018} or the use of pre-patterned Ta posts \cite{nguyen_novel_2017}, the use of ion-beam etching is still the most efficient, and is therefore the standard. After etching, the devices are encapsulated by a dielectric, for example SiN, to protect the exposed MTJ from oxidation and electrically isolate the bottom and top electrodes (shown with a light blue colour in the 2D illustration of [Figure \ref{fig:ch3-figure 1} (\textit{b})]). After the physical separation between electrodes, this dielectric is thinned down to expose the conductive hard-mask. The top electrode can then be deposited and patterned, allowing electrical connection to the junction ([Figure \ref{fig:ch3-figure 1} (\textit{b})]). At this stage, individual devices can be characterised, where the change in resistance is measured as a function of the applied field and/or voltage. 

\subsection{Electrical and magnetic characterisation of ultra-small MTJ}

\begin{figure}[h]
    \centering
    \includegraphics[width=1\linewidth]{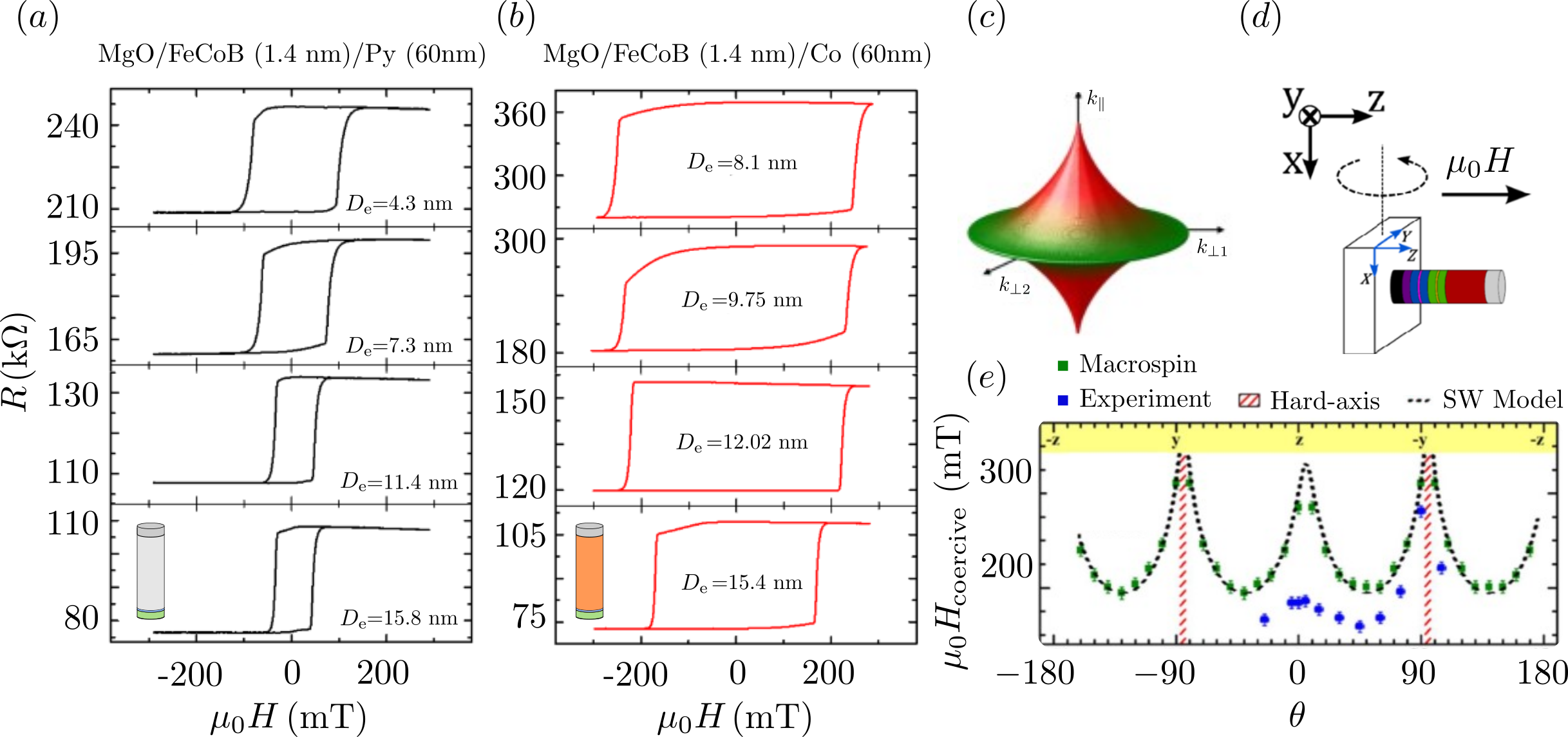}
    \caption{Dependence of the resistance on the applied field for different storage layers with  (\textit{a}) \SI{60}{\nano\meter} Py and (\textit{b}) \SI{60}{\nano\meter} Co for different values of electrical diameter. Reproduced from \cite{perrissin_highly_2018}, with permission from the Royal Society of Chemistry. (\textit{c}) 3D revolution Stoner-Wohlfarth astroid and (\textit{d}) the description of the geometry used for the dependence of the switching field on the applied field orientation. (\textit{e}) Measured switching fields for different applied field orientations (blue) and comparison with macrospin simulations (green and dashed line). Used with permission of IOP publishing, LTS, from \cite{perrissin_perpendicular_2019}; permission conveyed through Copyright Clearance Center, Inc.}
    \label{fig:ch3-figure 2}
\end{figure}

In Figure \ref{fig:ch3-figure 2} (\textit{a} and \textit{b}) there are examples of resistance versus field hysteresis loops for devices with large aspect ratio, and different storage layer materials (NiFe and Co). The expected effect of the shape anisotropy on the coercivity of the storage layer appears clearly on these, as larger magnetisation saturation results in larger coercivity. This increase of coercivity for small diameter, assuming the electrical diameter is indicative of the physical diameter, is consistent with the expectedly larger positive contribution of the shape anisotropy. Although the behaviour of the magnetisation reversal by field mimics the nature of a macrospin, further measurements are necessary to clarify this behaviour. A way to do so is by measuring the angular dependence of the switching field ($\mu_0H_\textrm{SW}$) and comparing it to the 3D Stoner-Wohlfarth astroid, as proposed in \cite{perrissin_perpendicular_2019}. For a system with rotation symmetry around the easy-axis, the 3D astroid forms a surface of revolution, as illustrated in Figure \ref{fig:ch3-figure 2} (\textit{c}), characterised by an easy-axis orientation and an isotropic hard-plane parallel to the substrate. The angular dependence of the switching field is measured by rotating the sample relative to the applied field in the \textit{YZ} and \textit{XY} planes (as illustration, the \textit{YZ}-plane geometry is shown in Figure \ref{fig:ch3-figure 2} (\textit{d})). Combining both measurements allows the hard-plane to be identified and, consequently, the easy-axis direction. Figure \ref{fig:ch3-figure 2} (\textit{e}) shows the experimental data for the \textit{YZ}-plane. Assuming that the magnetic easy-axis is along the pillar axis, its deviation from the normal to the substrate provides the pillar tilt and azimuthal angles  $(\theta, \varphi)$, calculated as 
\begin{equation}
    \textbf{k}_\parallel = \sin\theta \cos\varphi \textbf{X} + \sin\theta \sin\varphi \textbf{Y} + \cos\theta \textbf{Z}\;,
\end{equation}
where $\textbf{k}_\parallel$ is a unit vector along the easy axis of the pillar.

Additionally, examination of the experimental data allows to assess whether the system follows the macrospin approximation for a second-order uniaxial anisotropy, which follows \cite{stoner_mechanism_1948, perrissin_perpendicular_2019, jinnai_coherent_2021}
\begin{equation}
    \frac{\mu_0 H_\textrm{SW}(\varepsilon)}{\mu_0 H_\textrm{SW}(\varepsilon = 0)} = \left[\sin^{2/3}\varepsilon + \cos^{2/3}\varepsilon\right]^{-3/2},
    \label{eq:section3-SW}
\end{equation}
where $\varepsilon$ is the angle between the applied field and the easy-axis orientation of the magnetisation in the storage layer and $\mu_0 H_\textrm{SW}$ the switching field, defined as the field at which the magnetisation reverses, for different field orientations. To make sure that the storage layer is correctly simulated, we make use of the calculated cylindrical tilt angles $(\theta, \varphi)$. The switching field is computed using macrospin simulations (green points), which follows Equation \ref{eq:section3-SW} (dashed lines). Nonetheless, these simulations were realised considering that the layer behaves coherently. By comparing with the experimental results along an easy-axis orientation, although the trend is similar, there is a large reduction in the switching field for directions away from the hard axis orientation. This indicates that the state deviates from the perfectly coherent state, at least just before switching. This may be expected due to its large aspect-ratio, involving end states such as "C" or curling, and is a standard feature measured and understood in finite-size wires depending on their diameter \cite{wang_magnetic_2008, shtrikman_coercive_1959, frei_critical_1957}. These results follow nicely the previous discussion regarding the non-reversal dynamics at larger aspect-ratio. Indeed, it was concluded that, for a macrospin reversal it is necessary to reduce the aspect-ratio of the storage layer. 

\begin{figure}[h]
    \centering
    \includegraphics[width=1\linewidth]{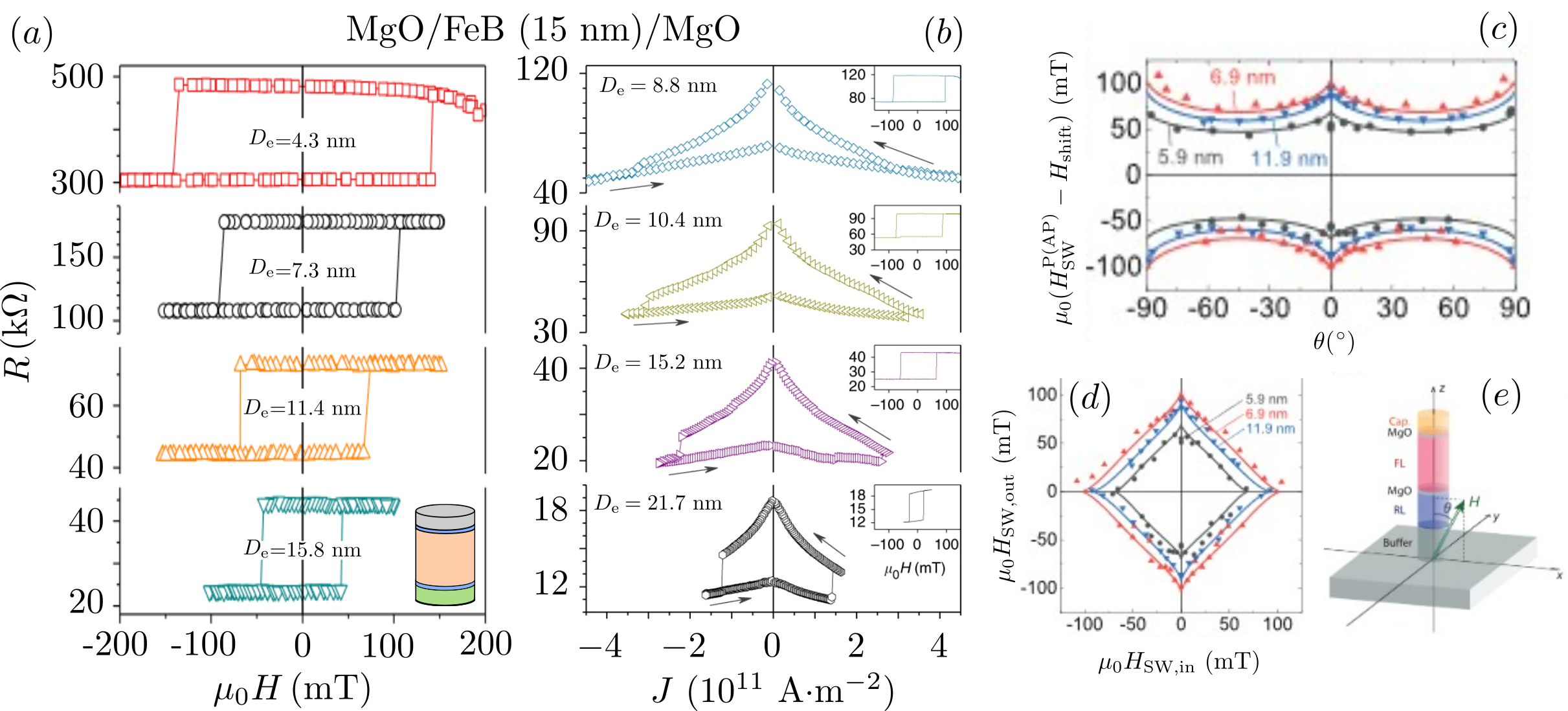}
    \caption{Dependence of the resistance on the (\textit{a}) applied field and (\textit{b}) current density for double-MgO storage layers with  \SI{15}{\nano\meter} FeB for different values of electrical diameter. Figure modified from \cite{watanabe_shape_2018}, reprinted with permission from Springer Nature. (\textit{c}) Switching field versus applied field angle. (\textit{d}) Astroid curves for different measured electrical diameters and (\textit{e}) Schematic representation of the measured device. Reprinted from \cite{jinnai_coherent_2021} with the permission of AIP publishing.}
    \label{fig:ch3-figure 3}
\end{figure}

The first experimental demonstration of STT-driven dynamics in such high aspect-ratio pillars was reported by K. Watanabe \textit{et al.} \cite{watanabe_shape_2018} for a thinner storage layer (\SI{15}{\nano\meter}) than the one reported by N. Perrissin \textit{et al.} (\SI{60}{\nano\meter}) \cite{perrissin_highly_2018}. The latter has a larger magnetic volume compared with the tunnel interface,  therefore requires a larger amount of angular momentum per unit surface of the input current for reversal, consistent with the discussions in section \ref{section-micromagnetics}. Furthermore, from the measured resistance of the parallel state $R_\textrm{p}$ in figure \ref{fig:ch3-figure 3} (\textit{a}), it appears that the resistance versus area products ($R\times A$) are significantly different for the two cases, with \SI{4.5}{\ohm\micro\meter\squared} reported for the \SI{15}{\nano\meter}-thick FeB storage layer and 12.9 and \SI{13.5}{\ohm\micro\meter\squared} for the \SI{60}{\nano\meter} thick NiFe and Co storage layers, respectively. Consequently, at a given voltage, the lower $R\times A$ enables an approximately three times larger current density to be injected before achieving the voltage breakdown of the tunnel barrier. Therefore, the combination of the smaller magnetic volume and lower $R\times A$ explains the easier STT-driven switching in devices with an electrical diameter $D_\textrm{e}$ (calculated from the respective $R_\textrm{p} \times A$) as small as \SI{8.8}{\nano\meter}, as shown in Figure \ref{fig:ch3-figure 3} (\textit{b}).

Additional experiments clarifying the coherent reversal of these  structures with moderate aspect-ratio were reported by B. Jinnai \textit{et al} \cite{jinnai_coherent_2021} through Stoner-Wohlfarth measurements. It is observed that the angular dependence follows equation \ref{eq:section3-SW}, which was not the situation for the non-uniform reversal of very large aspect-ratio pillars. 

\subsection{Temperature tolerance of the shape anisotropy}
\begin{figure}[b]
    \centering
    \includegraphics[width=1\linewidth]{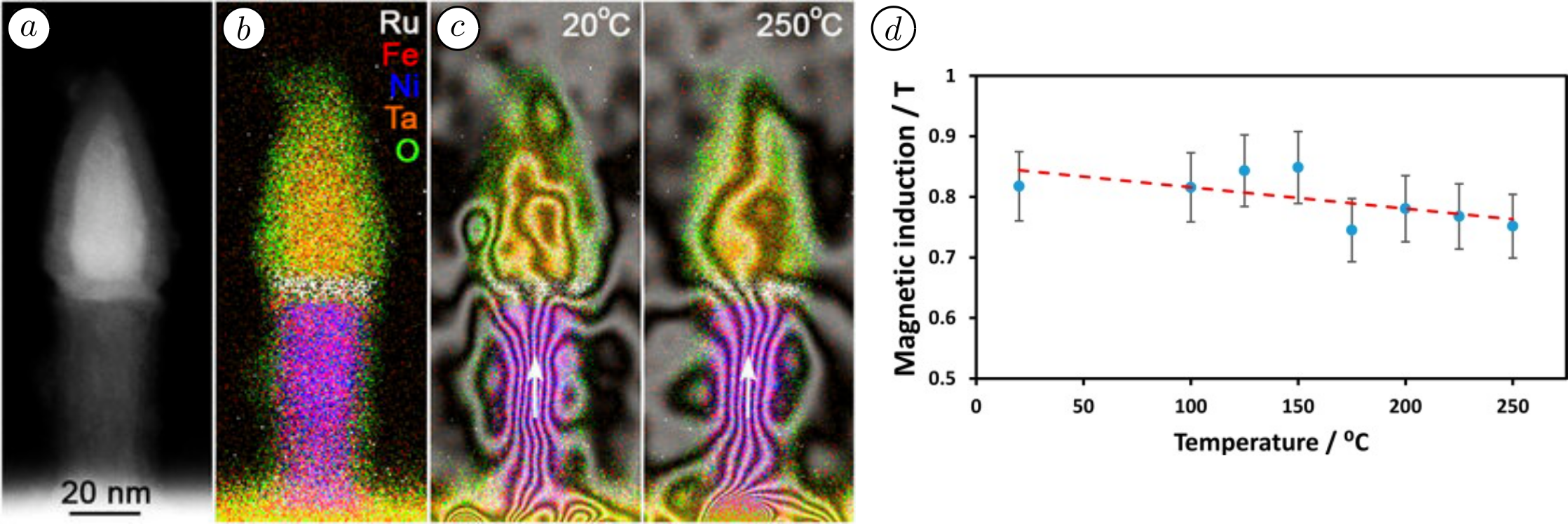}
    \caption{(\textit{a}) HAADF STEM image of a high aspect-ratio nano-pillar with a \SI{60}{\nano\meter} NiFe storage layer and (\textit{b}) associated EDX chemical map. (\textit{c}) Magnetic induction contours flowing along the major axis of the NiFe storage layer, for 20 and 250$\degree$C. Figure extracted from \cite{almeida_off-axis_2022}, with the permission of AIP Publishing.}
    \label{fig:ch3-figure 4}
\end{figure}
Besides maintaining a sizeable energy barrier down to very small diameters, another expected benefit of a high aspect-ratio storage layer is its tolerance to high operating temperature. Indeed, for the shape-anisotropy contribution, the energy barrier scales as $M_\textrm{s}^2$. By contrast, the Callen-Callen law predicts that the temperature dependence of the interfacial anisotropy scales with a power $n>3$ because of enhanced thermal fluctuations in low dimensions, which is the case in the ultrathin limit \cite{poulopoulos_1999}. In conventional perpendicular MTJ, as the main source of anisotropy comes from the surface anisotropy, the decay of stability versus temperature is acute, as it is further balanced by the in-plane favoured shape anisotropy (it self robust against temperature) \cite{watanabe_shape_2018, perrissin_highly_2018, lequeux_thermal_2020, lequeux_psa-stt-mram_2021, igarashi_temperature_2021}. A series of investigations have been performed on large aspect-ratio pillars to confirm this expectation experimentally \cite{lequeux_thermal_2020, lequeux_psa-stt-mram_2021, igarashi_temperature_2021, almeida_direct_2021}. Off-axis electron holography has been used for that purpose, as it provides a spatially resolved and quantitative measurement of the magnetic induction within and around the thick storage layer. For the case of a \SI{60}{\nano\meter} thick storage layer (shown in Figure \ref{fig:ch3-figure 4}) the measured value of magnetisation induction,  disentangled from  the demagnetising and stray fields, is $\mu_0M_\textrm{s} = 1.14\textrm{T}$. During the in-situ heating experiments, perpendicular orientation of the magnetisation is maintained up to 250$\degree$C, at which point the measured magnetic induction only reduces slightly. This observed decrease is close to the one expected for bulk Py \cite{almeida_off-axis_2022, almeida_quantitative_2022}. This confirms that $M_\textrm{s}$, and consequently, the shape anisotropy contribution to the energy barrier, allows a good resilience against high temperatures and follows approximately the bulk-material behaviour.

\begin{figure}[h]
    \centering
    \includegraphics[width=1\linewidth]{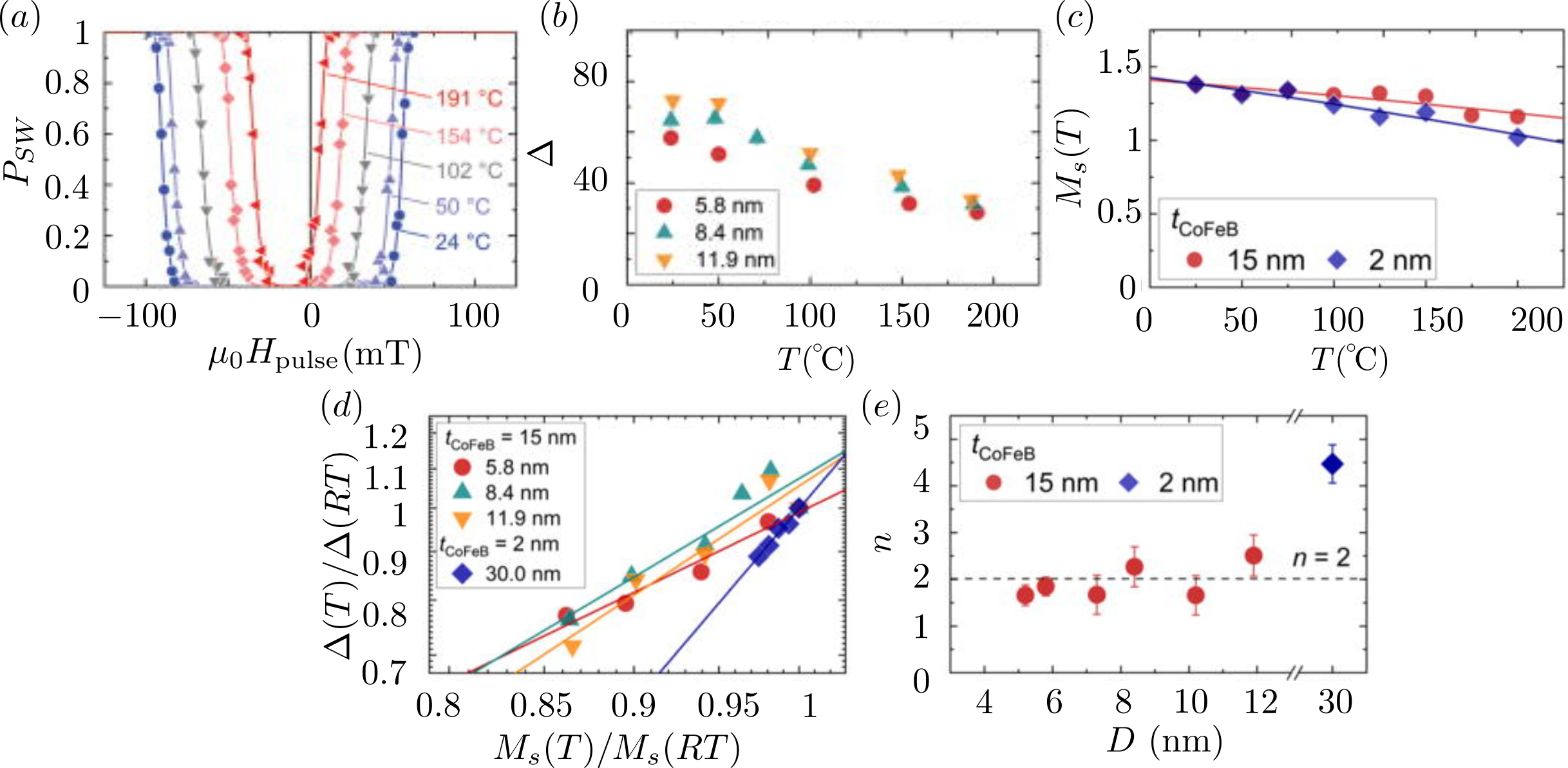}
    \caption{(\textit{a}) Switching probability as a function of the applied magnetic field at different temperatures.(\textit{b}) Temperature dependence of the extracted thermal stability factor. (\textit{c}) Saturation magnetisation as a function of temperature, results obtained from magnetometry. Scaling of the energy barrier with the saturation magnetisation, used to determine the exponent \textit{n}. (\textit{e}) Extracted scaling exponent \textit{n}. Reprinted from \cite{igarashi_temperature_2021} with the permission of AIP publishing.}
    \label{fig:ch3-figure 5}
\end{figure}

Results from Jinnai and Igarashi\textit{ et al} \cite{jinnai_coherent_2021, igarashi_temperature_2021}, allow to confirm directly the resilience of thermal stability in thick, but lower aspect-ratio pillars. The switching mechanism is close to coherent (as inferred from comparison with the Stoner-Wohlfarth astroid), and therefore a macrospin model can be trustfully applied to extract the energy barrier. The thermal stability factor can then derived through probability of switching measurements \cite{igarashi_temperature_2021} or random telegraph noise measurements \cite{jinnai_coherent_2021}, such as proven to be valid experimentally for single-domain particles \cite{wernsdorfer_experimental_1997}. The thermal stability can be then be derived from the probability of switching measurements for both P(AP) situations are shown at different temperatures as a function of an applied field for 50 events in Figure \ref{fig:ch3-figure 5} (\textit{a}). These experimental results were then fitted to the macrospin dependence of the switching probability on the applied field \cite{igarashi_temperature_2021}

\begin{equation}
    P^{\textrm{P(AP)}}_\textrm{SW} = 1 - \exp\left\{-\frac{\tau}{\tau_0}\exp\left[-\Delta\left(1\mp\frac{\mu_0H_\textrm{pulse} - \mu_0H_\textrm{shift}}{\HK}\right)^2\right]\right\}
\end{equation}
where $\tau_0$ is the inverse of the attempt frequency and was assumed to be \SI{1}{\nano\second}, $\tau$ is the pulse duration (of \SI{1}{\second} in this situation) and $\mu_0\textrm{H}_\textrm{shift}$ the shift field coming from an uncompensated reference layer observed, for instance, in Figure \ref{fig:ch3-figure 3} ($a$). The fitting to the experimental results allows to extract quantities such as the thermal stability $\Delta$ (Figure \ref{fig:ch3-figure 5} (\textit{b})) and the effective anisotropy field $\HK$. It is observed that, even though the stability decreases with increasing temperature, there is still a sizeable stability at temperatures as large as 190$\degree$C. To confirm that the shape anisotropy is responsible for this moderate decrease, and compare it with the decrease of effective field arising from interface anisotropy in the case of a flat p-MTJ, the decay of magnetisation with temperature was measured (Figure \ref{fig:ch3-figure 5} (\textit{c})). The thicker film (\SI{15}{\nano\meter}) exhibits a weaker reduction of $M_\textrm{s}$ with increasing temperature, consistent with the bulk-like behaviour expected \cite{almeida_quantitative_2022, sato_temperature-dependent_2018}. The relationship between the temperature-dependent anisotropy energy density and $M_\textrm{s}$ is analysed in the logarithmic plot shown in Figure \ref{fig:ch3-figure 5} (\textit{d}), using the power law expression \cite{callen_present_1966} 
\begin{equation}
    \frac{E_\textrm{B}(T)}{E_\textrm{B}(T^*)} = \left[\frac{M_\textrm{s}(T)}{M_\textrm{s}(T^*)}\right]^n\;,
\end{equation}
in which the exponent \textit{n} depends on the type of anisotropy, related to both dimensionality and symmetry, and $T^*$ is a reference temperature. For the case of usual p-MTJ, \textit{n} values are usually measured above 3 \cite{igarashi_magnetic-field-angle_2017, enobio_evaluation_2018} in patterned devices, and around 2 - 3 in thin film measurements \cite{sato_temperature-dependent_2018}. For the situation of the shape anisotropy MTJ (red data points in Figure \ref{fig:ch3-figure 5} (\textit{e})), this value is centred around 2, as expected from the dominant contribution of the shape anisotropy ($\propto \Ms^2$) to the energy barrier. 

The smaller \textit{n} for thick pillars confirms the advantage of the shape anisotropy MTJ for operation at elevated temperatures. This is relevant for applications in which the operating temperature exceeds that of commercial use (around 70$\degree$C), such as industrial (around 85$\degree$C), automotive (around 125$\degree$C) and military (around 150$\degree$C). It is also relevant for solder reflow applications (around 260$\degree$C), in which the information is loaded prior to the reflow soldering. 

\section{Perspectives for technological deployment at low pitch}
\label{section-perspectives}
In the previous sections we reviewed the benefits of a high aspect-ratio storage layer. Based on both simulations and experiments, we showed that it is possible to reverse these thick layers using a magnetic field as well as with spin-transfer torque. Additionally, these allow to expand the operation range of magnetic memory devices to higher-temperature applications, thanks to the robust resilience of the saturation magnetisation to temperature. Nonetheless, achieving thermal stability through such a large aspect-ratio can introduce substantial drawbacks in switching. Additional limitations arise in a technological deployment at extreme low pitch, from both a fabrication standpoint and from device-to-device magnetic inter-cell crosstalk, which is expected to be larger for the high aspect-ratio storage layer. Both these limitations are inherent to high density MTJ, but they are a bottleneck for the high aspect-ratio devices. We first focus on the impact of the neighbouring stray field in a device array and its impact on the stability factor. We will then conclude with an approach to overcome both non-uniform reversal and large stray field, using a multilayered MTJ structure. 

\subsection{Stray field in densely packed structures}

\begin{figure}[h!]
    \centering
    \includegraphics[width=1\linewidth]{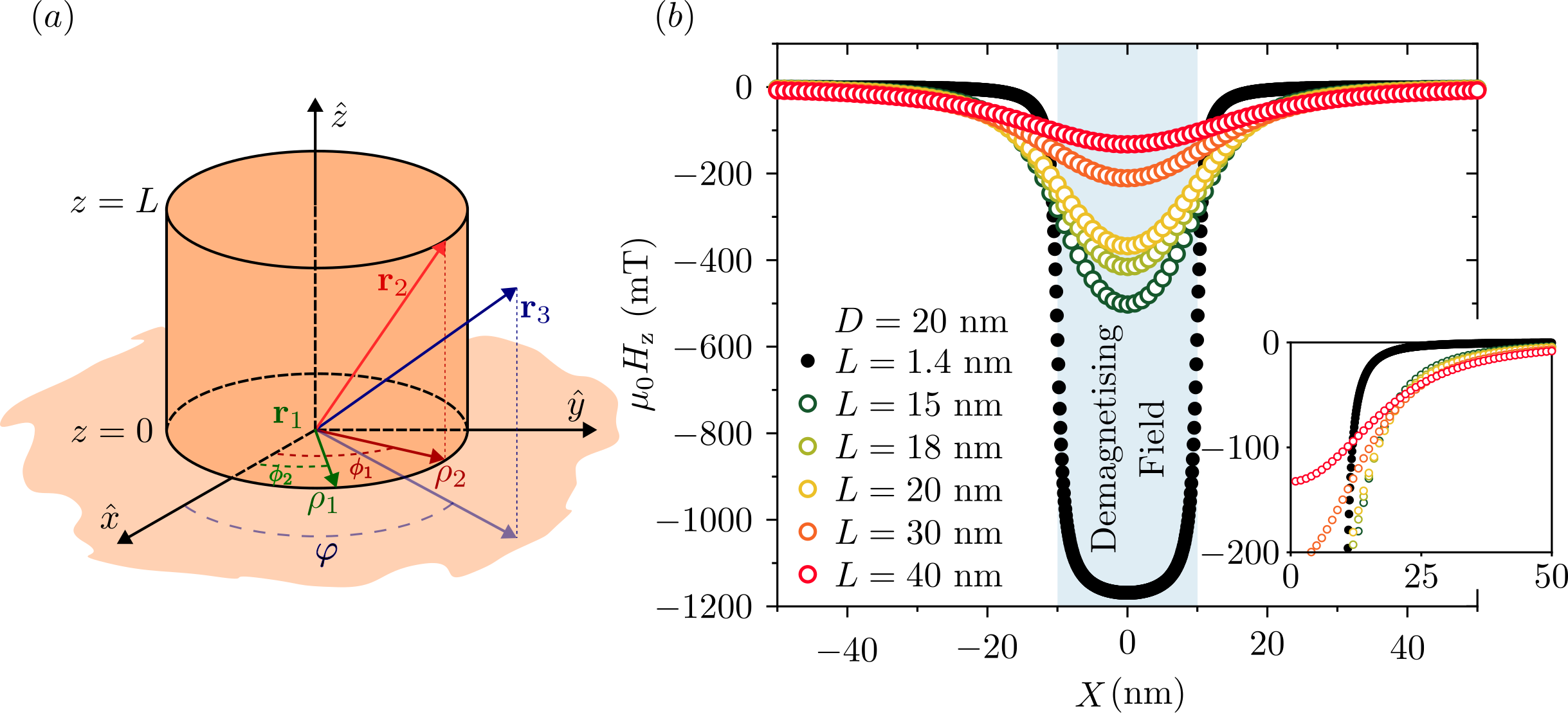}
    \caption{($a$) Schematics of the framework of cylindrical coordinates used to compute the potential vector and ($b$) magnetic field strength at mid thickness, for increasing aspect ratio at a fixed diameter of \SI{20}{nm}. The region of the demagnetising field is shown with a bluish colour. Inset of the figure shows a close-up of the decay of the magnetic stray field in space.}
    \label{fig:ch4-figure 1}
\end{figure}

The stray field affecting the storage layer of magnetic tunnel junctions needs to be evaluated and possibly compensated, to ensure the proper operation as a memory. A first source of stray field is internal to the MRAM cell, arising from an uncompensated reference layer, which can be optimised through material and device fabrication \cite{veiga_control_2023}. When the distance between neighbouring devices becomes small, the effect of the field emanating from the neighbouring storage layers becomes non-negligible, increasing as the pitch (distance between the centre of two neighbouring devices) decreases. Although not particularly detrimental for conventional perpendicular MTJ at a reasonable pitch of around twice the device diameter, as dipolar fields are short-ranged in 2D systems, it can be critical at extreme low pitch \cite{wu_impact_2020}. This cross-talk is expected to be more pronounced in high aspect-ratio storage layers, due to the larger magnetic volume and the slower lateral decay of the stray field in systems with larger dimensionality \cite{fruchart_high_1998}. To evaluate the increase in stray field compared with the usual perpendicular MTJ, we make use of a Fourier transform formalism \cite{taniguchi_analytical_2018, cacoilo_dipole-coupled_2024}. The magnetic field components $H_x$, $H_{y}$ and $H_{z}$ are computed through the magnetic scalar potential $\varphi_m$ as 

\begin{equation}
    \textbf{H} = \left(-\frac{\partial \varphi_\textrm{m}}{\partial_\textrm{x}}, -\frac{\partial \varphi_\textrm{m}}{\partial_\textrm{y}}, -\frac{\partial \varphi_\textrm{m}}{\partial_\textrm{z}}\right)  
    \label{eq:section4:field}
\end{equation}
where

\begin{equation}
    \varphi_\textbf{m} (\textbf{r}_3)= M_\textrm{s}\int_{\textrm{S}_1} G(\textbf{r}_3 - \textbf{r}_1) \textrm{d}\textbf{r}_1- M_\textrm{s}\int_{\textrm{S}_2} G(\textbf{r}_3 - \textbf{r}_2) \text{d}\textbf{r}_2,  
\end{equation}
with $S_1$ and $S_2$ the bottom and top surfaces, respectively, of the magnetic cylinder, $\textbf{r}$ an arbitrary point in space and $G(\textbf{r}-\textbf{r}') = 1/(4\pi|\textbf{r}-\textbf{r}'|)$ the associated Green function. These integrals are expressed in cylindrical coordinates through the system described in Figure \ref{fig:ch4-figure 1} ($a$)

\begin{equation}
    \varphi_\textrm{m} = M_\textrm{s}\int_0^{R_0} r\textrm{d}\rho\int_0^{2\pi} \rho \textrm{d}\phi \frac{1}{\sqrt{r^2 + (Z-L)^2}} -  M_\textrm{s}\int_0^{R_0} r\textrm{d}\rho\int_0^{2\pi} \rho \textrm{d}\phi \frac{1}{\sqrt{r^2 + Z^2}}
\end{equation}
where $r^2 = R^2 + \rho^2 - R\rho\cos(\phi-\varphi)$ is obtained for the cylindrical system, \textit{L} is the thickness of the cylinder, $R_0$ the radius of the cylinder and $M_\textrm{s}$ its spontaneous magnetisation. This integral is then calculated for any given value along $Z$ and $r$. Using the Gegenbauer’s addition theorem for the Bessel function \cite{taniguchi_analytical_2018}, we can expand the integrand as
\begin{equation}
    \frac{1}{\sqrt{r^2 + (Z-L)^2}} = \int_0^\infty e^{-k|Z-L|}\mathcal{J}_0(kr)~\textrm{d}k ,
\end{equation}
where $\mathcal{J}_0(kr)$ is a first order Bessel function. 

\begin{figure}[h!]
    \centering
    \includegraphics[width=1\linewidth]{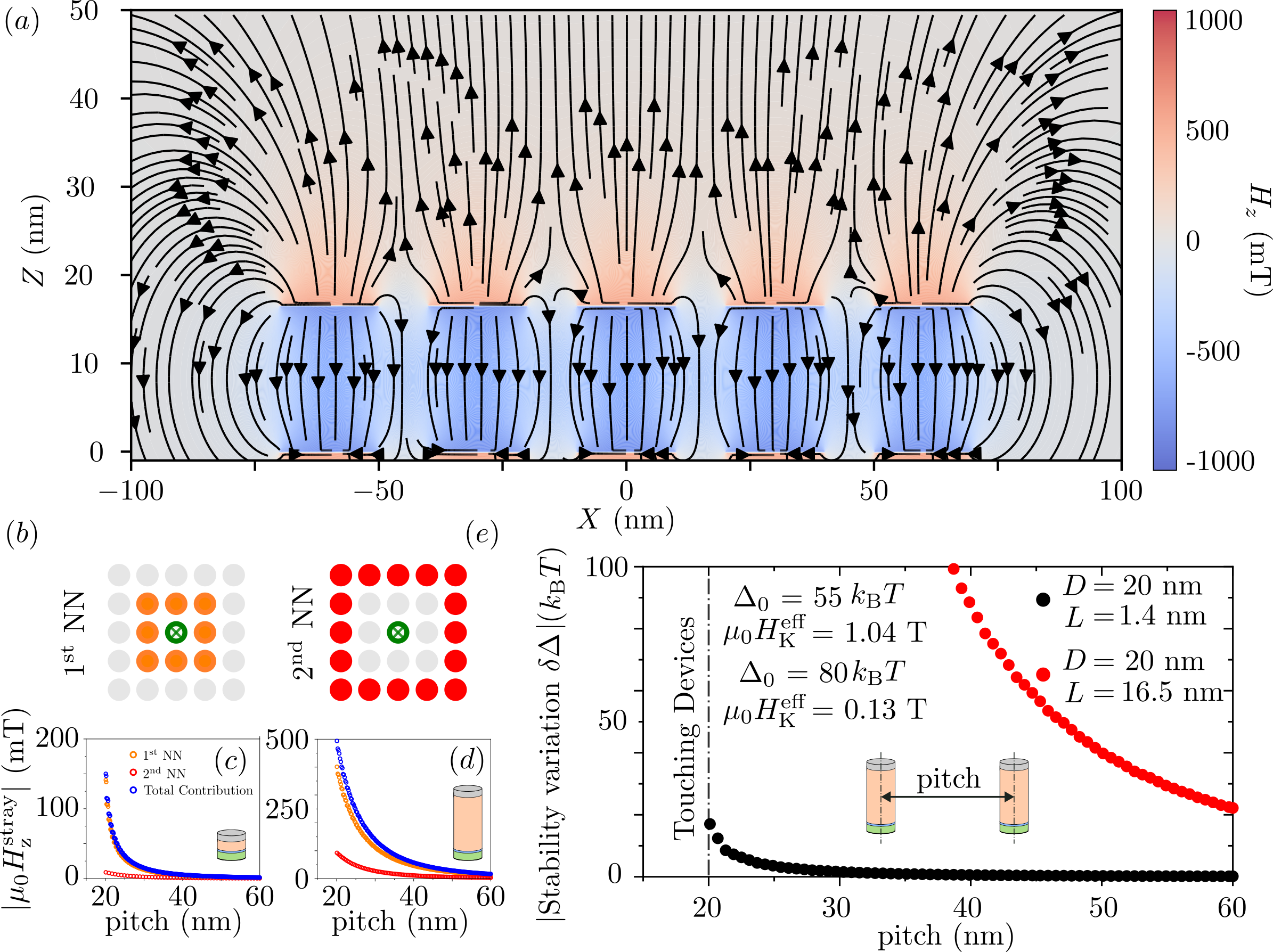}
    \caption{(\textit{a}) 2D plot of the magnetic field lines of 5 perpendicularly-magnetised cylinders with diameter of \SI{20}{nm} and a thickness of \SI{16.5}{nm}. The colour map is related to the perpendicular component of the magnetic field. (\textit{b}) Schematic of the first or first two rings of neighbours, acting on the central pillar. Magnitude of the average stray field arising for (\textit{c}) a flat MTJ pillar (\SI{20}{nm} diameter and \SI{1.4}{nm} thickness) and (\textit{d}) for a high aspect-ratio MTJ (\SI{20}{nm} diameter and \SI{16.5}{nm} thickness). (\textit{e}) Dependence of the variation in thermal stability factor on the pitch between neighbouring devices for the case of a flat storage layer (black dots) and high aspect-ratio storage layer (red dots).}
    \label{fig:ch4-figure 2}
\end{figure}

Through this modification and Equation \ref{eq:section4:field} it is possible to express both radial and perpendicular components of the magnetic field generated by a magnetised cylinder as

\begin{equation}
    H_r(r, z) = 2\pi R_0 M_s\int_0^\infty dk \left[ \textrm{sign}(Z - L)e^{-k|Z-L|} - \textrm{sign}(Z)e^{-k|Z|}\right]\mathcal{J}_0(kr)\mathcal{J}_1(kR_0),
\end{equation}
where $\mathcal{J}_x$ are Bessel functions of the order \textit{x}. These integrals can be further simplified through some transformations, leading to a somewhat straightforward calculation of the different components of the magnetic field (refer to reference \cite{taniguchi_analytical_2018} for further details). 

In Figure \ref{fig:ch4-figure 1} ($b$), the perpendicular component of the magnetic field is displayed across the middle of a pillar. A striking feature is the decay of the stray field for the case of a flat storage layer, compared to that of a larger aspect-ratio. For the smallest aspect-ratio, the stray field decay is sharp, while for the larger aspect-ratio, it extends considerably farther from the pillar, even in relative values \cite{cacoilo_dipole-coupled_2024}. In a second step, we can compute the stray field acting on a central magnetic storage layer and arising from its neighbours. As in two dimensions the extension of the stray field is short-ranged, it is possible to evaluate its impact based on a finite-size magnetic array, which we chose to be $5\times5$ [stray field lines are shown as an example in Figure \ref{fig:ch4-figure 2} ($a$)]. In Figure \ref{fig:ch4-figure 2} ($c$) we can see the effect of the stray field on the pitch between neighbouring devices [given by the schematics of Figure \ref{fig:ch4-figure 2} ($b$)] for the situation of the flat MTJ and ($d$) for the situation of the high aspect-ratio MTJ. For the flat MTJ, the stray field is significant when the devices are in contact but becomes negligible at a pitch above \SI{30}{nm}. Moreover, the impact of the second neighbours is insignificant, so considering the impact of the first neighbours is enough \cite{wu_impact_2020}. On the other hand, for the larger aspect-ratio pillars, a stronger effect of the stray field is observed, and the impact of the second nearest neighbours cannot be neglected \cite{cacoilo_dipole-coupled_2024}. Further rows of neighbours can be safely neglected thanks to the 1/$r^2$ decay of the stray field. The stray field in a dense array of large aspect-ratio storage layers is significant, and it is expected to affect the stability factor of the device. The stability factor is increased (reduced) if the average stray field points in the same (opposite) direction of the pillar magnetisation as \cite{khvalkovskiy_basic_2013}:
\begin{equation}
    \Delta_{\kbt}^\textrm{stray} = \Delta_{\kbt}^\textrm{0} \left( 1 \pm \left|\frac{\mu_0H_z^\textrm{stray}}{\HK}\right| \right)^2
    \label{eq:ch4:stability_hz}
\end{equation}
where $\HK$ is the effective anisotropy field of the magnetic element and $\mu_0H_z^\textrm{stray}$ the volume-averaged perpendicular stray field felt by the central magnetic device. This effect can be observed in Figure \ref{fig:ch4-figure 2} (\textit{e}), where the magnitude of the variation of the stability factor ($\delta\Delta = |\Delta_{\kbt}^{\textrm{stray}} - \Delta_{\kbt}^\textrm{0}|$) is displayed as a function of the pitch between devices, for the case of a flat MTJ and the high aspect-ratio MTJ. 
\begin{figure}[h!]
    \centering
    \includegraphics[width=1\linewidth]{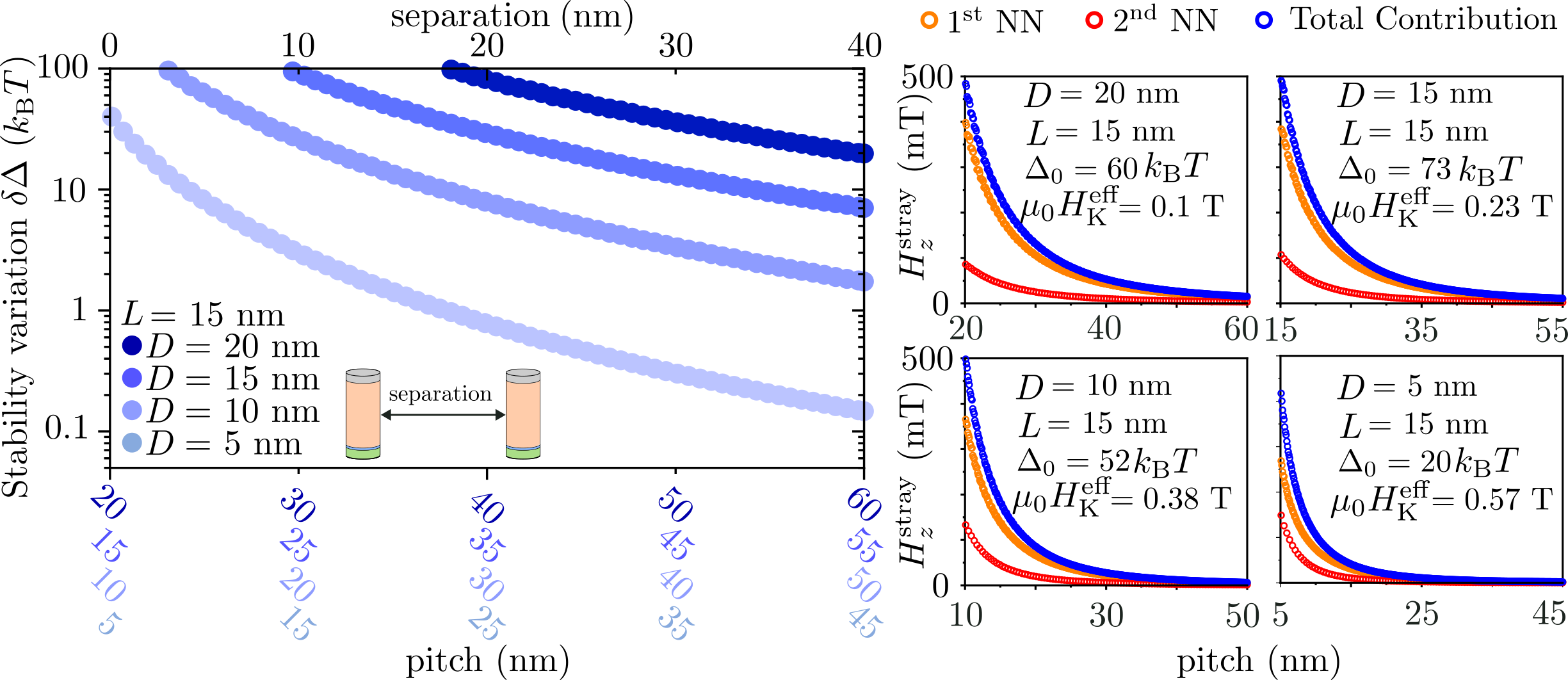}
    \caption{Variation in thermal stability as a function of the pitch for different aspect-ratio pillars at a fixed thickness of 15 nm and each individual dependence of the perpendicular component of the stray field for increasing pitch.}
    \label{fig:ch4-figure 3}
\end{figure}

The practical consequence for the inter-cell stray field can then be evaluated from the criterion of maximum allowance on variability of stability $\delta\Delta$ (reduction or enhancement). Taking the example here of a high-aspect-ratio storage layer with $\Delta_0 = 80\kbt$, we see that for a pitch of \SI{60}{\nano\meter} we still have a variation of $\delta\Delta\approx20\kbt$, which is detrimental for applications (either in terms of storage, but also because of asymmetry in switching currents). By comparison, the flat storage layer achieves $\delta\Delta \ll 5\kbt$ at a reasonable pitch of \SI{30}{\nano\meter}. High-aspect ratio MTJs might then require a reasonable device-device spacing to preserve their intrinsic properties, bringing a limitation in areal density. This strong stray-field sensitivity comes from the the combination of the long-range dipolar field generated and the lower $\HK$, which increases substantially the ratio $\mu_0H_z^\textrm{stray}/\HK$ in equation \ref{eq:ch4:stability_hz}. 

The sensitivity of the energy barrier to device-device interactions can be reduced in two ways: by decreasing the stray field generated by the neighbouring cells, or by increasing the effective anisotropy field of the storage layer. Reducing the thickness of the pillar for a fixed diameter can reduce its magnetic volume and thus its dipolar fields. However, when the perpendicular shape anisotropy is the main contributor to the energy barrier, this will also decrease the storage layer intrinsic thermal stability. In figure \ref{fig:ch4-figure 3} we examine a second alternative, in which the thickness is kept constant, but the diameter is reduced (therefore increasing its aspect-ratio). Despite the reduction in magnetic volume, we see that the first-nearest-neighbour changes only marginally because at smaller diameter, the pillars are closer to each other. At this stage, the relative contribution of the second-nearest neighbours becomes more significant. Through the increase in aspect-ratio, the increase in $\HK$ reduces the ratio $\mu_0H_z^\textrm{stray}/\HK$ and makes the device's thermal stability more resilient to inter-cell stray fields. However, the thermal stability remains limited to around 20$\kbt$, and simply increasing the thickness does not  necessarily restore the stability, as above a certain critical thickness the reversal becomes non-uniform and the energy barrier is capped by the domain-wall energy. A promising route is therefore to increase $\HK$ without relying exclusively on the large aspect-ratio. Multiple ferromagnet/insulator interfaces can be introduced along the storage layer to provide multiple surface anisotropies \cite{nishioka_novel_2019, jinnai_fast_2021, igarashi_single-nanometer_2024}, reducing the total magnetic thickness of the storage layer. This simultaneously reduces the generated dipolar field because of its lower magnetic volume and aspect-ratio, but also reduces the sensitivity to inter-cell stray fields through a significant increase in $\HK$. These multilayered storage layers may then have an adequate thermal stability at extremely low diameters, with a more robust operation in a densely packed array. 

While the multilayered structure reduces the cross-talk by increasing $\HK$, another strategy is to directly reduce the dipolar field generated by each device. To achieve this, a complementary approach is the dipole-coupled perpendicular shape anisotropy MTJ \cite{cacoilo_dipole-coupled_2024}. In this design, the storage layer (core) is surrounded by a magnetic shell whose magnetisation aligns antiparallel to that of the storage layer thanks to the large dipolar coupling. This strong compensation reduces the outbound stray field, being concentrated to the now core-shell coupled system. At the same time, this coupled system offers higher thermal stability than an isolated storage layer, allowing to reduce its combined thickness. This architecture mitigates two limitations of the single storage layer: the non-uniform reversal and its associated long switching time and large switching voltages, and the cross-talk in dense arrays. Its fabrication is, however, more demanding than that of an isolated MTJ, as it requires precise dimensional control and conformal deposition of a thin magnetic shell around an electrically isolated MTJ. 

\subsection{Multilayered storage layer as a prospect for high density}
\begin{figure}[b!]
    \centering
    \includegraphics[width=1\linewidth]{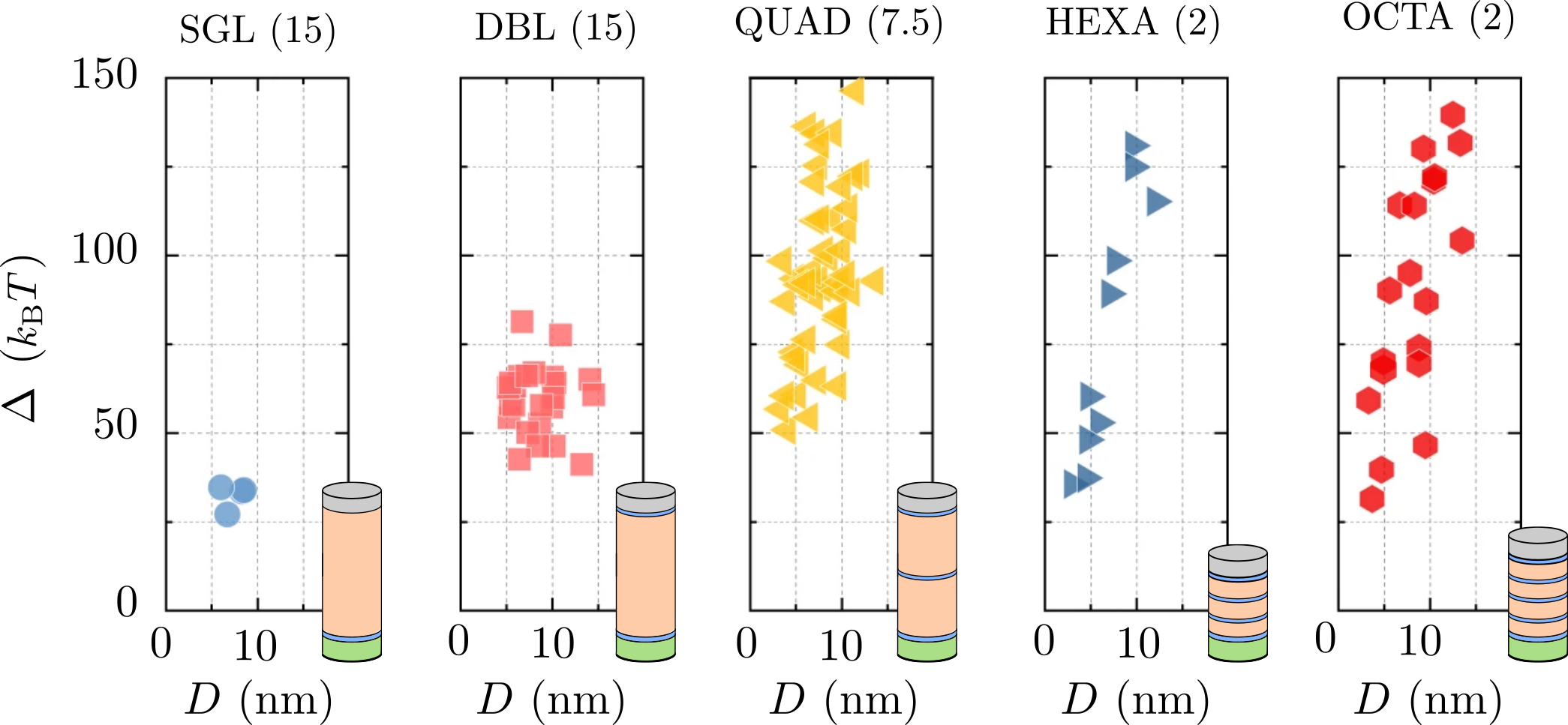}
    \caption{Dependence of the thermal stability factor $\Delta$ on the measured electrical diameter $D_\textrm{e}$ for different storage layer designs containing one (SGL), two (DBL), four (QUAD), six (HEXA) and eight (OCTA) ferromagnet/insulator interfaces. The indicated thickness corresponds to the thickness of each continuous magnetic layer. Modified from \cite{igarashi_single-nanometer_2024}, with the permission of Springer Nature.}
    \label{fig:ch4-figure 4}
\end{figure}
As previously introduced, an approach to obtain large stability at lower diameter is the use of an enhanced surface anisotropy, reducing the dependence of the stability on the aspect-ratio of the storage layer. This approach was reported by Nishioka et al \cite{nishioka_novel_2019, nishioka_novel_2020}, intercalating an MgO thin spacer in the storage layer. This thin layer introduces two additional FeCoB/MgO interfaces, while keeping both CoFeB coupled, therefore enhancing the total contribution of the surface anisotropy to the thermal stability of the now composite storage layer. By doing so, it is possible to increase the stability at lower nodes without significantly increasing the switching current. At this point, several considerations regarding the spacer MgO layer need to be addressed, such as the decrease of TMR due to an increase in series resistance, and the increase in $\RA$, which can impact the switching current. These contributions can be reduced by using a thinner MgO barrier layer, which allows for a smaller series resistance, lowering the impact on the TMR of the device. 

Nonetheless, this reduction still needs to be coupled with a sizeable surface anisotropy from the insertion layer. This approach was later used in the context of the high aspect-ratio MTJ \cite{jinnai_high-performance_2020, jinnai_fast_2021, igarashi_single-nanometer_2024}, as a means to overcome the limitation of the non-uniform reversal. Additionally, through the use of thicker magnetic layers, the positive shape anisotropy remains, alongside an increase in the effective anisotropy field. In addition to this design, it is also possible to make use of more than one insertion layers to continuously increase the effect of the surface anisotropy throughout the storage layer. 

In Figure \ref{fig:ch4-figure 4}, the stability for different storage layer structures is presented as a function of the measured electrical diameter \cite{igarashi_single-nanometer_2024}. Different acronyms are used, which relate to the number of FeCo/MgO interfaces: SGL (single layer) is related to a single interface, DBL (double layer) to a storage layer with a capping MgO, QUAD to 4 FeCo(B)/MgO interfaces, HEXA to six interfaces and OCTA to 8 interfaces. It is observed that, for the situation of a single storage layer, there is a significant enhancement of stability when a second MgO barrier is used as a capping layer. Using an additional insertion MgO layer, we can further increase the stability of the device at the same diameter range, although the spread between devices increases. Further improvements can be done using additional insertions through the storage layer, in which the consequent layers are coupled through dipolar and direct exchange coupling throughout the insertion layers, such as the HEXA and OCTA structures \cite{jinnai_high-performance_2020, jinnai_fast_2021, igarashi_single-nanometer_2024}. These allow for smaller magnetic volume with enhanced stability, which following the micromagnetic simulations discussed in section \ref{section-micromagnetics}, is associated with faster STT-switching at lower applied voltage. 

\begin{figure}[h!]
    \centering
    \includegraphics[width=1\linewidth]{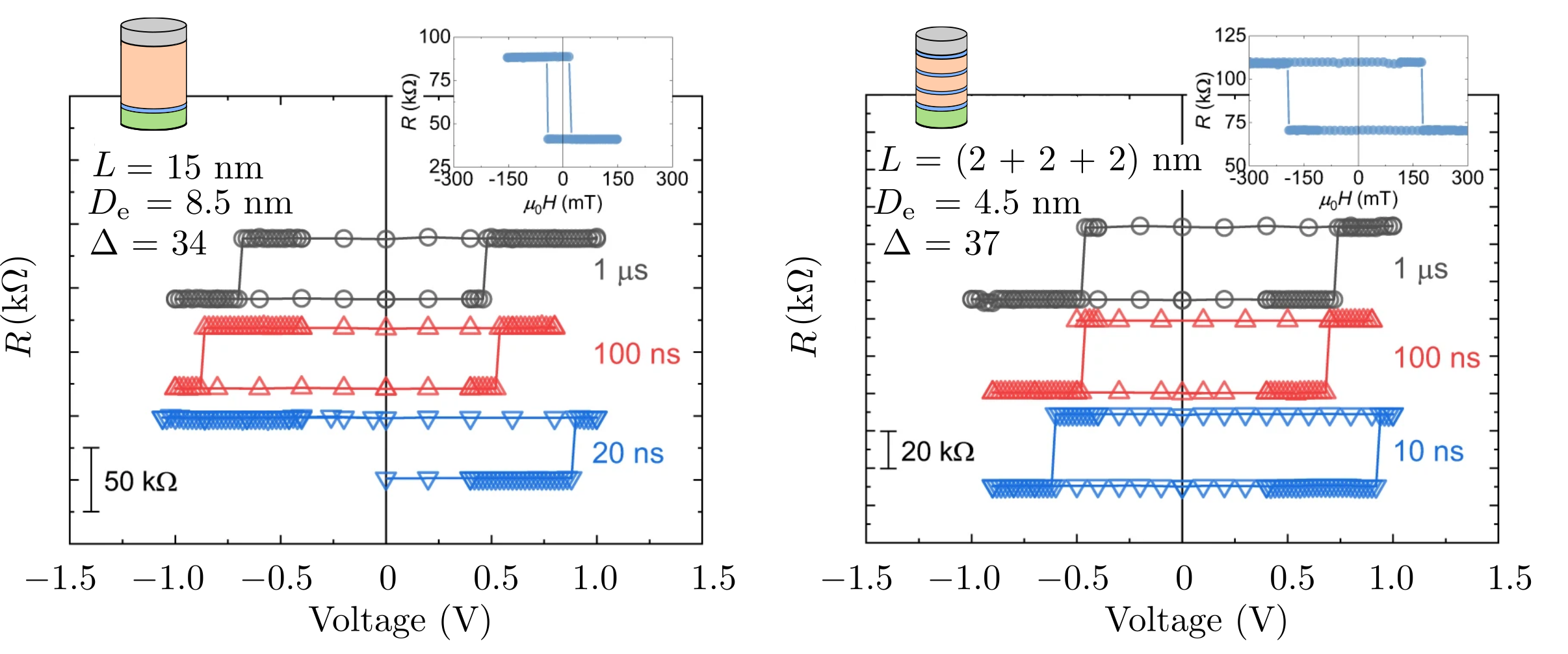}
    \caption{MTJ resistance as a function of the applied voltage for the situation of a single CoFeB/MgO interface and six CoFeB/MgO interfaces (HEXA structure), at different voltage pulses. Figure modified from \cite{igarashi_single-nanometer_2024}, with the permission of Springer Nature.}
    \label{fig:ch4-figure 5}
\end{figure}

To assess the advantage for STT writing of using layered structures to boost $\HK$ while keeping the aspect-ratio moderate, magnetisation reversal as a function of the applied voltage is shown in Figure \ref{fig:ch4-figure 5}, for the case of a single storage layer and a multilayered storage layer with 2 MgO insertions (HEXA structure). Although the diameter of the HEXA structure is reduced to that of the single layer, the value of stability is maintained. Additionally, thanks to the increased anisotropy field, a smaller magnetic thickness is necessary, which allows to reverse the magnetisation at lower applied voltages and shorter pulses (up to \SI{10}{\nano\second}). Compared with the single-interface high aspect-ratio storage layers, these multilayered structures exhibit shorter switching times because their enhanced $\HK$ reduces the characteristic switching time $\tau_\textrm{D}$ \cite{cacoilo_spin-torque-triggered_2021, jinnai_fast_2021, igarashi_single-nanometer_2024}. Moreover, thanks to the larger effective anisotropy field and smaller magnetic volume, this multilayered MTJ has prospects for very large density, with a projected density of around \SI{100}{Gbit\per\centi\meter\squared} \cite{shinoda_pitch_2024}. 
\newpage
\section{Conclusion}
\label{section-conclusion}

The use of a storage layer with a vertical aspect-ratio shows promising results for a high stability and high density STT-MRAM at sub-\SI{20}{nm} diameters. However, for this 3D design to be a realistic candidate among the different emerging non-volatile technologies, several challenges need to be addressed, such as the capping in stability and longer reversal times brought by non-uniform reversal and the increased inter-cell stray field at large bit densities. In this review, micromagnetics and analytics have been used to explain and motivate the presented experimental results. In detail, the vertical domain-wall limits the stability that we can obtain in the device if we only resort to the geometry of the storage layer. Additionally, both switching times and the voltage necessary for switching increase significantly for larger aspect-ratios, which is not directly proportional to the intrinsic stability of the device. These consequences can be somewhat overcome by reducing the aspect-ratio of the storage layer, for instance through an increase in the magnetisation saturation. Nonetheless, in matters related to scalability, using a multilayered structure stands out, where the original source of perpendicular anisotropy from the FeCo(B)/MgO interface can be further multiplied. Doing so, not only does the stability increase, but the switching voltage and switching times are significantly reduced, matching that of the DRAM technology (order of tens of ns). This is attributed to both the smaller magnetic volume we need to reverse, thanks to the smaller aspect-ratio, but also because of the increased perpendicular anisotropy field. These types of structures become notably interesting at lower nodes and tight pitch, as the variations of stability with the neighbouring stray fields from the magnetic array are small for layers with large effective anisotropy field. These different contributions hold promise in advancing high-density STT-MTJ technology across a spectrum of applications. Notably, we give compelling arguments for the use of the storage layer with vertical aspect-ratio, thanks to its small temperature dependence and the multilayered structure for high density STT-MTJ thanks to the smaller switching voltages and faster switching times. At this point, novel fabrication techniques become necessary, as the reliable and reproducible fabrication of ultra-small dimensions and tight pitch is not a trivial task. Nonetheless, as soon as they catch up, we are confident that this design will become competitive with commercial STT-MTJ technology. 

\section*{Data availability statement}
No new data were created or analysed in this study. 

\section*{Acknowledgments}
The authors would like to thank Natalia Boscolo-Meneguolo, Nicolas Perrissin, Liliana Buda-Prejbeanu, R. C. Sousa and Bernard Diény for discussion. This work was partly supported by Samsung Electronics Co., Ltd. (IO190709-06540-02), the ERC Advanced Grant MAGICAL NO. 669204. and ANR M-bed-RAM (ANR-23-CE24-0016). N. Caçoilo acknowledges financial support from the CFR thesis funding by CEA. The devices from SPINTEC were fabricated by the French RENATECH network and supported by its technical team, implemented at the Upstream Technological Platform in Grenoble PTA  (ANR-22-PEEL-0015). S. Fukami acknowledges support from MEXT X-NICS (JPJ011438), JSPS Kakenhi (24H00039 and 24H02235) and JST-ASPIRE JPMJAP2322. 

\section*{Conflict of interest}
All the authors have no conflicts of interest to disclose.

\printbibliography[heading=bibintoc]

@article{apalkov_magnetoresistive_2016,
	title = {Magnetoresistive {Random} {Access} {Memory}},
	volume = {104},
	doi = {10.1109/JPROC.2016.2590142},
	number = {10},
	journal = {Proceedings of the IEEE},
	author = {Apalkov, D. and Dieny, B. and Slaughter, J. M.},
	year = {2016},
	pages = {1796--1830},
}

@article{locatelli_spin-torque_2014,
	title = {Spin-torque building blocks},
	volume = {13},
	doi = {10.1038/nmat3823},
	journal = {Nature Materials},
	author = {Locatelli, N. and Cros, V. and Grollier, J.},
	year = {2014},
	pages = {11--20},
}

@article{dieny_opportunities_2020,
	title = {Opportunities and challenges for spintronics in the microelectronics industry},
	volume = {3},
	doi = {10.1038/s41928-020-0461-5},
	journal = {Nature Electronics},
	author = {Dieny, B. and Prejbeanu, I. L. and Garello, K. and Gambardella, P. and Freitas, P. and Lehndorff, R. and Raberg, W. and Ebels, U. and Demokritov, S. O. and Akerman, J. and Deac, A. and Pirro, P. and Adelmann, C. and Anane, A. and Chumak, A. V. and Hirohata, A. and Mangin, S. and Valenzuela, Sergio O. and Onbaşlı, M. Cengiz and d’Aquino, M. and Prenat, G. and Finocchio, G. and Lopez-Diaz, L. and Chantrell, R. and Chubykalo-Fesenko, O. and Bortolotti, P.},
	year = {2020},
	pages = {446--459},
}

@article{sun_spin-torque_2013,
	title = {Spin-torque switching efficiency in {CoFeB}-{MgO} based tunnel junctions},
	volume = {88},
	doi = {10.1103/PhysRevB.88.104426},
	journal = {Physical Review B},
	author = {Sun, J. Z. and Brown, S. L. and Chen, W. and Delenia, E. A. and Gaidis, M. C. and Harms, J. and Hu, G. and Jiang, Xin and Kilaru, R. and Kula, W. and Lauer, G. and Liu, L. Q. and Murthy, S. and Nowak, J. and O’Sullivan, E. J. and Parkin, S. S. P. and Robertazzi, R. P. and Rice, P. M. and Sandhu, G. and Topuria, T. and Worledge, D. C.},
	year = {2013},
	pages = {104426},
}

@inproceedings{endoh_recent_2020,
	title = {Recent {Progresses} in {STT}-{MRAM} and {SOT}-{MRAM} for {Next} {Generation} {MRAM}},
	doi = {10.1109/VLSITechnology18217.2020.9265042},
	booktitle = {{IEEE} {Symposium} on {VLSI} {Technology}},
	author = {Endoh, Tetsuo and Honjo, Hiroaki and Nishioka, Koichi and Ikeda, Shoji},
	year = {2020},
	pages = {1--2},
}

@inproceedings{choe_recent_2023,
	title = {Recent {Technology} {Insights} on {STT}-{MRAM}: {Structure}, {Materials}, and {Process} {Integration}},
	doi = {10.1109/IMW56887.2023.10145822},
	booktitle = {{IEEE} {International} {Memory} {Workshop}},
	author = {Choe, Jeongdong},
	year = {2023},
	pages = {1--4},
}

@article{krizakova_spin-orbit_2022,
	title = {Spin-orbit torque switching of magnetic tunnel junctions for memory applications},
	volume = {562},
	doi = {10.1016/j.jmmm.2022.169692},
	journal = {Journal of Magnetism and Magnetic Materials},
	author = {Krizakova, Viola and Perumkunnil, Manu and Couet, Sébastien and Gambardella, Pietro and Garello, Kevin},
	year = {2022},
	pages = {169692},
}

@article{ikeda_tunnel_2008,
	title = {Tunnel magnetoresistance of 604\% at {300K} by suppression of {Ta} diffusion in {CoFeB/MgO/CoFeB} pseudo-spin-valves annealed at high temperature},
	volume = {93},
	doi = {10.1063/1.2976435},
	journal = {Applied Physics Letters},
	author = {Ikeda, S. and Hayakawa, J. and Ashizawa, Y. and Lee, Y. M. and Miura, K. and Hasegawa, H. and Tsunoda, M. and Matsukura, F. and Ohno, H.},
	year = {2008},
	pages = {082508},
}

@article{scheike_631_2023,
	title = {631\% room temperature tunnel magnetoresistance with large oscillation effect in {CoFe}/{MgO}/{CoFe}(001) junctions},
	volume = {122},
	doi = {10.1063/5.0145873},
	journal = {Applied Physics Letters},
	author = {Scheike, Thomas and Wen, Zhenchao and Sukegawa, Hiroaki and Mitani, Seiji},
	year = {2023},
	pages = {112404},
}

@article{butler_spin-dependent_2001,
	title = {Spin-dependent tunneling conductance of {Fe}{\textbar}{MgO}{\textbar}{Fe} sandwiches},
	volume = {63},
	journal = {Physical Review B},
	author = {Butler, W. H. and Zhang, X.-G. and Schulthess, T. C. and MacLaren, J. M.},
	year = {2001},
	pages = {054416},
    doi = {10.1103/PhysRevB.63.054416}
}

@article{mathon_theory_2001,
	title = {Theory of tunneling magnetoresistance of an epitaxial {Fe}/{MgO}/{Fe}(001) junction},
	volume = {63},
	doi = {10.1103/PhysRevB.63.220403},
	journal = {Physical Review B},
	author = {Mathon, J. and Umerski, A.},
	year = {2001},
	pages = {220403},
}

@article{brown_thermal_1963,
	title = {Thermal {Fluctuations} of a {Single}-{Domain} {Particle}},
	volume = {130},
	doi = {10.1103/PhysRev.130.1677},
	journal = {Physical Review},
	author = {Brown, William Fuller},
	year = {1963},
	pages = {1677--1686},
}

@article{mangin_current-induced_2006,
	title = {Current-induced magnetization reversal in nanopillars with perpendicular anisotropy},
	volume = {5},
	doi = {10.1038/nmat1595},
	journal = {Nature Materials},
	author = {Mangin, S. and Ravelosona, D. and Katine, J. A. and Carey, M. J. and Terris, B. D. and Fullerton, Eric E.},
	year = {2006},
	pages = {210--215},
}

@article{nistor_ptcooxide_2009,
	title = {Pt/{Co}/oxide and oxide/{Co}/{Pt} electrodes for perpendicular magnetic tunnel junctions},
	volume = {94},
	doi = {10.1063/1.3064162},
	journal = {Applied Physics Letters},
	author = {Nistor, L. E. and Rodmacq, B. and Auffret, S. and Dieny, B.},
	year = {2009},
	pages = {012512},
}

@article{mizunuma_mgo_2009,
	title = {{MgO} barrier-perpendicular magnetic tunnel junctions with {CoFe/Pd} multilayers and ferromagnetic insertion layers},
	volume = {95},
	doi = {10.1063/1.3265740},
	journal = {Applied Physics Letters},
	author = {Mizunuma, K. and Ikeda, S. and Park, J. H. and Yamamoto, H. and Gan, H. and Miura, K. and Hasegawa, H. and Hayakawa, J. and Matsukura, F. and Ohno, H.},
	year = {2009},
	pages = {232516},
}

@article{ikeda_perpendicular-anisotropy_2010,
	title = {A perpendicular-anisotropy {CoFeB}–{MgO} magnetic tunnel junction},
	volume = {9},
	doi = {10.1038/nmat2804},
	journal = {Nature Materials},
	author = {Ikeda, S. and Miura, K. and Yamamoto, H. and Mizunuma, K. and Gan, H. D. and Endo, M. and Kanai, S. and Hayakawa, J. and Matsukura, F. and Ohno, H.},
	year = {2010},
	pages = {721--724},
}

@article{worledge_spin_2011,
	title = {Spin torque switching of perpendicular {Ta/CoFeB/MgO}-based magnetic tunnel junctions},
	volume = {98},
	doi = {10.1063/1.3536482},
	journal = {Applied Physics Letters},
	author = {Worledge, D. C. and Hu, G. and Abraham, David W. and Sun, J. Z. and Trouilloud, P. L. and Nowak, J. and Brown, S. and Gaidis, M. C. and O’Sullivan, E. J. and Robertazzi, R. P.},
	year = {2011},
	pages = {022501},
}

@article{yang_first-principles_2011,
	title = {First-principles investigation of the very large perpendicular magnetic anisotropy at {Fe/MgO} and {Co/MgO} interfaces},
	volume = {84},
	doi = {10.1103/PhysRevB.84.054401},
	journal = {Physical Review B},
	author = {Yang, H. X. and Chshiev, M. and Dieny, B. and Lee, J. H. and Manchon, A. and Shin, K. H.},
	year = {2011},
	pages = {054401},
}

@article{baumann_origin_2015,
	title = {Origin of {Perpendicular} {Magnetic} {Anisotropy} and {Large} {Orbital} {Moment} in {Fe} {Atoms} on {MgO}},
	volume = {115},
	doi = {10.1103/PhysRevLett.115.237202},
	journal = {Physical Review Letters},
	author = {Baumann, S. and Donati, F. and Stepanow, S. and Rusponi, S. and Paul, W. and Gangopadhyay, S. and Rau, I. G. and Pacchioni, G. E. and Gragnaniello, L. and Pivetta, M. and Dreiser, J. and Piamonteze, C. and Lutz, C. P. and Macfarlane, R. M. and Jones, B. A. and Gambardella, P. and Heinrich, A. J. and Brune, H.},
	year = {2015},
	pages = {237202},
}

@article{de_person_magnetic_2007,
	title = {Magnetic coupling between high magnetization perpendicular electrodes in an epitaxial {FePt/MgO/FePt} magnetic tunnel junction},
	volume = {76},
	doi = {10.1103/PhysRevB.76.184402},
	journal = {Physical Review B},
	author = {de Person, P. and Warin, P. and Jamet, M. and Beigne, C. and Samson, Y.},
	year = {2007},
	pages = {184402},
}

@article{barabash_first-principles_2009,
	title = {First-principles determination of low-temperature order and ground states of {Fe}-{Ni}, {Fe}-{Pd}, and {Fe}-{Pt}},
	volume = {80},
	doi = {10.1103/PhysRevB.80.220201},
	journal = {Physical Review B},
	author = {Barabash, Sergey V. and Chepulskii, Roman V. and Blum, Volker and Zunger, Alex},
	year = {2009},
	pages = {220201},
}

@article{chaves-oflynn_thermal_2015,
	title = {Thermal {Stability} of {Magnetic} {States} in {Circular} {Thin}-{Film} {Nanomagnets} with {Large} {Perpendicular} {Magnetic} {Anisotropy}},
	volume = {4},
	doi = {10.1103/PhysRevApplied.4.024010},
	journal = {Physical Review Applied},
	author = {Chaves-O’Flynn, Gabriel D. and Wolf, Georg and Sun, Jonathan Z. and Kent, Andrew D.},
	year = {2015},
	pages = {024010},
}

@article{bouquin_size_2018,
	title = {Size dependence of spin-torque switching in perpendicular magnetic tunnel junctions},
	volume = {113},
	doi = {10.1063/1.5055741},
	journal = {Applied Physics Letters},
	author = {Bouquin, Paul and Rao, Siddharth and Kar, Gouri Sankar and Devolder, Thibaut},
	year = {2018},
	pages = {222408},
}

@article{slonczewski_current-driven_1996,
	title = {Current-driven excitation of magnetic multilayers},
	volume = {159},
	doi = {10.1016/0304-8853(96)00062-5},
	journal = {Journal of Magnetism and Magnetic Materials},
	author = {Slonczewski, J. C.},
	year = {1996},
	pages = {L1--L7},
}

@article{berger_emission_1996,
	title = {Emission of spin waves by a magnetic multilayer traversed by a current},
	volume = {54},
	doi = {10.1103/PhysRevB.54.9353},
	journal = {Physical Review B},
	author = {Berger, L.},
	year = {1996},
	pages = {9353--9358},
}

@article{ralph_spin_2008,
	title = {Spin transfer torques},
	volume = {320},
	number = {7},
	journal = {Journal of Magnetism and Magnetic Materials},
	author = {Ralph, D. C. and Stiles, M. D.},
	year = {2008},
	pages = {1190--1216},
    doi = {10.1016/j.jmmm.2007.12.019}
}

@article{oh_bias-voltage_2009,
	title = {Bias-voltage dependence of perpendicular spin-transfer torque in asymmetric {MgO}-based magnetic tunnel junctions},
	volume = {5},
	doi = {10.1038/nphys1427},
	journal = {Nature Physics},
	author = {Oh, {Se-Chung} and Park, {Seung-Young} and Manchon, Aurélien and Chshiev, Mairbek and Han, Jae-Ho and Lee, Hyun-Woo and Lee, Jang-Eun and Nam, Kyung-Tae and Jo, Younghun and Kong, Yo-Chan and Dieny, Bernard and Lee, Kyung-Jin},
	year = {2009},
	pages = {898--902},
}

@article{timopheev_respective_2015,
	title = {Respective influence of in-plane and out-of-plane spin-transfer torques in magnetization switching of perpendicular magnetic tunnel junctions},
	volume = {92},
	doi = {10.1103/PhysRevB.92.104430},
	journal = {Physical Review B},
	author = {Timopheev, A. A. and Sousa, R. and Chshiev, M. and Buda-Prejbeanu, L. D. and Dieny, B.},
	year = {2015},
	pages = {104430},
}

@article{thomas_perpendicular_2014,
	title = {Perpendicular spin transfer torque magnetic random access memories with high spin torque efficiency and thermal stability for embedded applications (invited)},
	volume = {115},
	doi = {10.1063/1.4870917},
	journal = {Journal of Applied Physics},
	author = {Thomas, Luc and Jan, Guenole and Zhu, Jian and Liu, Huanlong and Lee, Yuan-Jen and Le, Son and Tong, Ru-Ying and Pi, Keyu and Wang, Yu-Jen and Shen, Dongna and He, Renren and Haq, Jesmin and Teng, Jeffrey and Lam, Vinh and Huang, Kenlin and Zhong, Tom and Torng, Terry and Wang, Po-Kang},
	year = {2014},
	pages = {172615},
}

@article{sato_magnetic_2017,
	title = {Magnetic tunnel junctions with perpendicular easy axis at junction diameter of less than 20 nm},
	volume = {56},
	doi = {10.7567/JJAP.56.0802A6},
	number = {8},
	journal = {Japanese Journal of Applied Physics},
	author = {Sato, Hideo and Ikeda, Shoji and Ohno, Hideo},
	year = {2017},
	pages = {0802A6},
}

@article{yoshida_size_2019,
	title = {Size dependence of the thermal stability factor in a perpendicular {CoFeB}/{MgO} magnetic tunnel junction studied by micromagnetic simulations},
	volume = {58},
	doi = {10.7567/1347-4065/aafd92},
	language = {en},
	number = {SB},
	journal = {Japanese Journal of Applied Physics},
	author = {Yoshida, Chikako and Tanaka, Tomohiro and Ataka, Tadashi and Fujisaki, Jun and Shimizu, Koichi and Hirahara, Takao and Shitara, Hideyuki and Furuya, Atsushi and Uehara, Yuji},
	year = {2019},
	pages = {SBBB05},
}

@article{sato_perpendicular-anisotropy_2012,
	title = {Perpendicular-anisotropy {CoFeB}-{MgO} magnetic tunnel junctions with a {MgO}/{CoFeB}/{Ta}/{CoFeB}/{MgO} recording structure},
	volume = {101},
	doi = {10.1063/1.4736727},
	journal = {Applied Physics Letters},
	author = {Sato, H. and Yamanouchi, M. and Ikeda, S. and Fukami, S. and Matsukura, F. and Ohno, H.},
	year = {2012},
	pages = {022414},
}

@article{beleggia_computation_2003,
	title = {On the computation of the demagnetization tensor field for an arbitrary particle shape using a {Fourier} space approach},
	volume = {263},
	doi = {10.1016/S0304-8853(03)00238-5},
	journal = {Journal of Magnetism and Magnetic Materials},
	author = {Beleggia, M. and De Graef, M.},
	year = {2003},
	pages = {L1--L9},
}

@article{tandon_computation_2004,
	title = {On the computation of the demagnetization tensor for uniformly magnetized particles of arbitrary shape. {Part} {I}: {Analytical} approach},
	volume = {271},
	doi = {10.1016/j.jmmm.2003.09.011},
	journal = {Journal of Magnetism and Magnetic Materials},
	author = {Tandon, S. and Beleggia, M. and Zhu, Y. and De Graef, M.},
	year = {2004},
	pages = {9--26},
}

@article{beleggia_general_2005,
	title = {General magnetostatic shape–shape interactions},
	volume = {285},
	doi = {10.1016/j.jmmm.2004.09.004},
	journal = {Journal of Magnetism and Magnetic Materials},
	author = {Beleggia, M. and De Graef, M.},
	year = {2005},
	pages = {L1--L10},
}

@article{beleggia_demagnetization_2005,
	title = {Demagnetization factors for elliptic cylinders},
	volume = {38},
	doi = {10.1088/0022-3727/38/18/001},
	journal = {Journal of Physics D: Applied Physics},
	author = {Beleggia, M. and Graef, M. De and Millev, Y. T. and Goode, D. A. and Rowlands, G.},
	year = {2005},
	pages = {3333},
}

@article{watanabe_shape_2018,
	title = {Shape anisotropy revisited in single-digit nanometer magnetic tunnel junctions},
	volume = {9},
	doi = {10.1038/s41467-018-03003-7},
	journal = {Nature Communications},
	author = {Watanabe, K. and Jinnai, B. and Fukami, S. and Sato, H. and Ohno, H.},
	year = {2018},
	pages = {663},
}

@article{perrissin_highly_2018,
	title = {A highly thermally stable sub-20 nm magnetic random-access memory based on perpendicular shape anisotropy},
	volume = {10},
	doi = {10.1039/C8NR01365A},
	journal = {Nanoscale},
	author = {Perrissin, N. and Lequeux, S. and Strelkov, N. and Chavent, A. and Vila, L. and Buda-Prejbeanu, L. D. and Auffret, S. and Sousa, R. C. and Prejbeanu, I. L. and Dieny, B.},
	year = {2018},
	pages = {12187--12195},
}

@article{brown_criterion_1957,
	title = {Criterion for {Uniform} {Micromagnetization}},
	volume = {105},
	doi = {10.1103/PhysRev.105.1479},
	journal = {Physical Review},
	author = {Brown, William Fuller},
	month = mar,
	year = {1957},
	pages = {1479--1482},
}

@article{buda_micromagnetic_2002,
	title = {Micromagnetic simulations of magnetisation in circular cobalt dots},
	volume = {24},
	doi = {10.1016/S0927-0256(02)00184-2},
	journal = {Computational Materials Science},
	author = {Buda, L. D. and Prejbeanu, I. L. and Ebels, U. and Ounadjela, K.},
	year = {2002},
	pages = {181--185},
}

@article{abo_definition_2013,
	title = {Definition of {Magnetic} {Exchange} {Length}},
	volume = {49},
	doi = {10.1109/TMAG.2013.2258028},
	journal = {IEEE Transactions on Magnetics},
	author = {Abo, Gavin S. and Hong, Yang-Ki and Park, Jihoon and Lee, Jaejin and Lee, Woncheol and Choi, Byoung-Chul},
	year = {2013},
	pages = {4937--4939},
}

@incollection{stano_magnetic_2018,
	title = {Magnetic nanowires and nanotubes},
	volume = {27},
	author = {Stano, Michal and Fruchart, Olivier},
	year = {2018},
	doi = {10.1016/bs.hmm.2018.08.002},
	pages = {155--267},
    booktitle = {Handbook of magnetic materials},
    publisher = {Elseveir}
}

@article{forster_energy_2003,
	title = {Energy barrier and effective thermal reversal volume in columnar grains},
	volume = {267},
	doi = {10.1016/S0304-8853(03)00306-8},
	journal = {Journal of Magnetism and Magnetic Materials},
	author = {Forster, Hermann and Bertram, Neal and Wang, Xiaobin and Dittrich, Rok and Schrefl, Thomas},
	year = {2003},
	pages = {69--79},
}

@article{e_energy_2003,
	title = {Energy landscape and thermally activated switching of submicron-sized ferromagnetic elements},
	volume = {93},
	doi = {10.1063/1.1536737},
	journal = {Journal of Applied Physics},
	author = {E, Weinan and Ren, Weiqing and Vanden-Eijnden, Eric},
	year = {2003},
	pages = {2275--2282},
}

@article{chaves-oflynn_stability_2010,
	title = {Stability of {$2\pi$} {Domain} {Walls} in {Ferromagnetic} {Nanorings}},
	volume = {46},
	doi = {10.1109/TMAG.2010.2045484},
	journal = {IEEE Transactions on Magnetics},
	author = {Chaves-O'Flynn, G. D. and Bedau, D. and Vanden-Eijnden, E. and Kent, A. D. and Stein, D. L.},
	year = {2010},
	pages = {2272--2274},
}

@article{carilli_truncation-based_2015,
	title = {Truncation-based energy weighting string method for efficiently resolving small energy barriers},
	volume = {143},
	doi = {10.1063/1.4927580},
	journal = {The Journal of Chemical Physics},
	author = {Carilli, Michael F. and Delaney, Kris T. and Fredrickson, Glenn H.},
	year = {2015},
	pages = {054105},
}

@article{grollier_magnetic_2011,
	title = {Magnetic domain wall motion by spin transfer},
	volume = {12},
	doi = {10.1016/j.crhy.2011.03.007},
	journal = {Comptes Rendus Physique},
	author = {Grollier, Julie and Chanthbouala, A. and Matsumoto, R. and Anane, A. and Cros, V. and Nguyen van Dau, F. and Fert, Albert},
	year = {2011},
	pages = {309--317},
}

@article{chshiev_analytical_2015,
	title = {Analytical description of ballistic spin currents and torques in magnetic tunnel junctions},
	volume = {92},
	doi = {10.1103/PhysRevB.92.104422},
	journal = {Physical Review B},
	author = {Chshiev, M. and Manchon, A. and Kalitsov, A. and Ryzhanova, N. and Vedyayev, A. and Strelkov, N. and Butler, W. H. and Dieny, B.},
	year = {2015},
	pages = {104422},
}

@article{thiaville_micromagnetic_2005,
	title = {Micromagnetic understanding of current-driven domain wall motion in patterned nanowires},
	volume = {69},
	doi = {10.1209/epl/i2004-10452-6},
	journal = {EPL (Europhysics Letters)},
	author = {Thiaville, A. and Nakatani, Y. and Miltat, J. and Suzuki, Y.},
	year = {2005},
	pages = {990},
}

@article{cacoilo_spin-torque-triggered_2021,
	title = {Spin-{Torque}-{Triggered} {Magnetization} {Reversal} in {Magnetic} {Tunnel} {Junctions} with {Perpendicular} {Shape} {Anisotropy}},
	volume = {16},
	doi = {10.1103/PhysRevApplied.16.024020},
	journal = {Physical Review Applied},
	author = {Caçoilo, N. and Lequeux, S. and Teixeira, B.M.S. and Dieny, B. and Sousa, R.C. and Sobolev, N.A. and Fruchart, O. and Prejbeanu, I.L. and Buda-Prejbeanu, L.D.},
	year = {2021},
	pages = {024020},
}

@inproceedings{sato_comprehensive_2013,
	title = {Comprehensive study of {CoFeB}-{MgO} magnetic tunnel junction characteristics with single- and double-interface scaling down to {1X} nm},
	doi = {10.1109/IEDM.2013.6724550},
	booktitle = {{IEEE} {International} {Electron} {Devices} {Meeting}},
	author = {Sato, H. and Yamamoto, T. and Yamanouchi, M. and Ikeda, S. and Fukami, S. and Kinoshita, K. and Matsukura, F. and Kasai, N. and Ohno, H.},
	year = {2013},
	pages = {3.2.1--3.2.4},
}

@article{sun_spin-current_2000,
	title = {Spin-current interaction with a monodomain magnetic body: {A} model study},
	volume = {62},
	doi = {10.1103/PhysRevB.62.570},
	journal = {Physical Review B},
	author = {Sun, J. Z.},
	year = {2000},
	pages = {570--578},
}

@article{koch_time-resolved_2004,
	title = {Time-{Resolved} {Reversal} of {Spin}-{Transfer} {Switching} in a {Nanomagnet}},
	volume = {92},
	doi = {10.1103/PhysRevLett.92.088302},
	journal = {Physical Review Letters},
	author = {Koch, R. H. and Katine, J. A. and Sun, J. Z.},
	year = {2004},
	pages = {088302},
}

@article{garello_ultrafast_2014,
	title = {Ultrafast magnetization switching by spin-orbit torques},
	volume = {105},
	doi = {10.1063/1.4902443},
	journal = {Applied Physics Letters},
	author = {Garello, Kevin and Avci, Can Onur and Miron, Ioan Mihai and Baumgartner, Manuel and Ghosh, Abhijit and Auffret, Stéphane and Boulle, Olivier and Gaudin, Gilles and Gambardella, Pietro},
	year = {2014},
	pages = {212402},
}

@article{worledge_theory_2017,
	title = {Theory of {Spin} {Torque} {Switching} {Current} for the {Double} {Magnetic} {Tunnel} {Junction}},
	volume = {8},
	doi = {10.1109/LMAG.2017.2707331},
	journal = {IEEE Magnetics Letters},
	author = {Worledge, Daniel C.},
	year = {2017},
	pages = {1--5},
}

@article{sun_spin-transfer_2022,
	title = {Spin-transfer torque switched magnetic tunnel junction for memory technologies},
	volume = {559},
	doi = {10.1016/j.jmmm.2022.169479},
	journal = {Journal of Magnetism and Magnetic Materials},
	author = {Sun, Jonathan Z.},
	year = {2022},
	pages = {169479},
}

@article{bedau_spin-transfer_2010,
	title = {Spin-transfer pulse switching: {From} the dynamic to the thermally activated regime},
	volume = {97},
	doi = {10.1063/1.3532960},
	journal = {Applied Physics Letters},
	author = {Bedau, D. and Liu, H. and Sun, J. Z. and Katine, J. A. and Fullerton, E. E. and Mangin, S. and Kent, A. D.},
	year = {2010},
	pages = {262502},
}

@article{perrissin_perpendicular_2019,
	title = {Perpendicular shape anisotropy spin transfer torque magnetic random-access memory: towards sub-10 nm devices},
	volume = {52},
	doi = {10.1088/1361-6463/ab0de4},
	journal = {Journal of Physics D: Applied Physics},
	author = {Perrissin, N. and Gregoire, G. and Lequeux, S. and Tillie, L. and Strelkov, N. and Auffret, S. and Buda-Prejbeanu, L. D. and Sousa, R. C. and Vila, L. and Dieny, B. and Prejbeanu, I. L.},
	year = {2019},
	pages = {234001},
}

@article{takeuchi_nanometer-thin_2022,
	title = {Nanometer-thin {L1}-{MnAl} film with {B2}-{CoAl} underlayer for high-speed and high-density {STT}-{MRAM}: {Structure} and magnetic properties},
	volume = {120},
	doi = {10.1063/5.0077874},
	journal = {Applied Physics Letters},
	author = {Takeuchi, Yutaro and Okuda, Ryotaro and Igarashi, Junta and Jinnai, Butsurin and Saino, Takaharu and Ikeda, Shoji and Fukami, Shunsuke and Ohno, Hideo},
	year = {2022},
	pages = {052404},
}

@inproceedings{nishioka_novel_2019,
	title = {Novel {Quad} interface {MTJ} technology and its first demonstration with high thermal stability and switching efficiency for {STT}-{MRAM} beyond {2Xnm}},
	doi = {10.23919/VLSIT.2019.8776499},
	booktitle = {{Symposium} on {VLSI} {Technology}},
	author = {Nishioka, K. and Honjo, H. and Ikeda, S. and Watanabe, T. and Miura, S. and Inoue, H. and Tanigawa, T. and Noguchi, Y. and Yasuhira, M. and Sato, H. and Endoh, T.},
	year = {2019},
	pages = {T120--T121},
}

@inproceedings{jinnai_high-performance_2020,
	title = {High-{Performance} {Shape}-{Anisotropy} {Magnetic} {Tunnel} {Junctions} down to 2.3 nm},
	doi = {10.1109/IEDM13553.2020.9371972},
	booktitle = {{IEEE} {International} {Electron} {Devices} {Meeting} ({IEDM})},
	author = {Jinnai, B. and Igarashi, J. and Watanabe, K. and Funatsu, T. and Sato, H. and Fukami, S. and Ohno, H.},
	year = {2020},
	pages = {24.6.1--24.6.4},
}

@article{nishioka_enhancement_2021,
	title = {Enhancement of magnetic coupling and magnetic anisotropy in {MTJs} with multiple {CoFeB}/{MgO} interfaces for high thermal stability},
	volume = {11},
	doi = {10.1063/9.0000048},
	journal = {AIP Advances},
	author = {Nishioka, K. and Honjo, H. and Naganuma, H. and Nguyen, T. V. A. and Yasuhira, M. and Ikeda, S. and Endoh, T.},
	year = {2021},
	pages = {025231},
}

@inproceedings{kim_integration_2011,
	title = {Integration of 28nm {MTJ} for 8{--16Gb} level {MRAM} with full investigation of thermal stability},
	booktitle = {{Symposium} on {VLSI} {Technology} - {Digest} of {Technical} {Papers}},
	author = {Kim, Y. and Oh, S. C. and Lim, W. C. and Kim, J. H. and Kim, W. J. and Jeong, J. H. and Shin, H. J. and Kim, K. W. and Kim, K. S. and Park, J. H. and Park, S. H. and Kwon, H. and Ah, K.H. and Lee, J. E. and Park, S. O. and Choi, S. and Kang, H. K. and Chung, C.},
	year = {2011},
	pages = {210--211},
}

@inproceedings{sato_1t-1mtj_2018,
	title = {{1T}-{1MTJ} {Type} {Embedded} {STT}-{MRAM} with {Advanced} {Low}-{Damage} and {Short}-{Failure}-{Free} {RIE} {Technology} down to 32 nm$\phi$ {MTJ} {Patterning}},
	doi = {10.1109/IMW.2018.8388774},
	booktitle = {{IEEE} {International} {Memory} {Workshop}},
	author = {Sato, Hideo and Watanabe, Toshinari and Koike, Hiroki and Saito, Takashi and Miura, Sadahiko and Honjo, Hiroaki and Inoue, Hirofumi and Ikeda, Shoji and Noguchi, Yasuo and Tanigawa, Takaho and Yasuhira, Mitsuo and Ohno, Hideo and Kang, Song Yun and Kubo, Takuya and Takatsuki, Koichi and Yamashita, Koji and Yagi, Yasushi and Tamura, Ryo and Nishimura, Takuro and Murata, Koh and Endoh, Tetsuo},
	year = {2018},
	pages = {1--4},
}

@inproceedings{nguyen_novel_2017,
	title = {Novel approach for nano-patterning magnetic tunnel junctions stacks at narrow pitch: {A} route towards high density {STT}-{MRAM} applications},
	doi = {10.1109/IEDM.2017.8268517},
	booktitle = {{IEEE} {International} {Electron} {Devices} {Meeting} ({IEDM})},
	author = {Nguyen, V. D. and Sabon, P. and Chatterjee, J. and Tille, L. and Coelho, P. Veloso and Auffret, S. and Sousa, R. and Prejbeanu, L. and Gautier, E. and Vila, L. and Dieny, B.},
	year = {2017},
	pages = {38.5.1--38.5.4},
}

@article{igarashi_magnetic-field-angle_2017,
	title = {Magnetic-field-angle dependence of coercivity in {CoFeB}/{MgO} magnetic tunnel junctions with perpendicular easy axis},
	volume = {111},
	doi = {10.1063/1.5004968},
	journal = {Applied Physics Letters},
	author = {Igarashi, J. and Llandro, J. and Sato, H. and Matsukura, F. and Ohno, H.},
	year = {2017},
	pages = {132407},
}

@article{jinnai_coherent_2021,
	title = {Coherent magnetization reversal of a cylindrical nanomagnet in shape-anisotropy magnetic tunnel junctions},
	volume = {118},
	doi = {10.1063/5.0043058},
	journal = {Applied Physics Letters},
	author = {Jinnai, Butsurin and Igarashi, Junta and Watanabe, Kyota and Enobio, Eli Christopher I. and Fukami, Shunsuke and Ohno, Hideo},
	year = {2021},
	pages = {082404},
}

@inproceedings{lequeux_psa-stt-mram_2021,
	title = {{PSA}-{STT}-{MRAM} solution for extended temperature stability},
	doi = {10.1109/IMW51353.2021.9439609},
	booktitle = {{IEEE} {International} {Memory} {Workshop}},
	author = {Lequeux, Steven and Almeida, Trevor and Caçoilo, Nuno and Palomino, Alvaro and Prejbeanu, Ioan Lucian and Sousa, Ricardo C. and Cooper, David and Dieny, Bernard},
	year = {2021},
	pages = {1--4},
}

@article{igarashi_temperature_2021,
	title = {Temperature dependence of the energy barrier in {X}/{1X} nm shape-anisotropy magnetic tunnel junctions},
	volume = {118},
	doi = {10.1063/5.0029031},
	journal = {Applied Physics Letters},
	author = {Igarashi, Junta and Jinnai, Butsurin and Desbuis, Valentin and Mangin, Stéphane and Fukami, Shunsuke and Ohno, Hideo},
	year = {2021},
	pages = {012409},
}

@article{lequeux_thermal_2020,
	title = {Thermal robustness of magnetic tunnel junctions with perpendicular shape anisotropy},
	volume = {12},
	doi = {10.1039/C9NR10366J},
	journal = {Nanoscale},
	author = {Lequeux, S. and Perrissin, N. and Grégoire, G. and Tillie, L. and Chavent, A. and Strelkov, N. and Vila, L. and Buda-Prejbeanu, L. D. and Auffret, S. and Sousa, R. C. and Prejbeanu, I. L. and Russo, E. Di and Gautier, E. and Conlan, A. P. and Cooper, D. and Dieny, B.},
	year = {2020},
	pages = {6378--6384},
}

@article{almeida_direct_2021,
	title = {Direct observation of the perpendicular shape anisotropy and thermal stability of {STT}-{MRAM} nano-pillars examined by off-axis electron holography},
	volume = {27},
	doi = {10.1017/S1431927621007819},
	journal = {Microscopy and Microanalysis},
	author = {Almeida, Trevor and Lequeux, Steven and Palomino, Alvaro and Caçoilo, Nuno and Massebouef, Aurélien and Sousa, Richard and Fruchart, Olivier and Prejbeanu, Ioan-Lucian and Dieny, Bernard and Cooper, David},
	year = {2021},
	pages = {2170--2172},
}

@article{almeida_quantitative_2022,
	title = {Quantitative {Visualization} of {Thermally} {Enhanced} {Perpendicular} {Shape} {Anisotropy} {STT}-{MRAM} {Nanopillars}},
	volume = {22},
	doi = {10.1021/acs.nanolett.2c00597},
	journal = {Nano Letters},
	author = {Almeida, Trevor P. and Lequeux, Steven and Palomino, Alvaro and Sousa, Ricardo C. and Fruchart, Olivier and Prejbeanu, Ioan-Lucian and Dieny, Bernard and Masseboeuf, Aurélien and Cooper, David},
	year = {2022},
	pages = {4000--4005},
}

@article{almeida_off-axis_2022,
	title = {Off-axis electron holography for the direct visualization of perpendicular shape anisotropy in nanoscale {3D} magnetic random-access-memory devices},
	volume = {10},
	doi = {10.1063/5.0096761},
	journal = {APL Materials},
	author = {Almeida, Trevor P. and Palomino, Alvaro and Lequeux, Steven and Boureau, Victor and Fruchart, Olivier and Prejbeanu, Ioan Lucian and Dieny, Bernard and Cooper, David},
	year = {2022},
	pages = {061104},
}

@article{sato_temperature-dependent_2018,
	title = {Temperature-dependent properties of {CoFeB}/{MgO} thin films: {Experiments} versus simulations},
	volume = {98},
	doi = {10.1103/PhysRevB.98.214428},
	journal = {Physical Review B},
	author = {Sato, H. and Chureemart, P. and Matsukura, F. and Chantrell, R. W. and Ohno, H. and Evans, R. F. L.},
	year = {2018},
	pages = {214428},
}

@article{callen_present_1966,
	title = {The present status of the temperature dependence of magnetocrystalline anisotropy, and the l(l+1)2 power law},
	volume = {27},
	doi = {10.1016/0022-3697(66)90012-6},
	journal = {Journal of Physics and Chemistry of Solids},
	author = {Callen, H. B. and Callen, E.},
	year = {1966},
	pages = {1271--1285},
}

@article{enobio_evaluation_2018,
	title = {Evaluation of energy barrier of {CoFeB}/{MgO} magnetic tunnel junctions with perpendicular easy axis using retention time measurement},
	volume = {57},
	doi = {10.7567/JJAP.57.04FN08},
	journal = {Japanese Journal of Applied Physics},
	author = {Enobio, Eli Christopher Inocencio and Bersweiler, Mathias and Sato, Hideo and Fukami, Shunsuke and Ohno, Hideo},
	year = {2018},
	pages = {04FN08},
}

@article{veiga_control_2023,
	title = {Control of interface anisotropy for spin transfer torque in perpendicular magnetic tunnel junctions for cryogenic temperature operation},
	volume = {13},
	doi = {10.1063/9.0000512},
	journal = {AIP Advances},
	author = {Veiga, P. B. and Mora-Hernandez, A. and Dammak, M. and Auffret, S. and Joumard, I. and Vila, L. and Buda-Prejbeanu, Liliana D. and Prejbeanu, I. L. and Dieny, B. and Sousa, R. C.},
	year = {2023},
	pages = {025253},
}

@inproceedings{wu_impact_2020,
	title = {Impact of {Magnetic} {Coupling} and {Density} on {STT}-{MRAM} {Performance}},
	doi = {10.23919/DATE48585.2020.9116444},
	booktitle = {2020 {Design}, {Automation} {Test} in {Europe} {Conference} {Exhibition} ({DATE})},
	author = {Wu, L. and Rao, S. and Taouil, M. and Marinissen, E. J. and Kar, G. Sankar and Hamdioui, S.},
	year = {2020},
	pages = {1211--1216},
}

@article{fruchart_high_1998,
	title = {High coercivity in ultrathin epitaxial micrometer-sized particles with in-plane magnetization: {Experiment} and numerical simulation},
	volume = {57},
	doi = {10.1103/PhysRevB.57.2596},
	journal = {Physical Review B},
	author = {Fruchart, O. and Nozières, J.-P. and Kevorkian, B. and Toussaint, J.-C. and Givord, D. and Rousseaux, F. and Decanini, D. and Carcenac, F.},
	year = {1998},
	pages = {2596--2606},
}

@article{taniguchi_analytical_2018,
	title = {An analytical computation of magnetic field generated from a cylinder ferromagnet},
	volume = {452},
	doi = {10.1016/j.jmmm.2017.11.078},
	journal = {Journal of Magnetism and Magnetic Materials},
	author = {Taniguchi, Tomohiro},
	year = {2018},
	pages = {464--472},
}

@article{cacoilo_dipole-coupled_2024,
	title = {Dipole-coupled core-shell perpendicular-shape-anisotropy magnetic tunnel junction with enhanced write speed and reduced crosstalk},
	volume = {21},
	doi = {10.1103/PhysRevApplied.21.044034},
	journal = {Physical Review Applied},
	author = {Caçoilo, N. and Buda-Prejbeanu, L.D. and Dieny, B. and Fruchart, O. and Prejbeanu, I.L.},
	year = {2024},
	pages = {044034},
}

@article{khvalkovskiy_basic_2013,
	title = {Basic principles of {STT}-{MRAM} cell operation in memory arrays},
	volume = {46},
	doi = {10.1088/0022-3727/46/7/074001},
	journal = {Journal of Physics D: Applied Physics},
	author = {Khvalkovskiy, A. V. and Apalkov, D. and Watts, S. and Chepulskii, R. and Beach, R. S. and Ong, A. and Tang, X. and Driskill-Smith, A. and Butler, W. H. and Visscher, P. B. and Lottis, D. and Chen, E. and Nikitin, V. and Krounbi, M.},
	year = {2013},
	pages = {074001},
}

@inproceedings{jinnai_fast_2021,
	title = {Fast {Switching} {Down} to 3.5 ns in {Sub}-5-nm {Magnetic} {Tunnel} {Junctions} {Achieved} by {Engineering} {Relaxation} {Time}},
	doi = {10.1109/IEDM19574.2021.9720509},
	booktitle = {{IEEE} {International} {Electron} {Devices} {Meeting} ({IEDM})},
	author = {Jinnai, B. and Igarashi, J. and Shinoda, T. and Watanabe, K. and Fukami, S. and Ohno, H.},
	year = {2021},
	pages = {1--4},
}

@article{igarashi_single-nanometer_2024,
	title = {Single-nanometer {CoFeB}/{MgO} magnetic tunnel junctions with high-retention and high-speed capabilities},
	volume = {2},
	doi = {10.1038/s44306-023-00003-2},
	journal = {npj Spintronics},
	author = {Igarashi, Junta and Jinnai, Butsurin and Watanabe, Kyota and Shinoda, Takanobu and Funatsu, Takuya and Sato, Hideo and Fukami, Shunsuke and Ohno, Hideo},
	year = {2024},
	pages = {1--9},
}

@article{nishioka_novel_2020,
	title = {Novel {Quad}-{Interface} {MTJ} {Technology} and its {First} {Demonstration} {With} {High} {Thermal} {Stability} {Factor} and {Switching} {Efficiency} for {STT}-{MRAM} {Beyond} {2X} nm},
	volume = {67},
	doi = {10.1109/TED.2020.2966731},
	journal = {IEEE Transactions on Electron Devices},
	author = {Nishioka, K. and Honjo, H. and Ikeda, S. and Watanabe, T. and Miura, S. and Inoue, H. and Tanigawa, T. and Noguchi, Y. and Yasuhira, M. and Sato, H. and Endoh, T.},
	year = {2020},
	pages = {995--1000},
}

@article{shinoda_pitch_2024,
	title = {Pitch {Scaling} {Prospect} of {Ultra}-{Small} {Magnetic} {Tunnel} {Junctions} for {High}-{Density} {STT}-{MRAM}: {Effects} of {Magnetostatic} {Interference} {From} {Neighboring} {Bits}},
	volume = {45},
	doi = {10.1109/LED.2023.3345743},
	journal = {IEEE Electron Device Letters},
	author = {Shinoda, Takanobu and Igarashi, Junta and Jinnai, Butsurin and Fukami, Shunsuke and Ohno, Hideo},
	year = {2024},
	pages = {184--187},
}

@article{Natalia_spin_accumulation,
	title = {Simulation of current-driven magnetization switching in nanopillars with perpendicular shape anisotropy},
	volume = {112},
	doi = {10.1103/dl2x-qv7t},
	journal = {Phys. Rev. B},
	author = {Meneguolo, Natalia Boscolo and Fruchart, Olivier and Toussaint, Jean-Christophe and Fattouhi, Mouad and Buda-Prejbeanu, Liliana D. and Prejbeanu, Ioan-Lucian and Gusakova, Daria},
	year = {2025},
	pages = {014448},
}

@article{sato_properties_2014,
	title = {Properties of magnetic tunnel junctions with a {MgO}/{CoFeB}/{Ta}/{CoFeB}/{MgO} recording structure down to junction diameter of 11 nm},
	volume = {105},
	doi = {10.1063/1.4892924},
	number = {6},
	journal = {Applied Physics Letters},
	author = {Sato, H. and Enobio, E. C. I. and Yamanouchi, M. and Ikeda, S. and Fukami, S. and Kanai, S. and Matsukura, F. and Ohno, H.},
	year = {2014},
	pages = {062403},
}

@book{hubert_magnetic_1998,
  author    = {Hubert, Alex and Sch{\"a}fer, Rudolf},
  title     = {Magnetic Domains: The Analysis of Magnetic Microstructures},
  publisher = {Springer},
  address   = {Berlin, Heidelberg},
  year      = {1998},
  edition   = {1},
  pages     = {696},
  isbn      = {978-3-540-64108-7},
  doi       = {10.1007/978-3-540-85054-0}
}

@article{wang_magnetic_2008,
	title = {Magnetic behavior in an ordered Co nanorod array},
	volume = {19},
	issn = {0957-4484},
	url = {https://doi.org/10.1088/0957-4484/19/45/455703},
	doi = {10.1088/0957-4484/19/45/455703},
	pages = {455703},
	number = {45},
	journaltitle = {Nanotechnology},
	shortjournal = {Nanotechnology},
	author = {Wang, T and Wang, Y and Fu, Y and Hasegawa, T and Oshima, H and Itoh, K and Nishio, K and Masuda, H and Li, F S and Saito, H and Ishio, S},
	urldate = {2026-08-03},
	date = {2008-10},
	langid = {english},
}

@article{frei_critical_1957,
	title = {Critical Size and Nucleation Field of Ideal Ferromagnetic Particles},
	volume = {106},
	url = {https://link.aps.org/doi/10.1103/PhysRev.106.446},
	doi = {10.1103/PhysRev.106.446},
	pages = {446--455},
	number = {3},
	journaltitle = {Physical Review},
	shortjournal = {Phys. Rev.},
	publisher = {American Physical Society},
	author = {Frei, E. H. and Shtrikman, S. and Treves, D.},
	urldate = {2026-08-03},
	date = {1957-05-01},
}

@article{shtrikman_coercive_1959,
	title = {The coercive force and rotational hysteresis of elongated ferromagnetic particles},
	volume = {20},
	issn = {0368-3842, 2777-3442},
	url = {http://dx.doi.org/10.1051/jphysrad:01959002002-3028600},
	doi = {10.1051/jphysrad:01959002002-3028600},
	pages = {286--289},
	number = {2},
	journaltitle = {Journal de Physique et le Radium},
	shortjournal = {J. Phys. Radium},
	publisher = {Revue Générale de l'Electricité},
	author = {Shtrikman, S. and Treves, D.},
	urldate = {2026-08-03},
	date = {1959-02-01},
	langid = {english},
}

@article{poulopoulos_1999,
  author  = {Poulopoulos, P. and Baberschke, K.},
  title   = {Magnetism in Thin Films},
  journal = {Journal of Physics: Condensed Matter},
  volume  = {11},
  number  = {48},
  pages   = {9495--9515},
  year    = {1999},
  doi     = {10.1088/0953-8984/11/48/310}
}

@article{stoner_mechanism_1948,
	title = {A mechanism of magnetic hysteresis in heterogeneous alloys},
	volume = {240},
	issn = {0080-4614},
	url = {https://doi.org/10.1098/rsta.1948.0007},
	doi = {10.1098/rsta.1948.0007},
	number = {826},
	urldate = {2026-08-12},
	journal = {Philosophical Transactions of the Royal Society of London, Series A: Mathematical and Physical Sciences},
	author = {Stoner, Edmund Clifton and Wohlfarth, E. P.},
	year = {1948},
	pages = {599--642},
}

@article{wernsdorfer_experimental_1997,
	title = {Experimental {Evidence} of the {N}{\textbackslash}'eel-{Brown} {Model} of {Magnetization} {Reversal}},
	volume = {78},
	url = {https://link.aps.org/doi/10.1103/PhysRevLett.78.1791},
	doi = {10.1103/PhysRevLett.78.1791},
	number = {9},
	urldate = {2026-08-12},
	journal = {Physical Review Letters},
	publisher = {American Physical Society},
	author = {Wernsdorfer, W. and Orozco, E. Bonet and Hasselbach, K. and Benoit, A. and Barbara, B. and Demoncy, N. and Loiseau, A. and Pascard, H. and Mailly, D.},
	month = mar,
	year = {1997},
	pages = {1791--1794},
}

\end{document}